\documentclass{jpp}
\usepackage[colorlinks=true,linkcolor=blue,citecolor=blue]{hyperref}
\usepackage{graphicx}

\usepackage[utf8]{inputenc}
\usepackage[T1]{fontenc}
\usepackage{amsmath}
\usepackage{dsfont}
\usepackage[section]{placeins}
\usepackage{multicol}

\let\originalleft\left
\let\originalright\right
\renewcommand{\left}{\mathopen{}\mathclose\bgroup\originalleft}
\renewcommand{\right}{\aftergroup\egroup\originalright}

\newcommand{\tens}[1]{\mathsfbi{#1}}
\newcommand{\vect}[1]{{\boldsymbol{#1}}}
\newcommand{\uvect}[1]{\hat{\vect{#1}}}
\newcommand{\avgR}[1]{\left\langle #1 \right\rangle_R}
\newcommand{\avgr}[1]{\left\langle #1 \right\rangle_r}
\newcommand{\avgTheta}[1]{\left\langle #1 \right\rangle}
\newcommand{\avgT}[1]{\left\langle #1 \right\rangle_{\Delta t}}
\newcommand{\partd}[2]{\frac{\partial #1}{\partial #2}}
\newcommand{\partdat}[3]{\left. \partd{#1}{#2} \right|_{#3}}
\newcommand{\pbra}[2]{\left\{ #1, #2 \right\}}

\newcommand{\krho}{\left(\vect{k}\bcdot\vect{\rho}\right)}
\newcommand{\krhonb}{\vect{k}\bcdot\vect{\rho}}
\newcommand{\hok}{h_\vect{k}^{(0)}}
\newcommand{\tempvel}{\left(\frac{v^2}{\vti^2} - \frac{3}{2}\right)}

\newcommand{\dv}{d^3\vect{v}}
\newcommand{\rhoidelperpsq} {\rho_i^2\nabla_\perp^2}
\newcommand{\pt}{\partial_t}
\newcommand{\py}{\partial_y}
\newcommand{\px}{\partial_x}
\newcommand{\vt}{\kappa_T}
\newcommand{\deltaT}{T}
\newcommand{\delsq}{\nabla^2}
\newcommand{\zf}[1]{\overline{#1}}
\newcommand{\dw}[1]{#1'}
\newcommand{\order}[1]{\mathcal{O}\left(#1\right)}

\newcommand{\kperprhoisq}{k_\perp^2\rho_i^2}
\newcommand{\kperprhoisqq}{k_\perp^4\rho_i^4}

\newcommand{\vti}{v_{ti}}
\newcommand{\rhoivti}{\rho_i \vti}
\newcommand{\exb}{\(\vect{E}\times\vect{B}\) }
\newcommand{\del}{\vect{\nabla}}
\newcommand{\phinorm}{\varphi}
\newcommand{\phinonnorm}{\phi}
\newcommand{\vd}{\vect{V_\text{\textit{D}}}}
\newcommand{\ve}{\vect{V_\text{\textit{E}}}}
\newcommand{\re}[1]{{\text{Re}{#1}}}
\newcommand{\im}[1]{{\text{Im}{#1}}}
\newcommand{\vtcrit}{\vt^c}
\newcommand{\vtstatic}{\vt^\text{static}}
\newcommand{\vtsec}{\vt^\text{sec}}
\newcommand{\vk}{\vect{k}}
\newcommand{\circled}[1]{{\raisebox{.5pt}{\textcircled{\raisebox{-.9pt} {#1}}}}}
\newcommand{\intr}{{\int d^3 \vect{r} \ }}
\newcommand{\intR}{{\int d^3 \vect{R} \ }}
\newcommand{\intv}{{\int d^3 \vect{v} \ }}
\newcommand{\intrv}{{\int d^3 \vect{r}\int d^3 \vect{v} \ }}
\newcommand{\shf}[1]{\Tilde{#1}}

\shorttitle{Dimits regime of curvature-driven ITG turbulence}
\shortauthor{P. G. Ivanov et al.}

\title{Zonally dominated dynamics and Dimits threshold in curvature-driven ITG turbulence}

\author{Plamen~G.~Ivanov\aff{1,2,3}
  \corresp{\email{plamen.ivanov@physics.ox.ac.uk}},
  A.~A.~Schekochihin\aff{1,4},
  W.~Dorland\aff{1,5},
  A.~R.~Field\aff{3},
 \and F.~I.~Parra\aff{1,6}}

\affiliation{\aff{1}Rudolf Peierls Centre for Theoretical Physics, University of Oxford, Oxford OX1 3PU, UK
\aff{2}St John's College, Oxford OX1 3JP, UK
\aff{3}EURATOM/UKAEA Fusion Association, Culham Science Centre, Abingdon, OX14 3DB, UK
\aff{4}Merton College, Oxford OX1 4JD, UK
\aff{5}Department of Physics, University of Maryland, College Park, Maryland 20740, USA
\aff{6}Worcester College, Oxford OX1 2HB, UK}

\begin{document}

\maketitle

\begin{abstract}
	
The saturated state of turbulence driven by the ion-temperature-gradient instability is investigated using a two-dimensional long-wavelength fluid model that describes the perturbed electrostatic potential and perturbed ion temperature in a magnetic field with constant curvature (a \(Z\)-pinch) and an equilibrium temperature gradient. Numerical simulations reveal a well-defined transition between a finite-amplitude saturated state dominated by strong zonal-flow and zonal-temperature perturbations, and a blow-up state that fails to saturate on a box-independent scale. We argue that this transition is equivalent to the Dimits transition from a low-transport to a high-transport state seen in gyrokinetic numerical simulations \citep{dimits2000}. A quasi-static staircase-like structure of the temperature gradient intertwined with zonal flows, which have patch-wise constant shear, emerges near the Dimits threshold. The turbulent heat flux in the low-collisionality near-marginal state is dominated by turbulent bursts, triggered by coherent long-lived structures closely resembling those found in gyrokinetic simulations with imposed equilibrium flow shear \citep{vanwyk2016}. The break up of the low-transport Dimits regime is linked to a competition between the two different sources of poloidal momentum in the system --- the Reynolds stress and the advection of the diamagnetic flow by the \exb flow. By analysing the linear ITG modes, we obtain a semi-analytic model for the Dimits threshold at large collisionality.

\end{abstract}

\section{Introduction}
\label{intro}

Understanding the heat transport properties of magnetically confined plasmas is crucial for the design of successful tokamak experiments. Since the characteristic correlation length of the turbulence is small compared to the size of the tokamak, one normally assumes that the local heat transport depends only on local conditions, such as density, temperature, magnetic field and their gradients \citep[this view has been challenged; see][]{pradalier2010}. Existing research suggests that the dominant contribution to the heat flux in tokamaks arises from turbulence driven by microinstabilities, the most prominent of which is the ion-temperature-gradient instability \citep{waltz88, cowley91, kotschenreuter95_gf}. We use "ITG" and "ITG turbulence" as shorthand terms for this instability and the turbulence driven by it, respectively. As the name suggests, it is controlled by the gradient of ion temperature, which is a source of free energy for unstable microscale perturbations. It is then natural to investigate the dependence of the heat flux carried by the perturbations on the temperature gradient that drives those perturbations. Knowing the relationship between them, one can invert this relationship and find the heating power needed to support a given temperature gradient. 

Strongly driven ITG turbulence, i.e., ITG turbulence with a temperature gradient far above the linear-instability threshold, is believed to saturate via a "critically balanced" turbulent cascade \citep{barnes2011}: free energy stored in the equilibrium gradient is injected into perturbations by the instability and nonlinearly transferred (cascaded) to smaller scales, where it is thermalised via collisions. This is governed by the same kind of processes as the Kolmogorov cascade in hydrodynamic turbulence \citep{frisch1995}. This strongly turbulent saturated state supports vigorous turbulent transport of energy, so increasing the temperature gradient in such a system requires very substantial increases in heating power.

Na\"ively, one expects strong turbulence and high levels of transport to set in as soon as the temperature gradient exceeds the linear-instability threshold. However, there is numerical evidence for a low-transport regime with low levels of turbulence at temperature gradients larger than the linear threshold for the ITG instability but smaller than some nonlinear threshold above which strong turbulence and a high-transport state set in \citep{dimits2000}. Simulations have shown that the low-transport state below this threshold (to which we refer as the "Dimits state" and "Dimits threshold", respectively) is dominated by strong zonal flows (ZFs) --- Larmor-scale shear flows in the poloidal direction. These help regulate turbulence by shearing heat-carrying perturbations and hence reducing their amplitude. In this paper, we attempt to explain how the Dimits state is maintained and what leads to its eventual collapse.  

Despite being fairly well studied, many aspects of ZF physics, e.g., generation of ZFs from turbulence, stability of zonal fields, dependence of experimentally important quantities, like the heat flux, on basic plasma parameters (density, temperature, magnetic field and their gradients) in zonally dominated plasmas, remain far from being settled. One of the established paradigms is the primary-secondary-tertiary instability scenario \citep{rogersdorland2000, rogers2005}. Let us outline it here. The primary ITG instability feeds energy into a spectrum of linearly unstable modes that become nonlinearly unstable to zonal perturbations: this is the "secondary instability". Saturation is reached when the energy injection into ZFs is balanced by their slow viscous damping. Increasing the temperature gradient increases the primary drive, hence the secondary drive, hence the amplitude of ZFs. However, ZFs of large enough amplitude become nonlinearly unstable to a "tertiary instability", so they break up, transferring energy back into the ITG modes. The suppression due to zonal shear having been lost, fully developed turbulence ensues. In this scenario, the Dimits threshold is given by the threshold of the tertiary instability.

Even though we show that the tertiary instability determines important properties of the saturated state (e.g., poloidal spectra), we find that the Dimits threshold is not directly determined by the tertiary instability. The latter only works to excite turbulent perturbations that coexist with the ZFs. The way these perturbations interact with the ZFs via a mechanism akin to a generalised nonlinear secondary instability is what determines the Dimits threshold. In the low-collisionality regime, the interactions between turbulent perturbations and ZFs give rise to predator-prey-like oscillations familiar from past studies of ZF physics \citep[see, e.g.,][]{diamond2005,ricci2006,kobayashi2012}.

Recent progress has suggested that an entirely different scenario might need to be developed for turbulence with imposed background flow shear, applicable to tokamak plasmas made to rotate differentially. The work by \citet{vanwyk2017} has shown that close to marginality, the effect of the self-generated zonal shear is negligible compared to the equilibrium flow shear. The heat flux in this near-marginal state is dominated not by space-filling turbulence, but by localised, long-time-coherent, soliton-like, finite temperature and density perturbations travelling through the plasma \citep{vanwyk2016}. We call these structures "ferdinons", after Ferdinand van Wyk's name. As the temperature-gradient drive is increased, the number of ferdinons increases, they begin overlapping and interacting strongly, and the system enters a fully developed turbulent state. We do not investigate the case of imposed background flow shear in this paper, but we do find that locally generated zonal flows arrange themselves in regions of nearly constant shear. Structures closely resembling ferdinons are seen drifting through these sheared regions. This suggests that the formation of localised structures is a robust feature of sheared ITG turbulence as they are seen both in our simplified model (described below and in Section~\ref{sect_model}) and in more realistic 3D GK simulations. 

A comprehensive treatment of the problem of transition to, and saturation of, ITG turbulence requires the gyrokinetic (GK) framework in toroidal tokamak geometry \citep{friemanchen82, sugama96, sugama97, sugama98, abelgk1, catto2019}. However, its complexity makes it both analytically and numerically hard to treat. In this paper, we attempt the more modest task of tackling the problem in a simplified model for the dynamical evolution of the perturbations of electrostatic potential (or, equally well, density) and ion temperature in a tokamak plasma. The model is derived as an exact asymptotic limit of the underlying gyrokinetic equations in a physically realisable, if not necessarily most general, regime (see Section~\ref{sect_equations}). The approximations used are chosen to ensure that our model has a number of features that we consider essential: 1) a curvature-driven ion-temperature-gradient (ITG) instability, characteristic of tokamak plasmas; 2) an appropriate modified adiabatic electron response, which has been found to be crucial for capturing essential zonal-flow properties \citep[e.g., the correct ITG secondary instability: see][]{hammett93, rogersdorland2000}; 3) it is a two-field model linking the perturbations of the electrostatic potential and the ion temperature, rather than a one-field drift-wave model of the \citet{hasegawamima78} variety. A two-field model allows us to capture the important ITG linear instability, while keeping the equations simple enough to allow for an analytic treatment. 

As already mentioned, fully developed ITG turbulence is critically balanced and, therefore, 3D, so we cannot hope to capture that in a 2D model. Beyond the Dimits state, we find that our model fails to reach a saturated state --- perturbations grow exponentially and the box-sized perturbations eventually dominate the spectrum, regardless of the size of the integration domain. As we explain in Section~\ref{sect_blowup}, both the critical-balance argument and the constraints of the additional conserved quantities in 2D provide heuristic reasoning why developed homogeneous turbulence might not be able to saturate in 2D. If the Dimits transition is indeed a transition between an inhomogeneous, ZF-dominated state, where saturation is governed by the (fundamentally two-dimensional) ZFs, and a state of homogeneous, critically balanced turbulence, it appears natural that any 2D model that we use to describe the Dimits regime will be unable to capture the strongly turbulent state. However, the fact that our model is able to reach a well-defined saturated state in the Dimits regime lends it some credibility, whereas the fact that it (predictably) fails to saturate beyond the Dimits transition allows us to identify the transition itself in an unambiguous and sharp way, as a transition from a regime with a finite saturated state to one without.

There are two ways for turbulence to achieve saturation --- it can either cascade injected energy down to dissipation scales or, if it is internally driven by an instability, it can assemble itself in a configuration that suppresses that instability, i.e., the initial unstable equilibrium evolves into a new equilibrium with weaker instabilities. As our model contains both ZF and zonal-temperature perturbations, in principle it can accommodate the physics of two possible instability-suppression mechanisms: shearing of the turbulence by ZFs and modifying the background temperature gradient by zonal temperature perturbations in order to cancel the ITG drive. Neither of these can be done uniformly across the entire domain because we impose periodic boundary conditions in the radial direction. Interestingly, we find that the zonal perturbations arrange themselves in alternating wide regions of nearly-constant zonal shear, strong enough to suppress turbulence, and narrow regions of strong zonal-temperature gradient, which flattens the background temperature gradient and quenches the ITG instability. The resulting "staircase"-like overall radial temperature profiles are reminiscent of those seen in global and local flux-driven gyrokinetic simulations \citep{pradalier2010, difpradalier2017, villard2013, rath2016} and reported in experimental data \citep{difpradalier2015}. The resulting turbulent heat flux is significantly suppressed. The stability, and hence existence, of this zonal state is controlled by the background temperature gradient --- a large enough gradient renders the staircase configuration unstable and the system enters a fully developed turbulent state. In Section~\ref{sect_zfstab}, we link this behaviour to the mechanism through which the turbulence feeds the ZF, viz., the turbulent flux of poloidal momentum. By considering the linearly unstable ITG modes, we find a semi-analytical prediction of the Dimits threshold at high collisionality and high temperature gradient.

The rest of the paper is organised as follows. In Section~\ref{sect_model}, we describe our model, whose detailed derivation is given in Appendix~\ref{appendix_curvy_derivation}. Section~\ref{sect_nl} describes the nonlinear saturated state and in particular the zonally dominated state near the Dimits threshold. In Section~\ref{sect_zfstab}, we focus on the turbulent momentum flux of ITG turbulence subject to strong zonal shear, and the physics of the Dimits regime and its breakup beyond the Dimits threshold. Our results are summarised and conclusions are drawn in Section~\ref{sect_discussion}.

\section{ITG-Driven Dynamics in a \(Z\)-pinch}
\label{sect_model}

We consider the local dynamics of the perturbations of electrostatic potential and ion temperature of a 2D plasma (in the plane perpendicular to the magnetic field) in a \(Z\)-pinch magnetic geometry with an equilibrium temperature gradient. Our equations are derived in a highly collisional, cold-ion asymptotic limit of the electrostatic ion gyrokinetic equation. Their detailed derivation can be found in Appendix~\ref{appendix_curvy_derivation}. Here we present a summary of these equations, their physical motivation and key properties. If the reader wishes to skip directly to Section~\ref{sect_nl}, which contains the analysis of the saturated state, she may want to glance first at the model equations --- these are \eqref{curvy_phi} and \eqref{curvy_psi}.

\subsection{Magnetic Geometry}
\label{sect_magn_geometry}
\begin{figure}
  \centering
  \includegraphics[scale=0.3]{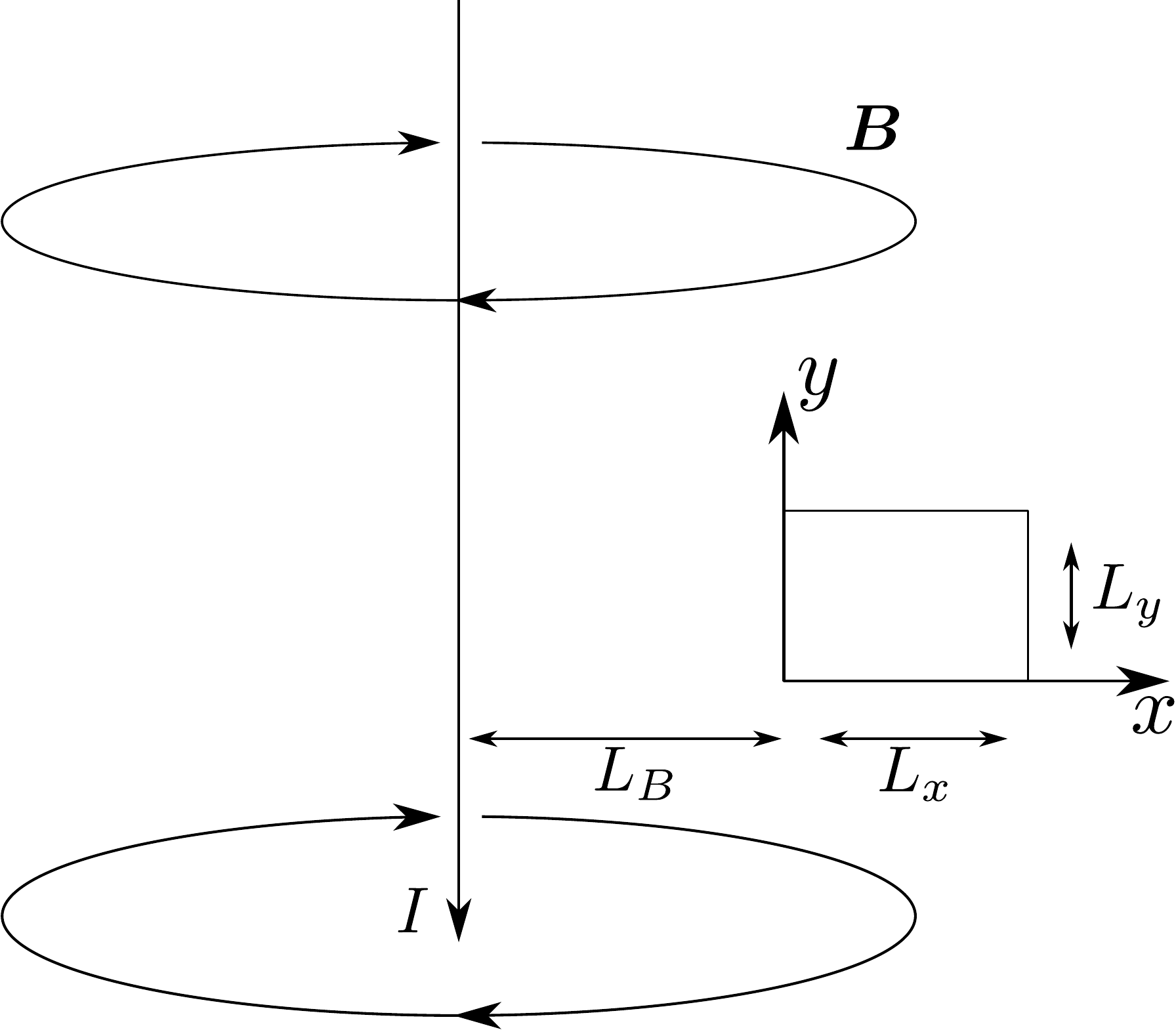}
  \caption{Illustration of the \(Z\)-pinch magnetic geometry.}.
\label{fig_zpinch}
\end{figure}

The magnetic geometry of constant magnetic curvature is chosen because it is the simplest one that enables an ITG instability by coupling the electrostatic potential and the temperature perturbations via the magnetic drift. The integration domain is positioned in the magnetic field of a line of current (\(Z\)-pinch\footnotemark) at radial distance \(L_B\) from the current line: see Figure~\ref{fig_zpinch}. \footnotetext{This simplification as a route to a minimal model of ion-scale turbulence goes back at least to \citet{ricci2006}.} We define the \(x\) and \(y\) axes as pointing radially outwards and parallel to the current, respectively. We assume \(L_B \gg L_x, L_y\), where \(L_x\) and \(L_y\) are the "radial" (\(x\)) and "poloidal" (\(y\)) sizes of the domain, respectively. Here we use the terms "radial" and "poloidal" to reflect the intended similarity of the domain to one positioned at the outboard midplane in a tokamak geometry. In that sense, we can think of the radial \(x\) coordinate as perpendicular to flux surfaces. These surfaces are parametrised by the poloidal \(y\) and field-parallel \(\uvect{b}\) coordinates. Here \(\vect{B} = B \uvect{b}\) is the magnetic field and the unit vectors \(\lbrace\uvect{x}, \uvect{y}, \uvect{b}\rbrace\) form a right-handed basis. In the 2D approximation employed here, all perturbed fields depend only on \(x\) and \(y\). The magnetic field of the \(Z\)-pinch with total current \(I\) is azimuthal around the current line (as shown on Figure~\ref{fig_zpinch}) and has magnitude \(B(x) = 2I/cx\). The radial gradient of this field is then
\begin{equation}
    \frac{1}{B} \frac{dB}{dx} = -\frac{1}{L_B}.
\end{equation}
This value is constant across the domain to lowest order in \(L_x/L_B \ll 1\). We define \(L_B\) to be the magnetic scale length. Similarly, we can define the ITG scale length
\begin{equation}
    \frac{1}{L_T} \equiv -\frac{1}{T_i} \frac{dT_i}{dx}.
\end{equation}
In a tokamak, \(L_B\) scales with the major radius of the device, while \(L_T\) scales with the minor radius. Here we will take the limit 
\begin{equation}
  L_B \gg L_T,
\end{equation}
equivalent to a large-aspect-ratio approximation in a tokamak geometry. We do this in order to ensure that the magnetic drift in the density equation is of the appropriate order [see~\eqref{phiEq}]. This drift is essential for the linear curvature-driven ITG instability that we aim to capture. 

\subsection{Electron Response}
\label{sect_eresponse}

The electron density is assumed to follow the modified adiabatic response 
\begin{equation}
    \label{eq_eresponse}
    \frac{\delta n_e}{n_e} = \frac{e (\phinonnorm - \zf{\phinonnorm})}{T_e},
\end{equation}
taking into account the fast parallel streaming of the electrons within the flux surfaces of a tokamak, in the small-mass-ratio limit \(m_e / m_i \to 0\) \citep{dorland93,hammett93}. Here \(\delta n_e\) and \(n_e\) are the perturbed and equilibrium electron density, respectively, \(\phinonnorm\) is the electric potential, \(T_e\) is the electron temperature, and
\begin{equation}
    \zf{\phinonnorm}(x) \equiv \frac{1}{L_y} \int dy \ \phinonnorm(x, y)
\end{equation}
is the poloidal (zonal) spatial average of the perturbed electric potential \(\phinonnorm\). We refer to zonally averaged fields as being "zonal". In the 2D approximation, the turbulent fields (e.g., \(\phinonnorm\)) are independent of \(z\), hence we do not need to integrate over the \(z\) direction. The difference 
\begin{equation}
	\label{eq_nonzonal_def}
	\dw{\phinonnorm}(x, y) \equiv \phinonnorm(x, y) - \zf{\phinonnorm}(x)
\end{equation}
is the "nonzonal" part of the field.

A cautious reader has spotted that there is no way to define flux surfaces in the magnetic geometry of a $Z$-pinch, as the magnetic field lines do not describe 2D surfaces, but rather close on themselves after one turn around the current axis. This problem can be fixed by demanding that the magnetic field be, in fact, sheared: \(\vect{B} = B_0 (\uvect{z} + x\uvect{y}/L_s)\), where \(L_s\) is the characteristic scale length of the magnetic shear. This is the magnetic field of a helimak \citep{gentle2008}. The field lines define cylindrically symmetric concentric flux surfaces and the electron parallel streaming mixes the azimuthal (\(z\)) and poloidal (\(y\)) directions. We can then take the limit \(L_s \to \infty\) \textit{after} performing the small-mass-ratio (\(m_e/m_i \to 0\)) expansion and eliminate magnetic shear from the ion equations, while retaining the flux-surface effect in the electron response.

\subsection{Cold-Ion Limit}
\label{sect_coldions}
The cold-ion limit allows us to simplify the gyroaveraging operator that appears in gyrokinetics. Its corresponding Fourier-space operator is a multiplication by the Bessel function
\begin{equation}
	\label{eq_j0}
    J_0\left(\frac{k_\perp v_\perp}{\Omega_i}\right) = 1 - \frac{1}{4} \kperprhoisq \frac{v_\perp^2}{\vti^2} + \order{\kperprhoisqq}.
\end{equation}
The square of the ion gyroradius \(\rho_i = \vti m_ic/ Z eB\) and the ion temperature \(T_i = m_i \vti^2 / 2\) are both proportional to the square of the ion thermal speed \(\vti\) (here \(m_i\) and \(Z\) are the ion mass and charge in units of \(e\), respectively). Thus, the cold-ion limit is equivalent to a long-wavelength expansion \(\order{k_\perp \rho_i} \ll 1\) with a finite sound radius \(\order{k_\perp \rho_s} \sim 1\), where \(k_\perp\) is the perpendicular (to the magnetic field) wavenumber, \(\rho_s = \rho_i / \sqrt{2\tau}\) is the sound radius, and \(\tau = T_i / ZT_e\) is the temperature ratio (\(T_e\) is assumed finite). The sound radius \(\rho_s\) is the natural normalisation for the microphysical length scales in the problem: see equations \eqref{phiEq} and \eqref{psiEq} below and their derivation in Appendix~\ref{appendix_curvy_derivation}.

 \subsection{Model Equations}
\label{sect_equations}

We take the density and temperature moments of the electrostatic ion gyrokinetic equation and adopt the high-collisionality, cold-ion, long-wavelength, large-aspect-ratio ordering
\begin{equation}
    \label{eq_ordering}
    \frac{\pt}{\nu_i} \sim \tau \sim \kperprhoisq \sim \frac{L_T}{L_B} \ll 1 \sim \frac{\phinorm}{\deltaT},
\end{equation}
where \(\phinorm \equiv Z_ie\phinonnorm / T_i \) is the normalised electric potential, \(\deltaT = \delta T / T_i\) is the normalised ion-temperature perturbation, and \(\nu_i\) is the ion-ion collision frequency. The equations that we obtain in Appendix~\ref{appendix_curvy_derivation} are
\begin{align}
	\label{phiEq}
	& \frac{\partial}{\partial t} \left( \tau \phinorm' - \frac{1}{2} \rhoidelperpsq \phinorm\right)- \frac{\rhoivti}{L_B} \frac{\partial}{\partial y} \left( \phinorm + \deltaT \right) + \frac{\rhoivti}{2L_T} \frac{\partial}{\partial y} \left( \frac{1}{2} \rhoidelperpsq \phinorm \right)
	\\& \quad + \frac{1}{2} \rhoivti \bigg( \pbra{\phinorm}{\tau\phinorm' - \frac{1}{2} \rhoidelperpsq \phinorm} + \frac{1}{2}\rho_i^2 \boldsymbol{\nabla_\perp} \bcdot \pbra{\boldsymbol{\nabla_\perp}\phinorm}{\deltaT} \bigg) \notag
	\\& \quad = - \frac{1}{2} \chi \rho_i^2 \nabla_\perp^4 (a\phinorm - b\deltaT), \notag
	\label{psiEq}
	\\ &\frac{\partial\deltaT}{\partial t}  + \frac{\rhoivti}{2L_T} \frac{\partial\phinorm}{\partial y}  + \frac{1}{2} \rhoivti \pbra{\phinorm}{\deltaT} = \chi \nabla_\perp^2 \deltaT,
\end{align}
where the Poisson bracket is defined by
\begin{equation}
	\label{eq_pbra_def}
    \pbra{f}{g} = \uvect{b} \bcdot \left( \del f \times \del g \right) = \frac{\partial f}{\partial x} \frac{\partial g}{\partial y} - \frac{\partial f}{\partial y} \frac{\partial g}{\partial x},
\end{equation}
and 
\begin{equation}
    \label{eq_chi_def}
    \chi \equiv \frac{8}{9} \sqrt{\frac{2}{\pi}} \nu_{i}\rho_i^2
\end{equation}
is the thermal diffusivity. The numerical factor in \eqref{eq_chi_def} and the constants \(a = 9/40\), \(b = 67/160\) in \eqref{phiEq} are specific to the Landau collision operator (see Appendix~\ref{appendix_col}). These agree with more general calculations of the collisional perpendicular viscosity, gyroviscosity and collisional heat flux \citep{mikhailovskii71, catto2004, catto2005}.

Let us discuss the physics content of \eqref{phiEq} and \eqref{psiEq}. Equation \eqref{psiEq} is the more obvious one --- it describes the advection of the total temperature (perturbations plus equilibrium) by the \exb drift \(\ve = c\uvect{b}\times\del\phinonnorm / B\), and the thermal diffusion perpendicular to the magnetic field. Indeed, \eqref{psiEq} can be rewritten as
\begin{equation}
	\label{eq_curvy_psi_advective}
    \frac{d}{dt} \left(\delta T + T_i\right) = \chi \nabla_\perp^2 \delta T,
\end{equation}
where the advective time derivative is
\begin{equation}
    \frac{d}{dt} \equiv \frac{\partial}{\partial t} + \ve \bcdot \nabla.
\end{equation}
The advection of the equilbrium temperature, \(\ve \bcdot \del T_i\) [the second term on the left-hand side of \eqref{psiEq}] is responsible for the injection of free energy (see Section~\ref{sect_cons}), causing the ITG instability.

Equation \eqref{phiEq} describes the time evolution of the sum of perturbed ion density and the vorticity of the \exb drift velocity:
\begin{equation}
\label{eq_density_response}
  \frac{\delta n_i}{n_i} + \frac{1}{n_i}\int \dv \ \left(\phinorm - \avgr{\avgR{\phinorm}}\right)F_i =
  \tau \phinorm' - \frac{1}{2} \rhoidelperpsq \phinorm + \order{\kperprhoisqq \phinorm},  
\end{equation}
where \(n_i\) and \(\delta n_i\) are the equilibrium and perturbed ion densities, respectively [see \eqref{eq_density_term1}]. Thus, \eqref{phiEq} can be thought of as both a perturbed-ion-density equation and as the curl of the perpendicular-momentum equation. The equality in \eqref{eq_density_response} follows from the quasineutrality condition \(Z\delta n_i = \delta n_e\), the electron response \eqref{eq_eresponse}, the approximation \eqref{eq_j0}, and the ordering \eqref{eq_ordering}.

The first term of \eqref{phiEq} is the time derivative of \eqref{eq_density_response}. The second term, \((\rhoivti / L_B) \py \left( \phinorm + \deltaT \right)\), is the magnetic drift (both curvature and \(\nabla B\)) of pressure perturbations. This, or rather the \(\py \deltaT\) part of it, is essential for the curvature-driven ITG instability. It appears in the density equation because the magnetic drift creates charge separation, and hence electrostatic potential, which is then coupled to the perturbed density via quasineutrality. The third term, \((-\rhoivti / 4L_T) \py \left( \rhoidelperpsq \phinorm \right)\), is a finite-Larmor-radius (FLR) term originating from the gyroaveraged \exb drift. It is the diamagnetic drift due to the equilibrium temperature gradient. The first of the nonlinear terms represents the advection of the quantity \eqref{eq_density_response} by the \exb drift. The second nonlinear term \(\boldsymbol{\nabla_\perp} \bcdot \pbra{\boldsymbol{\nabla_\perp}\phinorm}{\deltaT}\) is another FLR effect, which describes the \exb advection of diamagnetic momentum. This term provides a crucial source of turbulent poloidal momentum flux that destabilises the ZF profiles, destroying the ZF-dominated Dimits regime. The nature of this term and its role in the Dimits transition are discussed in detail in Section~\ref{sect_stresses}. Note that the nonlinear terms in \eqref{phiEq} and \eqref{psiEq} are equivalent to the nonlinearities appearing in the model analysed by \citet{rogersdorland2000} in the limit \eqref{eq_ordering}. Finally, the collisional terms in \eqref{phiEq} represent the viscous damping of the \exb flow and also couple the density and temperature perturbations. The latter coupling does not appear to be important for the results of this paper, but has been kept for consistency. 

To prepare \eqref{phiEq} and \eqref{psiEq} for numerical analysis and distil important parameters, we introduce normalised variables and fields

\begin{equation}
	\label{eq_normalisations}
    \begin{gathered}
	    \hat{t} \equiv \frac{2\rho_s\Omega_i}{L_B}t, \qquad
	    \hat{x} \equiv \frac{x}{\rho_s}, \qquad \hat{y} \equiv \frac{y}{\rho_s}, \\
	    \hat{\phinorm} \equiv \frac{\tau L_B\phinorm}{2\rho_s}  = \frac{\tau L_B}{2\rho_s} \frac{Z_ie\phinonnorm}{T_i}, \qquad 
	    \hat{\deltaT} \equiv \frac{\tau L_B\deltaT}{2\rho_s}  = \frac{\tau L_B}{2\rho_s} \frac{\delta T}{T_i}, \\
	    \vt \equiv \frac{\tau L_B}{2 L_T}, \qquad
	    \hat{\chi} \equiv \frac{L_B}{2\rho_s} \frac{\chi}{\Omega_i \rho_s^2},
    \end{gathered}
\end{equation}
where \(\Omega_i = \vti / \rho_i\) is the ion gyrofrequency. Dropping hats and subscripts (\(\del_\perp \mapsto \del\)), we obtain from \eqref{phiEq} and \eqref{psiEq} the following equations in normalised units:
\begin{align}
\label{curvy_phi}
	&\partial_t \left( \dw{\phinorm} - \delsq \phinorm\right) - \partial_y \left( \phinorm + \deltaT \right) + \vt \partial_y  \delsq \phinorm  +\pbra{\phinorm}{\phinorm' - \delsq \phinorm}  
	+ \del \bcdot \pbra{\del \phinorm}{\deltaT} \nonumber \\
	&\quad=-\chi \nabla^4 (a\phinorm - b\deltaT), \\
\label{curvy_psi}
&\partial_t\deltaT +\vt\partial_y \phinorm +\pbra{\phinorm}{\deltaT} = \chi \nabla^2 \deltaT.
\end{align}
These equations have two independent parameters: the normalised equilibrium temperature gradient, \(\vt\), and the normalised collisionality, \(\chi\)\footnotemark\footnotetext{\label{fn_cols}The reader might wonder what the experimentally relevant values of \(\chi\) are. Using the data from \citet{abelgk2}, we find \(\chi \approx 6\times10^{-4}\) for a deuterium plasma in JET. Thus, the low-collisionality regime is the one we expect to be of greater interest.}. There are two other parameters --- \(L_x\) and \(L_y\), the domain lengths in \(x\) and \(y\) --- but any physically relevant results must be independent of these if our equations are indeed a valid local model of the plasma. This turns out to be true for the saturated Dimits state. 

We solve \eqref{curvy_phi} and \eqref{curvy_psi} numerically in a doubly periodic box of size \(L_x\) and \(L_y\) using a pseudo-spectral algorithm. We integrate the linear terms implicitly in time, while the nonlinear terms are integrated explicitly using the Adams-Bashforth three-step method. This integration scheme is similar to the one implemented in the popular gyrokinetic code GS2 \citep{kotschenreuther95, dorland2000}.

\subsection{Relationship to Hasegawa-Mima Equation and Related Models}
\label{sect_hasegawamima}

It is easy to see that setting \(\vt = 0\) effectively decouples \eqref{curvy_psi} from \eqref{curvy_phi} --- taking an initial condition \(\deltaT(t = 0) = 0\) then leads to a trivial solution \(\deltaT(t) = 0\). In that case, \eqref{curvy_phi} reduces to 
\begin{equation}
\label{eq_hasegawamima}
\partial_t \left( \dw{\phinorm} - \delsq \phinorm\right) - \partial_y \phinorm +\pbra{\phinorm}{\phinorm' - \delsq \phinorm} = -a \chi \nabla^4 \phinorm,
\end{equation}
which is the well-known (modified) Charney-Hasegawa-Mima (mCHM) equation \citep{hasegawamima78} that includes the appropriate modified adiabatic electron response, with viscous damping. Even though we have considered a situation with no equilibrium density gradient, the magnetic drift provides a term identical to the one that would have arisen from the \exb advection of an inhomogeneous equilibrium density profile. This puts the model considered here in the same class of systems as those proposed by \citet{hasegawawakatani83}, \citet{terryhorton83} and others --- all essentially extensions of the Hasegawa-Mima equation with additional physics to account for microinstabilities in the plasma. 

As \eqref{eq_hasegawamima} is contained within the model considered in this paper, equations \eqref{curvy_phi} and \eqref{curvy_psi} should, in principle, capture the behaviour of the mCHM equation as well as additional temperature and ITG effects. There has recently been a significant effort to advance the understanding of \eqref{eq_hasegawamima} and its relatives \citep{parker2013, parker2014, parker2016, ruiz2016, ruiz2019, zhu2018, zhu2019, zhu2019_dimits, zhu2020jpp, zhou2019, plunk2017, stonge2017, majda2018, qi2019}. The mCHM equation does capture certain important phenomena, such as the generation of ZFs through a secondary instability (see Section~\ref{sect_sec}); however, its predictive capabilities for ITG turbulence are unclear. In particular, we shall find that the break up of the Dimits state of \eqref{curvy_phi} and \eqref{curvy_psi} is, in an essential way, governed by the behaviour of the temperature perturbations, which are absent in \eqref{eq_hasegawamima} (see Section~\ref{sect_zfstab}).

\subsection{Linear Physics of ITG Instability}
\label{sect_linear}

\begin{figure}
	\centering
	\includegraphics[scale=0.47]{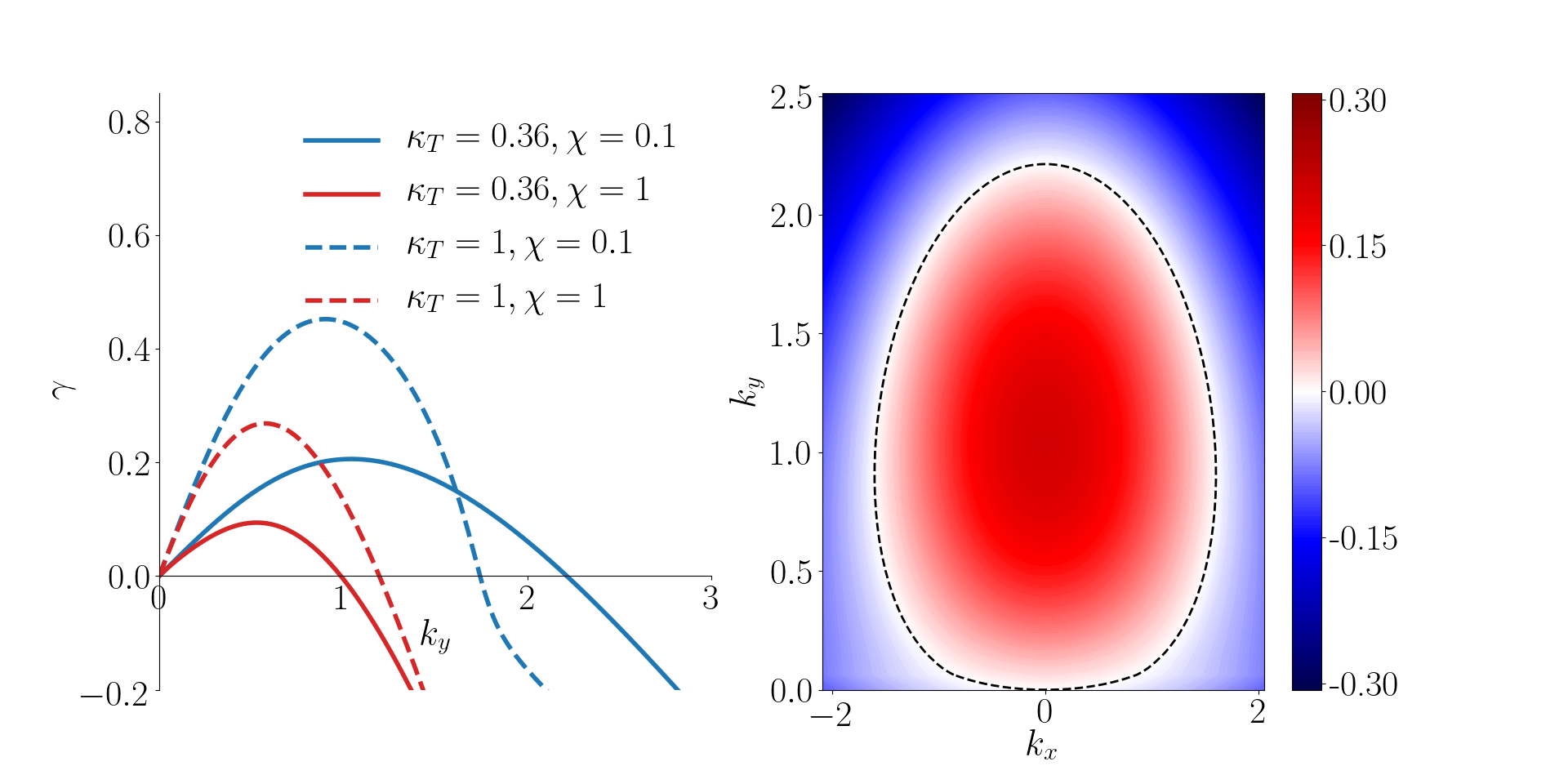}
	\caption{\textbf{Left:} Dependence of the growth rate \(\gamma_\vk\) on \(k_y\) for the streamer modes (\(k_x = 0\)). \textbf{Right:} Dependence of \(\gamma_\vk\) on \(k_x\) and \(k_y\) for \(\kappa_T = 0.36, \chi = 0.1\). The dashed line is the boundary between stable and unstable modes (\(\gamma_\vk = 0\)). In Section~\ref{sect_nl}, we will consider nonlinear simulations with these same parameters. }
	\label{fig_linear}
\end{figure}

Let us analyse the linear stability of \eqref{curvy_phi} and \eqref{curvy_psi}. Dropping the nonlinear terms, we look for Fourier modes \(\phinorm, \deltaT \propto \exp\left[ (\gamma_\vk - i\omega_\vk) t + i \vect{k}\bcdot\vect{r} \right]\), where \(\gamma_\vk\) and \(\omega_\vk\) are the real growth rate and frequency, respectively. Figure~\ref{fig_linear} shows \(\gamma_\vect{k}\) as a function of the wavenumber \(\vect{k}\). Qualitatively it resembles the growth rate of toroidal ITG modes in tokamaks \citep{horton81}. This is expected because the mechanism of the toroidal ITG instability is similar to that of the 2D curvature-driven ITG instability. The terms that give rise to the instability are the magnetic drift term \(-\py \deltaT\) in \eqref{curvy_phi} and the \exb advection of the equilibrium temperature \(\vt\py \phinorm\) in \eqref{curvy_psi}. We find that the fastest growing linear modes are radially extended across the entire box, i.e., they have \(k_x = 0\). Such modes are sometimes called "streamers". 

The exact dispersion relation is
\begin{align}
\label{eq_disp_relation}
(\gamma_\vk - i\omega_\vk)^2(&1+k^2) +(\gamma_\vk - i\omega_\vk)\left\lbrace -ik_y (1+\vt k^2) + \chi k^2 \left[1 + (1+a)k^2\right] \right\rbrace \nonumber \\  &+a\chi^2k^6 - \vt k_y^2 - ik_y\chi k^2 \left[1+\vt (1-b)k^2\right] = 0.
\end{align}
We can get a good qualitative idea of the properties of the instability by setting \(\chi=0\). Then the solution of \eqref{eq_disp_relation} is
\begin{equation}
    \label{eq_disp_relation_nodamping}
    \gamma_\vk - i\omega_\vk = \frac{ik_y(1+\vt k^2) \pm k_y\sqrt{4\vt(1+k^2) - (1+\vt k^2)^2}}{2(1+k^2)},
\end{equation}
so the growth rate of the unstable mode is 
\begin{equation}
	\label{eq_gammak_nodamping}
    \gamma_\vect{k} = \frac{k_y\sqrt{(2\sqrt{\vt} - 1 + \vt k^2)(2\sqrt{\vt} + 1 - \vt k^2)}}{2(1+k^2)}.
\end{equation}
To simplify further, consider \(\vt \gg 1 \gg \vt^{-1/4} \gg k\). Then
\begin{equation}
    \label{eq_itg_verysimple}
    \gamma_\vect{k} \approx k_y \sqrt{\vt}. 
\end{equation}
This expression, with the normalisations \eqref{eq_normalisations} undone, is the well-known "bad-curvature-instability" growth rate \citep{beer95}:
\begin{equation}
   \gamma_\vect{k} = \Omega_i \frac{\rho_i^2 k_y}{\sqrt{2\tau L_BL_T}}.
\end{equation}

Note that there is another, physically distinct, ITG instability usually referred to as the "slab ITG mode". This instability relies on coupling density and temperature through parallel-velocity perturbations, and so is naturally three-dimensional \citep{cowley91}. This mode is entirely absent from our 2D model.

Now let us return to the general dispersion. An important feature of the modes described by \eqref{eq_disp_relation} is the boundedness of the region of unstable wavenumbers in the \(\vect{k}\) plane (right panel of Figure~\ref{fig_linear}). This allows us to integrate \eqref{curvy_phi} and \eqref{curvy_psi} without the need for artificial dissipation. There are both collisionless and collisional mechanisms that lead to the suppression of the ITG instability. Let us consider these mechanisms.

\subsubsection{Collisionless Bounds on Unstable Wavenumbers}
\label{sect_unstable_region_cless}

It is easy to see that, in order to be positive, the collisionless growth rate \eqref{eq_gammak_nodamping} requires \(k < k_\text{max,FLR}\), where
\begin{equation}
\label{eq_maxk_flr}
k_\text{max,FLR}^2 = \frac{1 + 2\sqrt{\vt}}{\vt}.
\end{equation}
For \(\vt < 1/4\), \eqref{eq_gammak_nodamping} also gives a lower bound on the wavenumbers \(k\) of the unstable collisionless modes, viz., \(k > k_\text{min,FLR}\), where 
\begin{equation}
\label{eq_mink_flr}
k_\text{min,FLR}^2 = \frac{1 - 2\sqrt{\vt}}{\vt}.
\end{equation}
Adding collisions re-establishes the instability at low \(k\). We deem this to be an unimportant peculiarity of our model, thus we shall only consider \(\vt > 1/4\).

\subsubsection{Collisional Bounds on Unstable Wavenumbers}
\label{sect_unstable_region_col}

For nonzero (\(\chi > 0\)) collisionality, the term \(a\chi^2 k^6\) in \eqref{eq_disp_relation} dominates over the ITG term \(\vt k_y^2\) when \(k\) is large enough and gives strictly damped modes. To show this, let us simplify \eqref{eq_disp_relation} by writing it as

\begin{equation}
	\label{eq_disp_short}
	(\gamma_\vk - i\omega_\vk + A) (\gamma_\vk - i\omega_\vk + B - iC) - fAB + igAC = 0,
\end{equation}
where 
\begin{equation}
	\label{eq_disp_short_defs}
	A = \chi k^2, \quad B = \frac{a\chi k^4}{1+k^2}, \quad C = k_y \frac{1+\vt k^2}{1+k^2}, \quad f = \frac{\vt k_y^2}{a\chi^2k^6}, \quad g = \frac{b \vt k^2}{1 + \vt k^2}.
\end{equation}
The instability threshold is given by \(\gamma_\vk = 0\). The real and imaginary parts of \eqref{eq_disp_short} for \(\gamma_\vk = 0\) are
\begin{align}
\label{eq_marginaldisp_real}
&-\omega_\vk^2 -\omega_\vk C + (1-f)AB = 0, \\
\label{eq_marginaldisp_imag}
&-\omega_\vk (A+B) - (1-g)AC = 0.
\end{align}
Substituting into \eqref{eq_marginaldisp_real} the value of \(\omega_\vk\) derived from \eqref{eq_marginaldisp_imag}, and using \(A \neq 0\), we find
\begin{equation}
g(1-g)AC^2 + BC^2(1-g) + (1-f)B(A+B)^2 = 0.
\end{equation}
Since \(g \in (0, 1)\)\footnotemark\footnotetext{Note that for \(b < 0\) or \(b > 1\), there would be a collisional (\(\vt = 0\)) instability. No such instability exists in our model because the Landau collision operator gives \(g \in (0, 1)\).}, a necessary condition for instability is
\begin{equation}
f > 1 \implies a\chi^2k^6 < \vt k_y^2.
\end{equation}
Thus, the region of unstable modes is bounded by \(k < k_{\text{max,}\chi}\), where
\begin{equation}
\label{eq_maxk_col}
k_{\text{max,}\chi}^2 = \sqrt{\frac{\vt}{a\chi^2}}.
\end{equation}

\subsection{Conservation Laws}
\label{sect_cons}

Equations \eqref{curvy_phi} and \eqref{curvy_psi} have several conservations laws describing the time evolution of quantities that would be conserved in the absence of equilibrium gradients and dissipation:
\begin{align}
\label{eq_free_e_cons}
&\pt \int dxdy \ \frac{1}{2} \deltaT^2 = -\vt \int dxdy \ \deltaT \py \phinorm - \chi \int dxdy \ \left( \del \deltaT \right)^2, \\
\label{eq_phi_cons}
&\pt \int dxdy \ \frac{1}{2} \left[\dw{\phinorm}^2 + (\del \phinorm)^2\right] \\  &= -\int dxdy \  \deltaT \py \phinorm \nonumber - \chi \int dxdy \ (\nabla^2 \phinorm) \left( a\nabla^2 \phinorm - b \nabla^2 \deltaT  \right) , \\
\label{eq_phi_weird_cons}
&\pt \int dxdy \ \left[ \frac{1}{2} \dw{\phinorm}^2 + \deltaT \phinorm' + \frac{1}{2} \left(\del \deltaT + \del \phinorm\right)^2 \right] \nonumber \\ &=  -\chi \int dxdy \ \bigg[ \left(\del \dw{\phinorm}\right) \bcdot \left(\del \deltaT\right) + a \left( \nabla^2 \phinorm \right)^2 \nonumber \\&\quad+(a + 1 - b) \left( \nabla^2 \phinorm \right)\left( \nabla^2 \deltaT \right) + (1-b)\left( \nabla^2 \deltaT \right)^2 \bigg].
\end{align}
These conservation laws can be deduced directly from \eqref{curvy_phi} and \eqref{curvy_psi}: e.g., \eqref{eq_free_e_cons} is obtained by multiplying \eqref{curvy_psi} by \(\deltaT\) and integrating over \(x\) and \(y\). They are also particular cases of the conservation laws of the gyrokinetic equation. The conservation of the variance of \(T\), given by \eqref{eq_free_e_cons}, is the lowest-order version of the gyrokinetic free-energy budget. The other two conservation laws, \eqref{eq_phi_cons} and \eqref{eq_phi_weird_cons}, can be derived from the conservation of the two-dimensional gyrokinetic invariant \citep[see][]{schekochihingk2009, plunk2010}. This invariant is a function of velocity in the GK formalism. The model presented here is based only on two velocity moments of the distribution function, namely density and temperature, and so the two-dimensional invariant yields two independent conservation laws. More specifically, \eqref{eq_phi_cons} is a generalisation of the "electrostatic gyrokinetic invariant". The derivations of the three invariants of our system directly from the corresponding GK invariants can be found in Appendix~\ref{appendix_cons}.

Equations \eqref{eq_free_e_cons} and \eqref{eq_phi_cons} imply that a steady saturated state, i.e., \(\pt = 0\) for all averaged quantities, can be achieved only if appropriate balance between injection and dissipation terms is established:
\begin{align}
     \frac{\chi}{\vt} \int \frac{dxdy}{L_xL_y} \ \left( \del \deltaT \right)^2 = \chi \int \frac{dxdy}{L_xL_y} \ (\nabla^2 \phinorm) \left( a\nabla^2 \phinorm - b \nabla^2 \deltaT  \right) = Q,
\end{align}
where the total radial heat flux \(Q\) is\footnotemark\footnotetext{ \label{fn_hf} The dimensional ion heat flux \(Q_i = V^{-1} \int d^3\vect{r}\int d^3\vect{v} (\ve \bcdot \uvect{x}) (m_i v^2 / 2) \delta f_i \) \citep{barnes2011}, where \(V\) is the volume of integration and \(\delta f_i\) is the perturbed ion distribution function [see \eqref{eq_f_def} in Appendix~\ref{appendix_gk}], is related to \(Q\) via \(Q_i / Q = 3 n_i T_i \vti \left(\rho_i / L_B\right)^2 / \tau^{5/2} \sqrt{2}.\)}
\begin{equation}
    \label{eq_heatflux_def}
    Q = -\frac{1}{L_xL_y}\int dxdy \ \deltaT \py \phinorm.
\end{equation}
Thus, a saturated state would necessarily have a net positive "turbulent" (or "anomalous") heat flux \(Q > 0\). Note that the first term on the right-hand side of \eqref{eq_free_e_cons}, which represents injection of free energy, is \(\vt Q\). The turbulent heat flux is enabled by the turbulence excited by the ITG instability. 

Note as well that the linearly unstable modes have a positive radial heat flux. Indeed, from \eqref{eq_heatflux_def},
\begin{equation}
    Q = \sum_\vect{k} ik_y \deltaT_\vect{k} \phinorm_\vect{k}^* = \sum_\vect{k} i k_y |\phinorm_\vect{k}|^2 \frac{\deltaT_\vect{k}}{\phinorm_\vect{k}}.
\end{equation}
The relative phase of the temperature and potential perturbations can be obtained from \eqref{curvy_psi}:
\begin{equation}
    \label{eq_temp_phi_linear_ratio}
    \frac{\deltaT_\vect{k}}{\phinorm_\vect{k}} = \frac{-ik_y\vt}{\gamma_\vk - i\omega_\vk + \chi k^2},
\end{equation}
where \(\gamma_\vk - i\omega_\vk\) is the solution of the dispersion relation \eqref{eq_disp_relation}. Then
\begin{equation}
    Q = \sum_\vect{k} \vt k_y^2 |\phinorm_\vk|^2 \frac{\gamma_\vk + \chi k^2}{(\gamma_\vk + \chi k^2)^2 + \omega_\vk^2} > 0
\end{equation}
for the unstable modes, which have \(\gamma_\vk > 0\). 

Finally, the third conservation law \eqref{eq_phi_weird_cons} has some peculiar properties. First, neither the conserved quantity on the left-hand side nor the dissipation rate on the right-hand side is sign-definite. Secondly, all of the evolution is dissipative, i.e., this invariant is not injected by any equilibrium gradients and is constant in time if \(\chi = 0\).

\subsection{Secondary Instability}
\label{sect_sec}

Before we delve into the study of nonlinear saturation, let us show how ZFs can be generated from the linearly unstable ITG modes. Consider the stability of a streamer mode with \(k_x = 0, k_y = q\) (the "primary" mode) to infinitesimal "secondary" perturbations:
\begin{align}
	\label{eq_sec_fields1}
    &\phinorm = (\phinorm_q e^{i q y } + \text{c.c.}) + \delta \phinorm (x, y), \\
    \label{eq_sec_fields1_1}
    &\deltaT = (\deltaT_q e^{i q y } + \text{c.c.}) + \delta \deltaT (x, y).
\end{align}
A common way of analysing the secondary instability is to take a Galerkin truncation by considering only four Fourier modes \((k_x, k_y) = \lbrace (0, q), (p, 0), (p, \pm q) \rbrace\) and their complex conjugates: the \((0, q)\) mode is the primary streamer in \eqref{eq_sec_fields1} and \eqref{eq_sec_fields1_1} and the others are
\begin{align}
	\label{eq_sec_fields2}
    &\delta \phinorm =  \left(\delta\phinorm_+ e^{i q y } + \delta\phinorm_- e^{- i q y } + \delta\phinorm_0\right) e^{ipx} e^{\gamma_2 t} + \text{c.c.}, \\
    &\delta \deltaT = \left(\delta\deltaT_+ e^{i q y } + \delta\deltaT_- e^{- i q y } + \delta\deltaT_0\right) e^{ipx} e^{\gamma_2 t} + \text{c.c.},
\end{align}
where \(p\) is the radial wavenumber of the secondary perturbations, \(\delta\phinorm_0\) and \(\delta\deltaT_0\) are the zonal flow and temperature, and \(\delta\phinorm_\pm\) and \(\delta\deltaT_\pm\) are known as "sidebands". Substituting all this into \eqref{curvy_phi} and \eqref{curvy_psi} and linearising the nonlinear terms for \(\delta \phinorm \ll \phinorm_q\) and \(\delta \deltaT \ll \deltaT_q\), we obtain a closed set of equations. In order to keep things simple, we drop the linear terms in \eqref{curvy_phi} and \eqref{curvy_psi} --- this is valid when the amplitude of the primary mode is large enough, so that interactions with it are more important for the evolution of \(\delta \phinorm\) and \(\delta \deltaT\) than the effects of the equilibrium gradients and collisions. Observe that, due to the structure of the Poisson bracket \eqref{eq_pbra_def}, all nonlinear terms are proportional to \(pq\). Defining for convenience \(\gamma_2 \equiv \sqrt{2}pq |\phinorm_q| \hat{\gamma}_2\), we obtain the following equation for~\(\hat{\gamma}_2\):
\begin{align}
    \label{eq_sec_disp}
    &\left(\hat{\gamma}_2^2 + U\right)\left(\hat{\gamma}_2^2 + V\right) = W,
\end{align}
where
\begin{align}
	\label{eq_sec_defs}
    &U = 1 + \frac{q^2\re{(\deltaT_q / \phinorm_q)}}{1+p^2+q^2}, \nonumber \\
    &V = \frac{p^2 \im{(\deltaT_q / \phinorm_q)}^2 + p^2 \left[1 + \re{(\deltaT_q / \phinorm_q)}\right]^2 - \left(1+q^2\right)\left[1 + \re{(\deltaT_q / \phinorm_q)}\right]}{1 + p^2 + q^2}, \nonumber \\
    &W = \frac{p^2q^2}{(1+p^2+q^2)^2} \Big[|\deltaT_q|^2 / |\phinorm_q|^2 + 2 \re{(\deltaT_q / \phinorm_q)}\Big] \Big[1 + \re{(\deltaT_q / \phinorm_q)}\Big].
\end{align}
We see that the growth rate \(\gamma_2\) of the secondary instability depends both on the amplitudes of the primary fields \(\phinorm_q\) and \(\deltaT_q\), and on their relative phase. 

\subsubsection{No Temperature Perturbation}

If we set \(\deltaT_q = 0\), i.e., ignore the temperature perturbation of the primary streamer, \eqref{eq_sec_disp} gives the well-known dispersion relation for the secondary instability of the modified Hasegawa-Mima model \citep{rogersdorland2000,strintzi2007}:
\begin{equation}
    \label{eq_secondary_hm}
    \gamma_2^\text{HM} = pq |\phinorm_q| \sqrt{\frac{2(1+q^2-p^2)}{1+p^2+q^2}}.
\end{equation}
This form of the secondary instability has long been associated with the strong ZFs observed numerically in ITG turbulence \citep{hammett93}\footnotemark. We will show that the inclusion of the temperature perturbations can introduce qualitative and quantitative changes, and even suppress the secondary instability completely. \footnotetext{Especially in contrast with the much weaker ZFs observed in electron-temperature-gradient-driven (ETG) turbulence on electron scales \citep{jenko2000, strintzi2007}. However, this distinction between ITG and ETG turbulence has recently been challenged by \citet{colyer2017}, who found that the long-time saturated state of ETG turbulence is also dominated by ZFs, although the system does go through a streamer-dominated quasi-saturated state at earlier times.} 

\subsubsection{Long-Wavelength Limit}
\label{sect_sec_longwave}
To simplify \eqref{eq_sec_disp}, we can consider the long-wavelength limit \(p \ll 1\). Then \eqref{eq_sec_disp} gives
\begin{align}
	\label{eq_sec_disp_longwavelength}
	\left[\hat{\gamma}_2^2 + 1 + \frac{q^2\re{(\deltaT_q / \phinorm_q)}}{1+q^2}\right]\left[\hat{\gamma}_2^2 - 1 - \re{(\deltaT_q / \phinorm_q)}\right] = \order{p^2} \approx 0.
\end{align}
Thus, there are two independent branches of the secondary instability with instability conditions given by \(\re{(\deltaT_q / \phinorm_q)} + 1/q^2 < -1\) and \(\re{(\deltaT_q / \phinorm_q)} > -1\), respectively. The second branch is a modified form of the long-wavelength Hasegawa-Mima secondary instability \eqref{eq_secondary_hm}\footnotemark\footnotetext{\citet{plunk2017} found the same expression in the context of the "warm-ion" approximation, i.e., dropping the FLR terms in the nonzonal part of \eqref{curvy_phi}.}:
\begin{equation}
\label{eq_secondary_pqsmall}
\gamma_2 = pq |\phinorm_q| \sqrt{2\left(1+\re {\frac{\deltaT_q}{\phinorm_q}}\right)}.
\end{equation}
We observe that \eqref{eq_secondary_pqsmall} relies only on a handful of the nonlinear terms in \eqref{curvy_phi} and \eqref{curvy_psi}. Substituting \eqref{eq_sec_fields1} and \eqref{eq_sec_fields2} into \eqref{curvy_phi} and taking the limit \(p \ll 1\) gives us the following equations for \(\delta\phinorm_0, \delta\phinorm_+\) and \(\delta\phinorm_-\):
\begin{align}
	\label{eq_sec_phi0}
	&\hat{\gamma_2} \delta\phinorm_0 = \frac{1}{|\phinorm_q|\sqrt{2}} \left[ \delta\phinorm_+(\phinorm_q^* + \deltaT_q^*) - \delta\phinorm_-(\phinorm_q + \deltaT_q) \right],  \\
	\label{eq_sec_phi1}
	&\hat{\gamma_2} \delta\phinorm_+ = \frac{1}{|\phinorm_q|\sqrt{2}} \phinorm_q \delta\phinorm_0,  \\
	\label{eq_sec_phi-1}
	&\hat{\gamma_2} \delta\phinorm_- = -\frac{1}{|\phinorm_q|\sqrt{2}} \phinorm_q^* \delta\phinorm_0.
\end{align}
Substituting \eqref{eq_sec_phi1} and \eqref{eq_sec_phi-1} into \eqref{eq_sec_phi0} yields precisely \eqref{eq_secondary_pqsmall}. We do not consider the equations for the temperature perturbations because \(\delta\deltaT = 0\) is a consistent solution and it corresponds to \eqref{eq_secondary_pqsmall}. The terms on the right-hand side of \eqref{eq_sec_phi1} and \eqref{eq_sec_phi-1} arise from the zonal advection term \(\pbra{\zf{\phinorm}}{\dw{\phinorm} - \nabla^2\dw{\phinorm}}\) in \eqref{curvy_phi} and represent the tilting of the primary streamer by the ZF. The terms on the right-hand side of \eqref{eq_sec_phi0} are the poloidal \exb and diamagnetic flows caused by the interaction of the primary mode and the two sidebands \((k_x, k_y) = (p, \pm q)\). The quantity \(\re{(\deltaT_q / \phinorm_q)}\) controls the response of the primary mode to the zonal perturbation: \(\re{(\deltaT_q / \phinorm_q)} > -1\) yields an unstable ZF, while \(\re{(\deltaT_q / \phinorm_q)} < -1\) results in a stable, oscillatory perturbation. 

\begin{figure}
	\centering
	\includegraphics[scale=0.28]{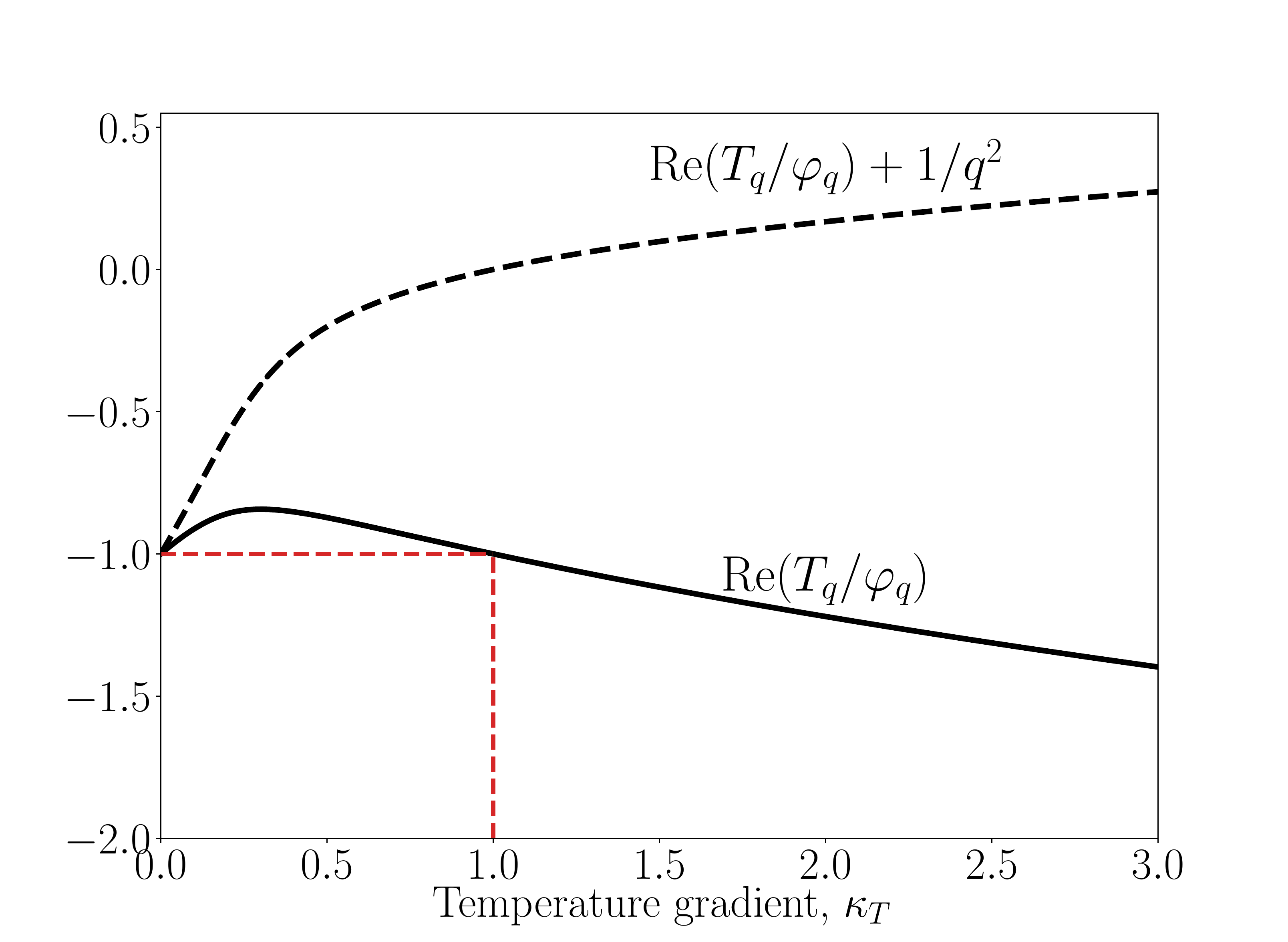}
	\caption{Temperature-gradient dependence of \(\re{(\deltaT_\vk/\phinorm_\vk)}\) (solid) and \(\re{(\deltaT_\vk/\phinorm_\vk)} + 1/q^2\) (dashed) for the most unstable collisionless (\(\chi = 0\)) mode. We find that \(\re{(\deltaT_\vk/\phinorm_\vk)} < -1\) for \(\vt > 1\) and \(\re{(\deltaT_\vk/\phinorm_\vk)} + 1/q^2 > -1\) for all \(\vt\). The secondary instability is present only for \(\vt < 1\). }
	\label{fig_re_collisionless}
\end{figure}

Let us consider the collisionless case (\(\chi = 0\)), where analytical progress is possible, and ask for what values of \(\vt\) the two modes described by \eqref{eq_sec_disp_longwavelength} are unstable. Let us take \((k_x, k_y) = (0, q)\) to be the linear mode with the largest growth rate. We then define the critical gradient \(\vtsec\) for the long-wavelength secondary instability of the fastest-growing streamer as the value of \(\vt\) at which \(\re{(\deltaT_q / \phinorm_q)} = -1\). Using \eqref{eq_disp_relation_nodamping} and the relationship \eqref{eq_temp_phi_linear_ratio} between \(\deltaT_q\) and \(\phinorm_q\), we obtain
\begin{equation}
\label{eq_collisionless_re}
\re{\frac{\deltaT_q}{\phinorm_q}} = -\frac{1 + \vt q^2}{2}.
\end{equation}
To determine \(q\), we seek the maximum of \(\gamma_\vect{k}\), as given by \eqref{eq_gammak_nodamping} for \(k_x = 0\) and \(k_y = q\). We find	
\begin{equation}
\label{eq_dgammak_dky}
\frac{\partial \gamma_\vect{k}}{\partial q} \propto \vt^2 q^6 + 3\vt^2q^4 - q^2 - 4\vt + 1 = 0,
\end{equation}
where the equality holds for the most unstable mode. As an equation for \(q^2\), \eqref{eq_dgammak_dky} is a cubic with only one positive solution for \(\vt > 1/4\). Substituting that solution into \eqref{eq_collisionless_re}, we find \(\re{(\deltaT_q/\phinorm_q)}\) as a function of \(\vt\). This relationship is given in Figure~\ref{fig_re_collisionless}. In particular, we obtain that \(\re{(\deltaT_q/\phinorm_q)} = -1\) at \(\vt = 1\), as can indeed be verified analytically from \eqref{eq_dgammak_dky} and \eqref{eq_collisionless_re}, and \(\re{(\deltaT_q/\phinorm_q)} < -1\) for \(\vt > 1\). We also find that \(\re{(\deltaT_q / \phinorm_q)} + 1/q^2 > -1\) always. Thus, for \(\vt > 1\), the most unstable collisionless ITG mode is stable to the secondary perturbations. Note that \eqref{eq_collisionless_re} depends crucially on the diamagnetic drift \(\vt\py \delsq\phinorm\) in \eqref{curvy_phi}. If we do not include the diamagnetic drift, we find that \(\re{(\deltaT_q / \phinorm_q)} = -1/2\) regardless of \(\vt\) and \(q\), and thus the collisionless secondary instability is never quenched.

\subsubsection{General Case}

Let us go back to the general secondary dispersion relation \eqref{eq_sec_disp}. Its solution is
\begin{align}
	\label{eq_sec_sol}
	\hat{\gamma}_2^2 = \frac{-(U+V) \pm \sqrt{(U-V)^2 + 4W}}{2},
\end{align}
where
\begin{align}
\label{eq_uplusv}
U + V = \frac{p^2\left[1 + \im{(\deltaT_q / \phinorm_q)}^2\right] + p^2 \left[1 + \re{(\deltaT_q / \phinorm_q)}\right]^2 - \re{(\deltaT_q / \phinorm_q)}}{1+p^2+q^2}.
\end{align}
We can use the primary dispersion relation \eqref{eq_disp_relation} to show that \(\re{(\deltaT_\vk / \phinorm_\vk)} < 0\), and hence \mbox{\(U+V>0\)}, for any unstable primary mode with wavenumber \(\vk\). Indeed, the real part of \eqref{eq_temp_phi_linear_ratio} is
\begin{equation}
\label{eq_re_tphi}
\re{\frac{\deltaT_\vk}{\phinorm_\vk}} = \frac{k_y\vt\omega_\vk}{|\gamma_\vk -i\omega_\vk + \chi k^2|^2} < 0
\end{equation}
if \(k_y \omega_\vk < 0\). Let us show that this is true. For \(k \ll 1\) and \(\vt > 1/4\) (the reasons for the latter are discussed at the end of Section~\ref{sect_unstable_region_cless}), the dispersion relation \eqref{eq_disp_relation} gives simply \(k_y\omega_\vk = -k_y^2 / 2 < 0\). Since the solutions to  \eqref{eq_disp_relation} are continuous functions of \(\vk\), if \(k_y\omega_\vk\) changes sign and becomes positive, then \(\omega_\vk = 0\) somewhere. However, if we set \(\omega_\vk = 0\), the imaginary part of \eqref{eq_disp_short} gives \(\gamma_\vk = (g - 1)A < 0\). Therefore, \(k_y\omega_\vk\) cannot change sign within the region of linear instability and so \(k_y\omega_\vk < 0\) for all linearly unstable modes.

\begin{figure}
	\centering
	\begin{center}
		\begin{tabular}{ cc } 
			\includegraphics[scale=0.47]{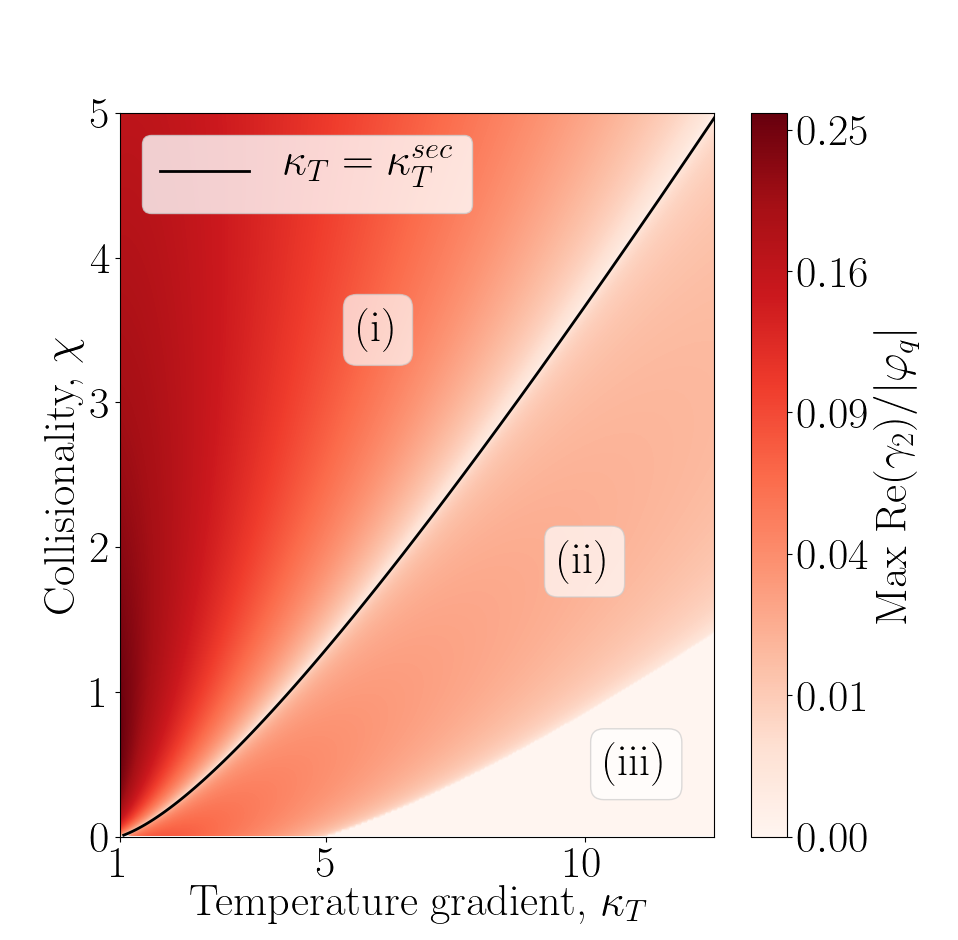} & \includegraphics[scale=0.47]{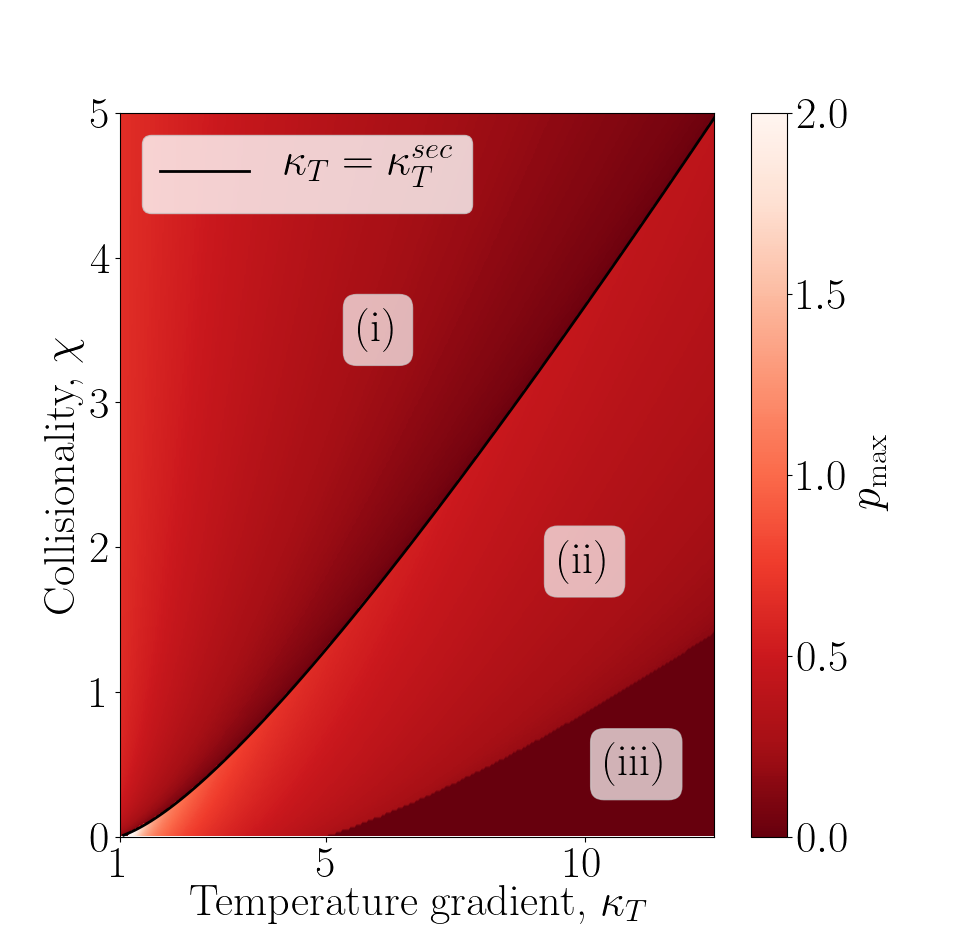}
		\end{tabular}
	\end{center}
	\caption{ \textbf{Left:} Secondary-instability growth rate \eqref{eq_sec_sol} of the most unstable streamer mode, maximised over all values of \(p\). The growth rate vanishes on the \(\re{\left(\deltaT_q / \phinorm_q\right)} = -1\) curve (shown in black). \textbf{Right:} Radial wavenumber \(p_\text{max}\) at which the maximum growth rate shown in the left panel is attained. A discontinuity in the most unstable wavenumber across the \(\re{\left(\deltaT_q / \phinorm_q\right)} = -1\) curve is evident. The absolute-stability region (iii), visible in the bottom right of both panels, where \(\gamma_2 = 0\), is the region where \((U-V)^2 + 4W\) is always positive and \(\gamma_2\) is purely imaginary. }
	\label{fig_sec_parspace}
\end{figure}

\begin{figure}
	\centering
	\includegraphics[scale=0.27]{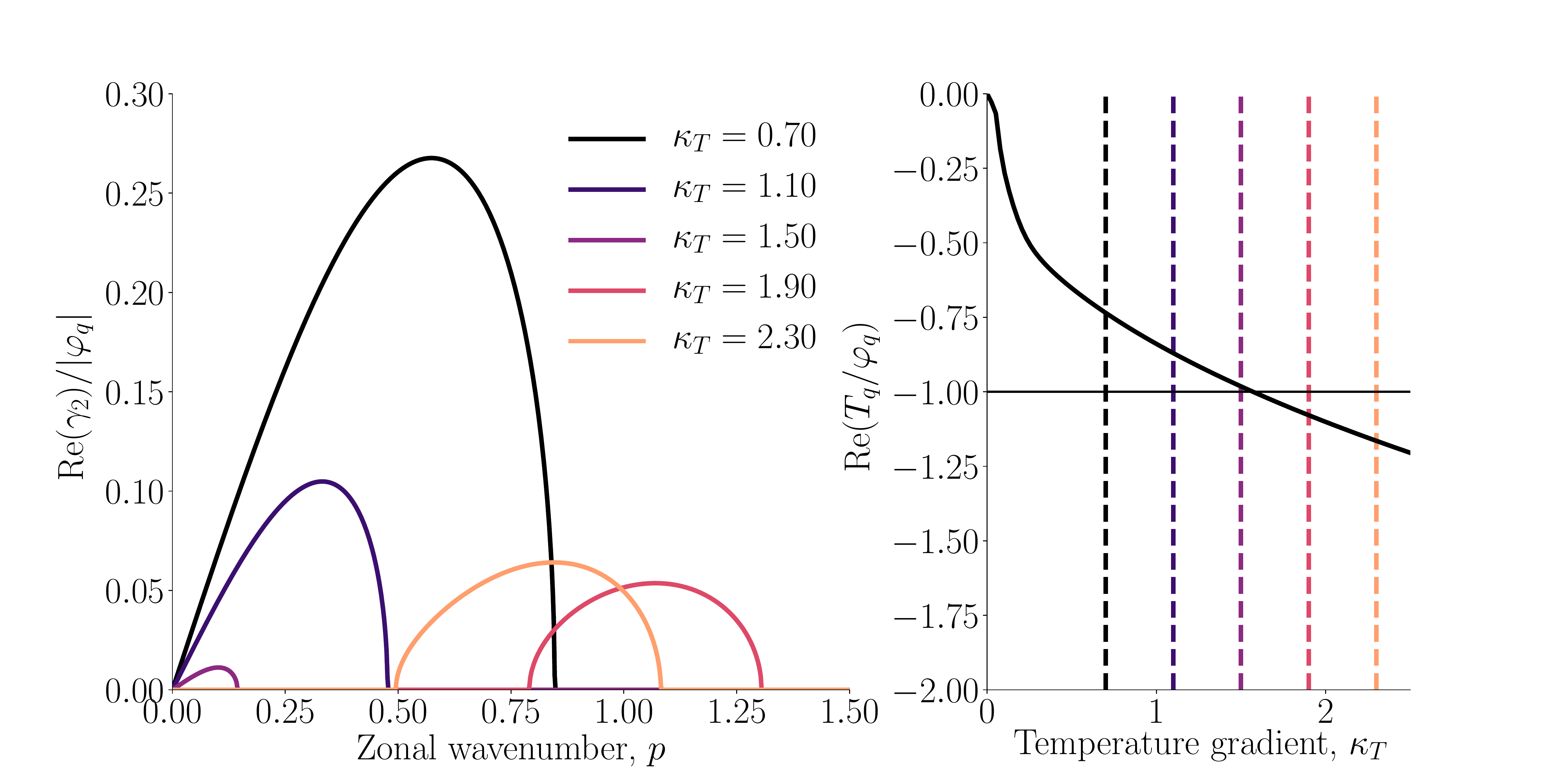}
	\caption{\textbf{Left:} Secondary-instability growth rate for \(\chi = 0.1\) and a number of values of \(\vt\) versus zonal (radial) wavenumber, as given by \eqref{eq_sec_disp}. The primary mode \(\phinorm_q, \deltaT_q\) is taken to be the most unstable one in every case. \textbf{Right:} \(\re{\left(\deltaT_k/\phinorm_k\right)}\) for the fastest-growing linear mode \(\vect{k} = (0, q)\) for \(\chi = 0.1\) as a function of \(\vt\). The dashed lines correspond to the same values of \(\vt\) as in the left panel.}
	\label{fig_sec}
	\label{fig_sec_phase_vs_vt}
\end{figure}

We now consider the solution \eqref{eq_sec_sol} assuming that the relationship between \(\phinorm_q\) and \(\deltaT_q\) is given by \eqref{eq_temp_phi_linear_ratio} with \(k_y = q\) and \(\gamma_\vect{k}\) and \(\omega_\vect{k}\) corresponding to the most unstable mode. This gives us \(\gamma_2\) as a function of \(\vt\), \(\chi\) and \(p\). Figure~\ref{fig_sec_parspace} shows the real part of \(\gamma_2\) maximised over \(p\) for each pair of equilibrium parameters \(\vt\) and \(\chi\), and the wavenumber \(p_\text{max}\) at which that maximum is attained. Let us discuss this figure. There are three distinct regions:
\begin{enumerate}
	\item \(\vt < \vtsec\), where \(\vtsec\) is defined as the value of \(\vt\) where \(\re{(\deltaT_q / \phinorm_q)} = -1\); in this region, \(\re{(\deltaT_q / \phinorm_q)} > -1\). Additionally, \(UV - W < 0\) for \(p = p_\text{max}\), so \(\hat{\gamma}_2^2\) given by \eqref{eq_sec_sol} is real and positive. The instability exists for arbitrarily small values of \(p\) (i.e., for an arbitrarily long wavelength of the ZF). Increasing the temperature gradient \(\vt\) towards \(\vtsec\) has a dramatic effect on the secondary instability of the most unstable mode: it diminishes both the growth rate and the region of zonal wavenumbers that go unstable. On the line \(\vt = \vtsec\), \(\hat{\gamma}_2\) is purely imaginary and there are no growing secondary modes, just like in the long-wavelength analysis of Section~\ref{sect_sec_longwave}. Indeed, substituting \(\re{\left(\deltaT_q / \phinorm_q\right)} = -1\) in \eqref{eq_sec_defs}, we obtain \(W = 0\) and \(U, V > 0\). Then, by \eqref{eq_sec_sol}, \(\hat{\gamma}_2^2 = -U\) or \(-V\). Figure~\ref{fig_sec} (\(\vt = 0.7, 1.1, 1.5\)) shows \(\gamma_2\) vs. \(p\) in region~(i). 
	\item \(\vt > \vtsec\). Increasing \(\vt\) past \(\vtsec\) changes the fastest-growing secondary mode discontinuously. The fastest-growing secondary mode now has \mbox{\(UV - W > 0\)} and \mbox{\((U-V)^2 + 4W < 0\)}. Hence the \(\hat{\gamma}_2^2\) given by \eqref{eq_sec_sol} is complex. In this region, there is always \(\hat{\gamma}_2\) with a positive real part. The peak-growth wavenumber \(p_\text{max}\) changes discontinuously across the \(\re{(\deltaT_q / \phinorm_q)} = -1\) line. For \(\vt > \vtsec\), the secondary instability does not extend to arbitrarily small \(p\) (Figure~\ref{fig_sec}, \(\vt = 1.9, 2.3\)), consistent with the discussion of the long-wavelength secondary instability in Section~\ref{sect_sec_longwave}.
	\item \(\vt > \vtsec\), but now \((U-V)^2 + 4W > 0\) for all values of \(p\), so \(\hat{\gamma}_2^2\) given by \eqref{eq_sec_sol} is real and negative. The location of this region of stability depends on the value of \(\im{(\deltaT_q / \phinorm_q)}\), as well as \(\re{(\deltaT_q / \phinorm_q)}\), and does not have a simple analytic form like the boundary between regions (i) and (ii). 
\end{enumerate}

This analysis of the secondary instability suggests that the system will fail to generate ZFs at a high enough \(\vt\). In what follows, we will indeed find that the zonally dominated Dimits regime ceases to exist when the temperature gradient exceeds a certain threshold, \(\vt > \vtcrit\). However, the na\"ive guess \(\vtcrit \approx \vtsec\), as given by the secondary-instability threshold of the most unstable streamer, does not yield satisfactory agreement with the observed threshold for the Dimits regime (see Section~\ref{sect_dimits_linearapprox}). The secondary-instability picture is incomplete because we must take into account not only whether ZFs can be generated by the ITG modes, but also whether the strong ZFs that support the Dimits regime are resilient to nonzonal perturbations. We shall pick up this topic in Section~\ref{sect_zfstab}.

\subsection{Tertiary Instability}
\label{sect_tert}

To study the stability of a zonal state, we consider infinitesimal ITG perturbations over a background of strong ZF and zonal temperature:
\begin{align}
	\label{eq_tert_ordering}
    &\phinorm = \zf{\phinorm} + \dw{\phinorm}, \ \dw{\phinorm} \ll \zf{\phinorm}, \nonumber \\
    &\deltaT = \zf{\deltaT} + \dw{\deltaT}, \ \dw{\deltaT} \ll \zf{\deltaT},
\end{align}
and linearise \eqref{curvy_phi} and \eqref{curvy_psi} to obtain evolution equations for \(\dw{\phinorm}\) and \(\dw{\deltaT}\). We refer to the ITG modes governed by these linearised equations as "tertiary modes", and to their linear instability as the "tertiary instability" (in truth, this is just the primary ITG instability but for an equilibrium state modified by the zonal fields). We will discover that this instability can seed turbulent perturbations in the Dimits regime, but is not solely responsible for the transition to strong turbulence (see Sections~\ref{sect_nl} and \ref{sect_zfstab}). Further discussion of the tertiary instability has been exiled to Appendix~\ref{appendix_tert}.

\section{Nonlinear Saturation and Zonal Staircase}
\label{sect_nl}

\begin{figure}
	\centering
	\includegraphics[scale=0.27]{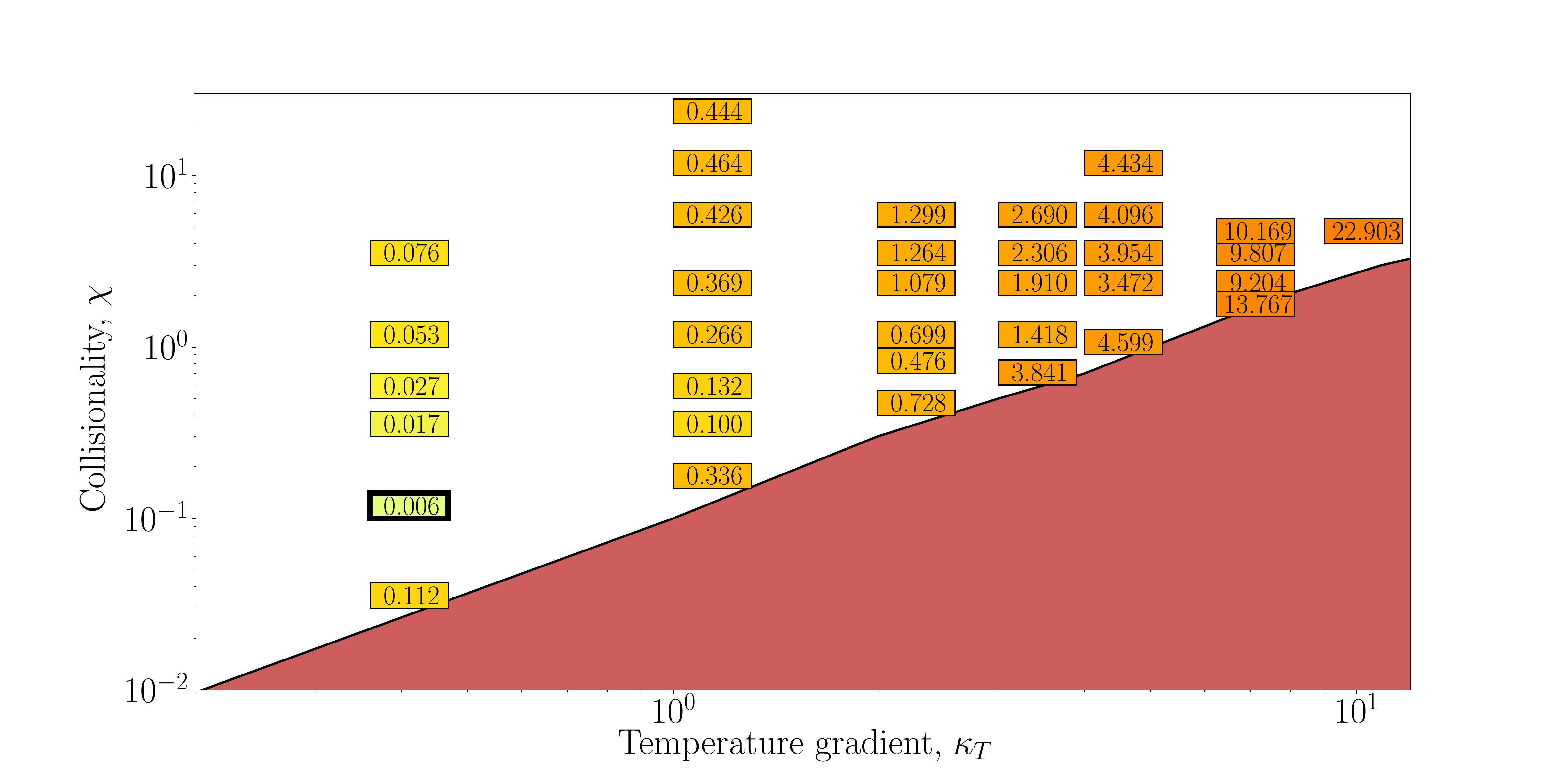}
	\includegraphics[scale=0.27]{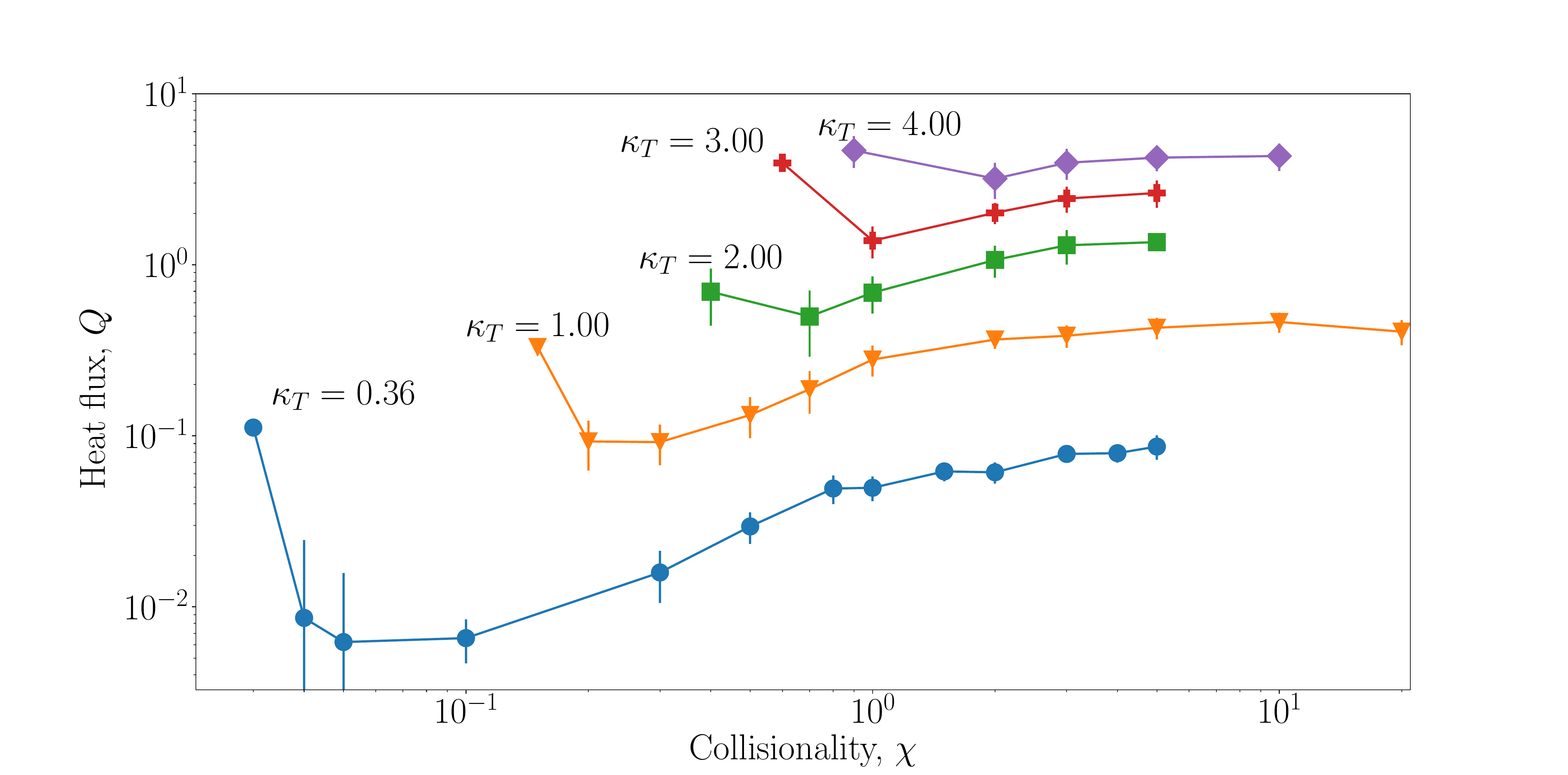} 
	\caption{ \textbf{Top:} Box-averaged heat flux \(Q\) as a function of \(\vt\) and \(\chi\). \(Q\) is defined in \eqref{eq_heatflux_def} and given here in units of \(3 n_i T_i \vti \left(\rho_i / L_B\right)^2 / \tau^{5/2} \sqrt{2}\); see also the footnote on page~\pageref{fn_hf}. The shaded (in red) region is beyond the Dimits threshold, where strong turbulence resides (see Section \ref{sect_blowup}). Bold-framed is the parameter point corresponding to SimL and SimH, viz., \(\vt = 0.36, \chi = 0.1\). \textbf{Bottom:} Box-averaged heat flux \(Q\) in the saturated state versus \(\chi\) for various \(\vt\).}
	\label{fig_parspace}
\end{figure}
\begin{figure}
	\centering
	\includegraphics[scale=0.27]{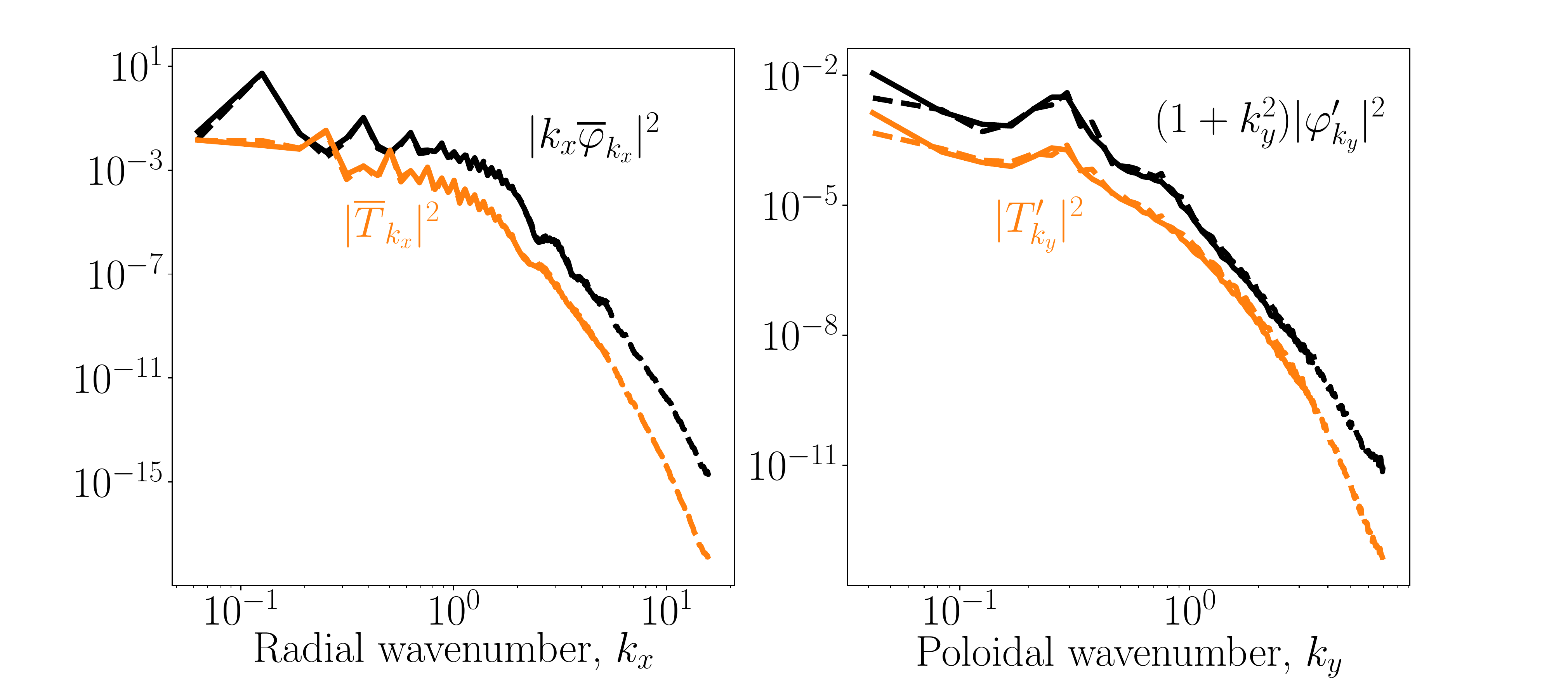}
	\caption{ Comparison of the spectra of turbulence for SimL (solid) and SimH (dashed), described in Section~\ref{sect_nl}. \textbf{Left:} ZF velocity \(|k_x\zf{\phinorm}_{k_x}|^2\) (black) and zonal temperature (orange) \(|\zf{\deltaT}_{k_x}|^2\) spectra. \textbf{Right:} Streamer (\(k_x = 0\)) contribution to \eqref{eq_phi_cons}, viz., \((1+k_y^2)|\dw{\phinorm}_{k_y}|^2\), (black) and temperature \(|\dw{\deltaT}_{k_y}|^2\) (orange). A clear peak at \(k_y \approx 0.25\) is seen. This corresponds to the dominant poloidal wavenumber in the ZF minima (see also Figure~\ref{fig_convection_zones_spectra} in Appendix~\ref{appendix_tert}). The fastest linearly growing streamer has \(k_y \approx 1\). }
	\label{fig_conv_spectra}
\end{figure}
\begin{figure}
	\centering
	\includegraphics[scale=0.26]{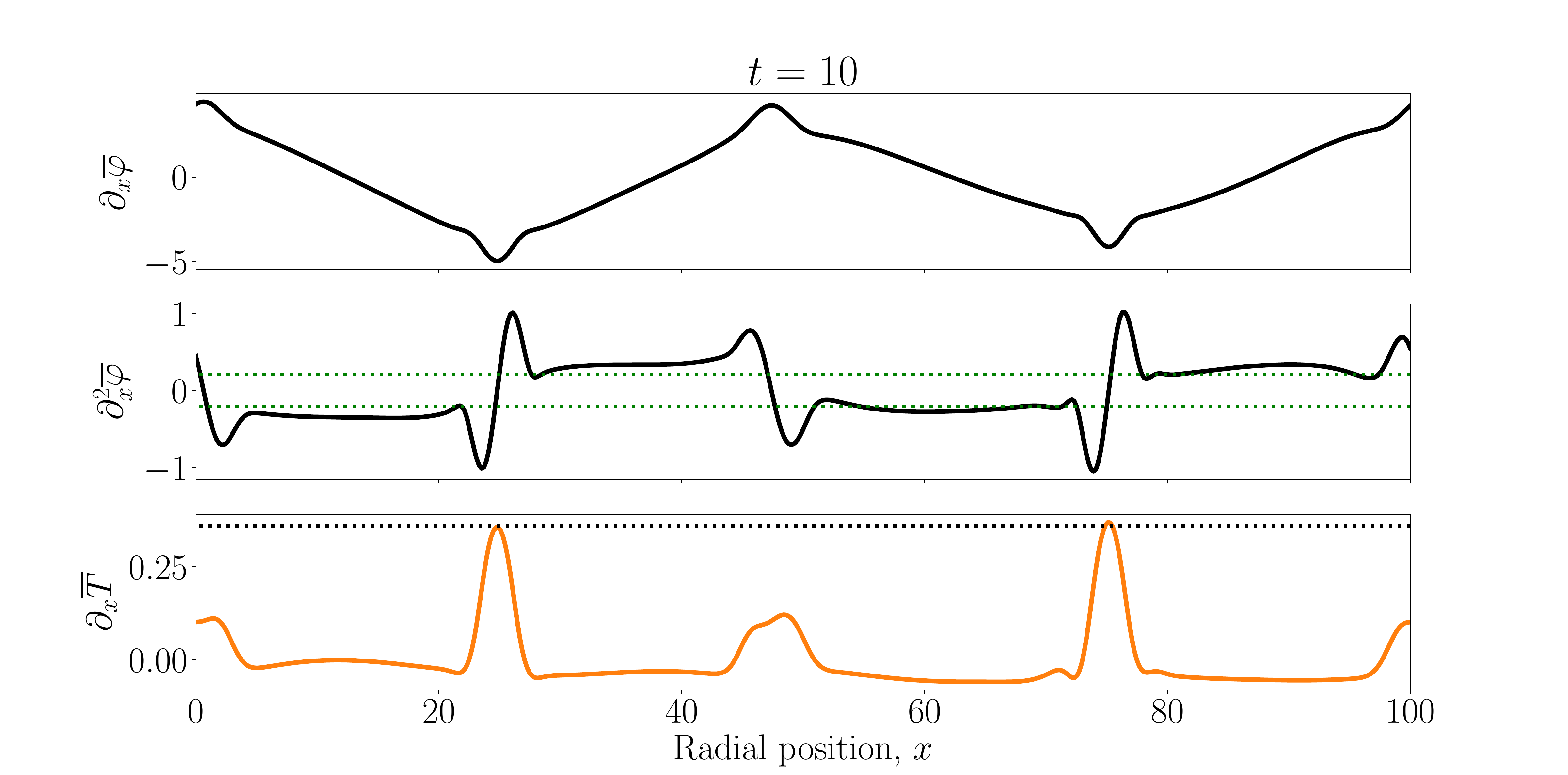}
	\includegraphics[scale=0.26]{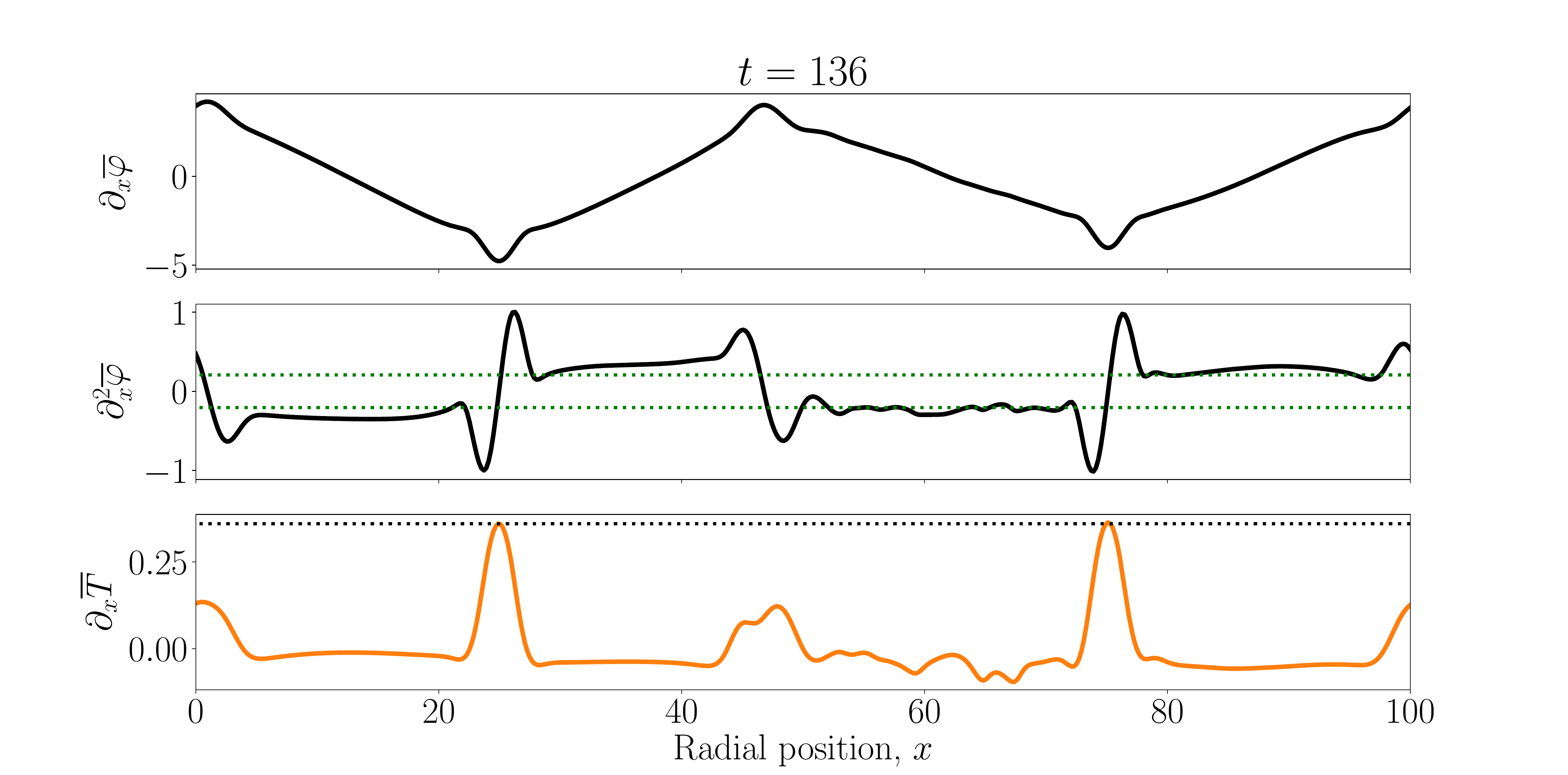}
	\includegraphics[scale=0.26]{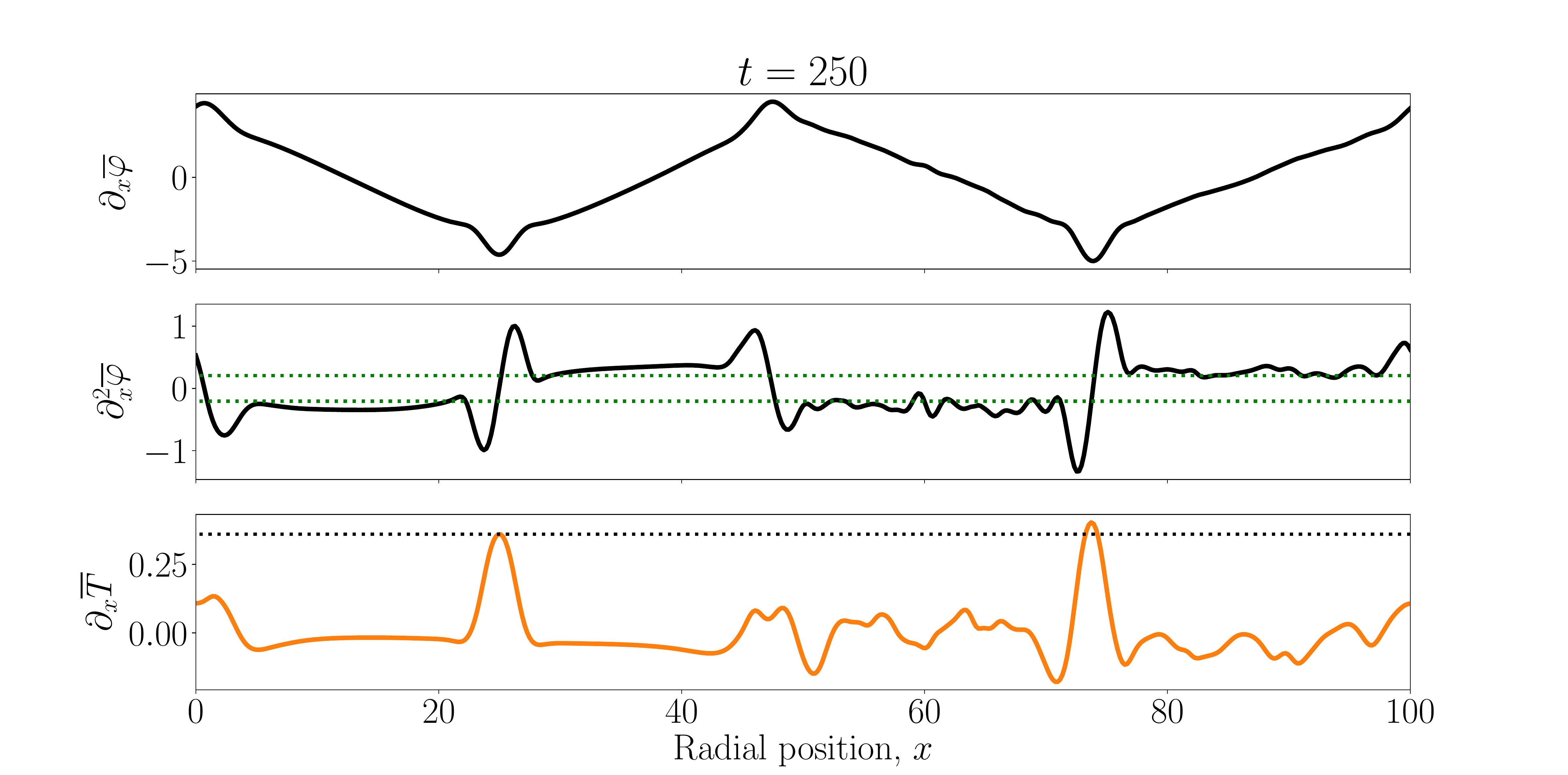}
	\caption{Radial profiles of ZF (\(\px \zf{\phinorm}\)), zonal shear (\(\px^2 \zf{\phinorm}\)) and zonal-temperature gradient (\(\px \zf{\deltaT}\)). The dotted green line corresponds to the largest linear ITG growth rate \(\pm \gamma_\text{max}\). The dotted black line shows the value of the equilibrium temperature gradient \(\vt\). Note that the turbulence that develops in the shear zones does not disturb the ZF and zonal shear significantly. See Figure~\ref{fig_snapshot_t_dimits} for 2D snapshots at these same times. The data is from SimH. The locations of the ZF extrema are determined by the initial conditions used, see Section~\ref{sect_zfscale}.}
	\label{fig_staircase_profiles}
\end{figure}
\begin{figure}
	\centering
	\includegraphics[scale=0.47]{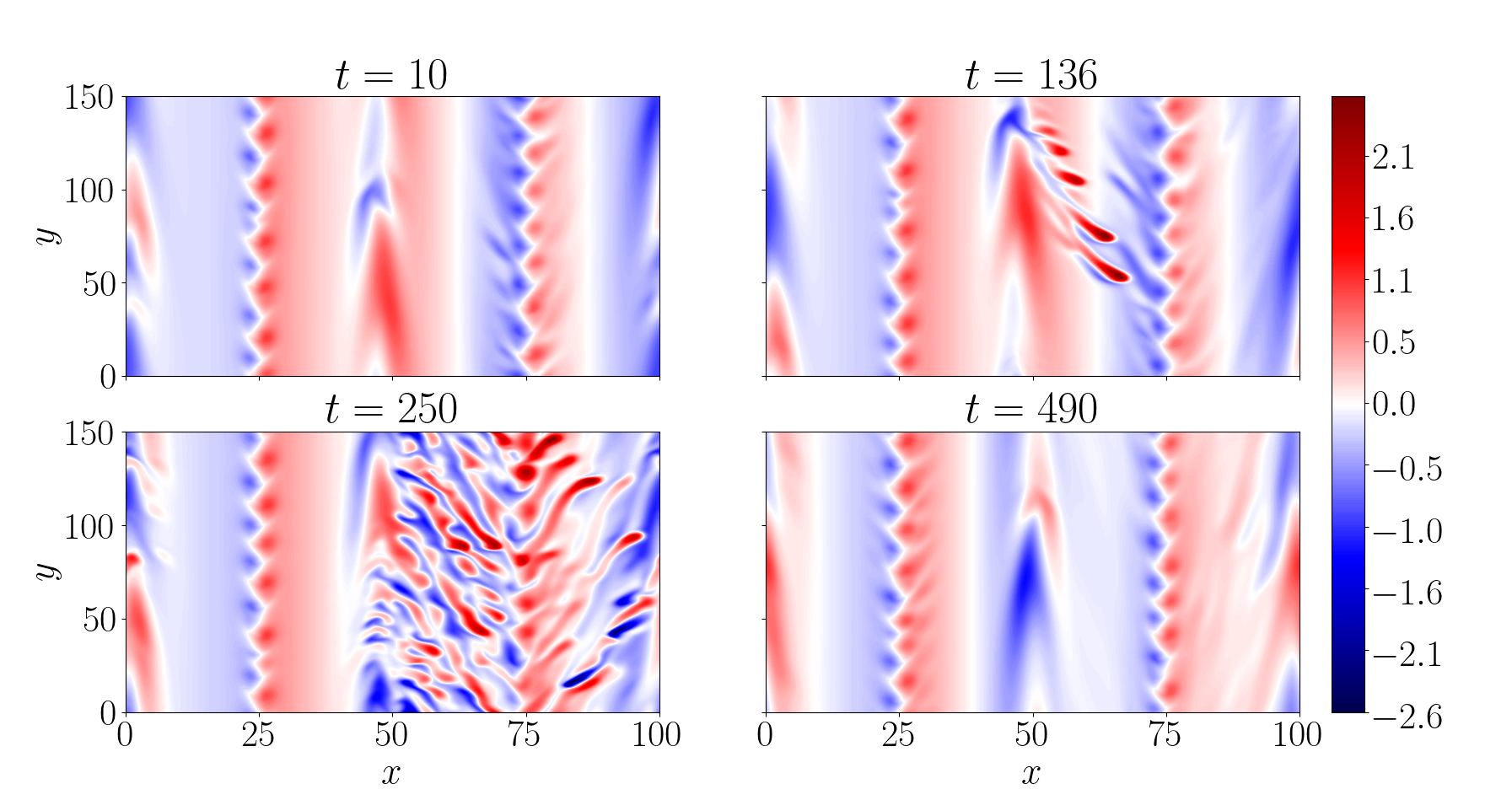}
	\caption{Snapshots of temperature perturbations in the Dimits state (a movie is available in the supplementary material). The data is from SimH. \textbf{Top left:} Quiescent, \(t = 10\); there is a ZF minimum at \(x = 24.5\) and a ZF maximum at \(x = 47\). \textbf{Top right:} Ferdinons visible around \(x = 65\), \(t = 136\). A zoomed-in version can be found in Figure~\ref{fig_ferd_closeup}. \textbf{Bottom left:} Turbulent burst, \(t = 250\). \textbf{Bottom right:} Relaxation back to the zonal staircase after the burst, \(t = 490\). The full time history of SimH can be found in Figure~\ref{fig_burst_heatflux} and the radial profiles of the zonal fields and heat flux are shown in Figure~\ref{fig_staircase_profiles}. }
	\label{fig_snapshot_t_dimits}
\end{figure}
\begin{figure}
	\centering
	\includegraphics[scale=0.47]{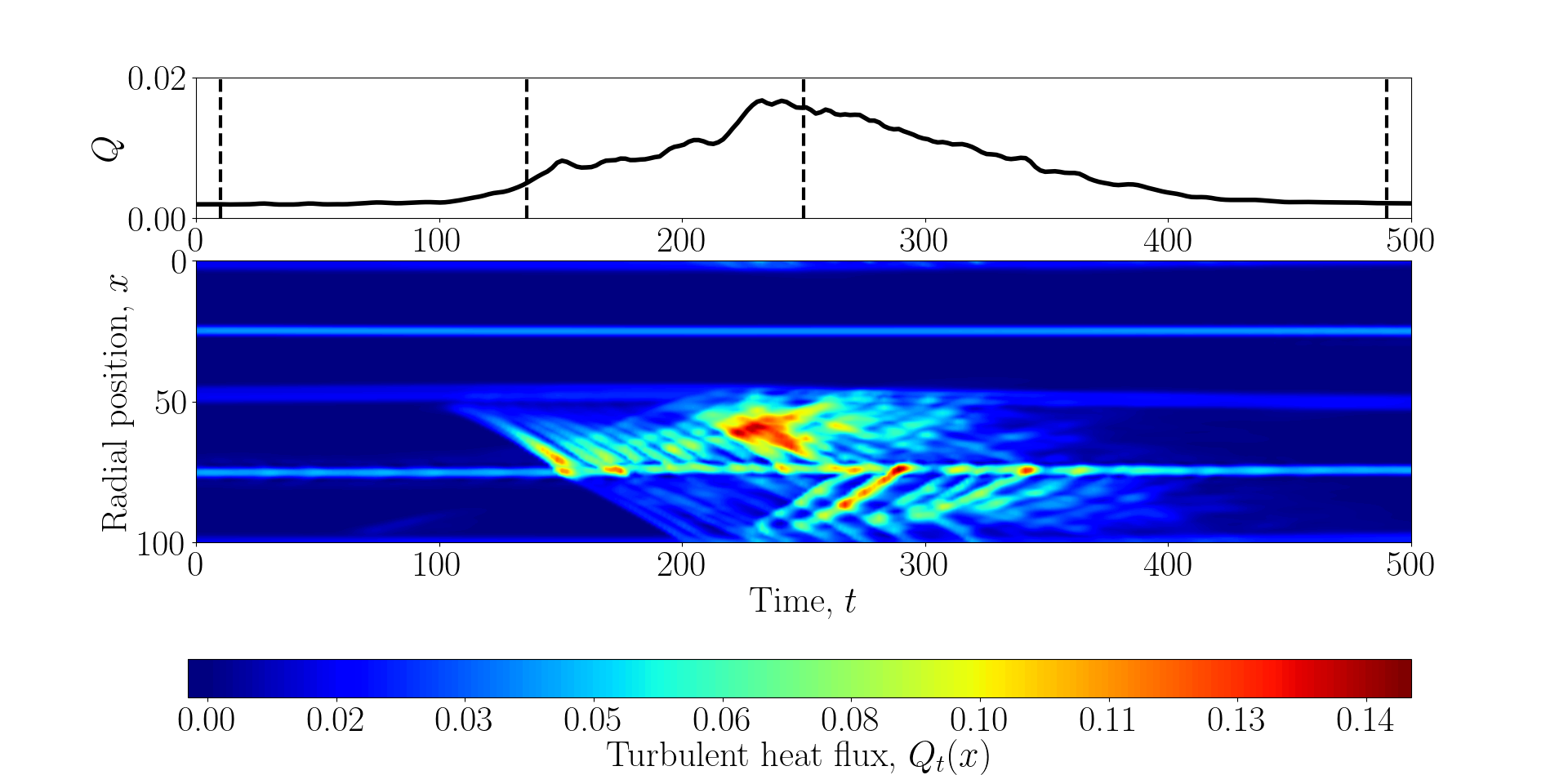}
	\caption{\textbf{Top:} Time evolution of the total heat flux \(Q\) during a turbulent burst for SimH. The dashed lines correspond to the times used for Figures~\ref{fig_staircase_profiles} and~\ref{fig_snapshot_t_dimits}. \textbf{Bottom:} Time trace of the local (integrated only over \(y\)) radial turbulent heat flux \(Q_t(x) = -\zf{T\py\phinorm}\) as a function of radial position. A turbulent burst in the right half of the domain is clearly visible for \(t \in [100, 400]\). The linear streaks correspond to radially drifting ferdinons (see Section~\ref{sect_ferds}). }
	\label{fig_burst_heatflux}
\end{figure}
\begin{figure}
	\centering
	\includegraphics[scale=0.47]{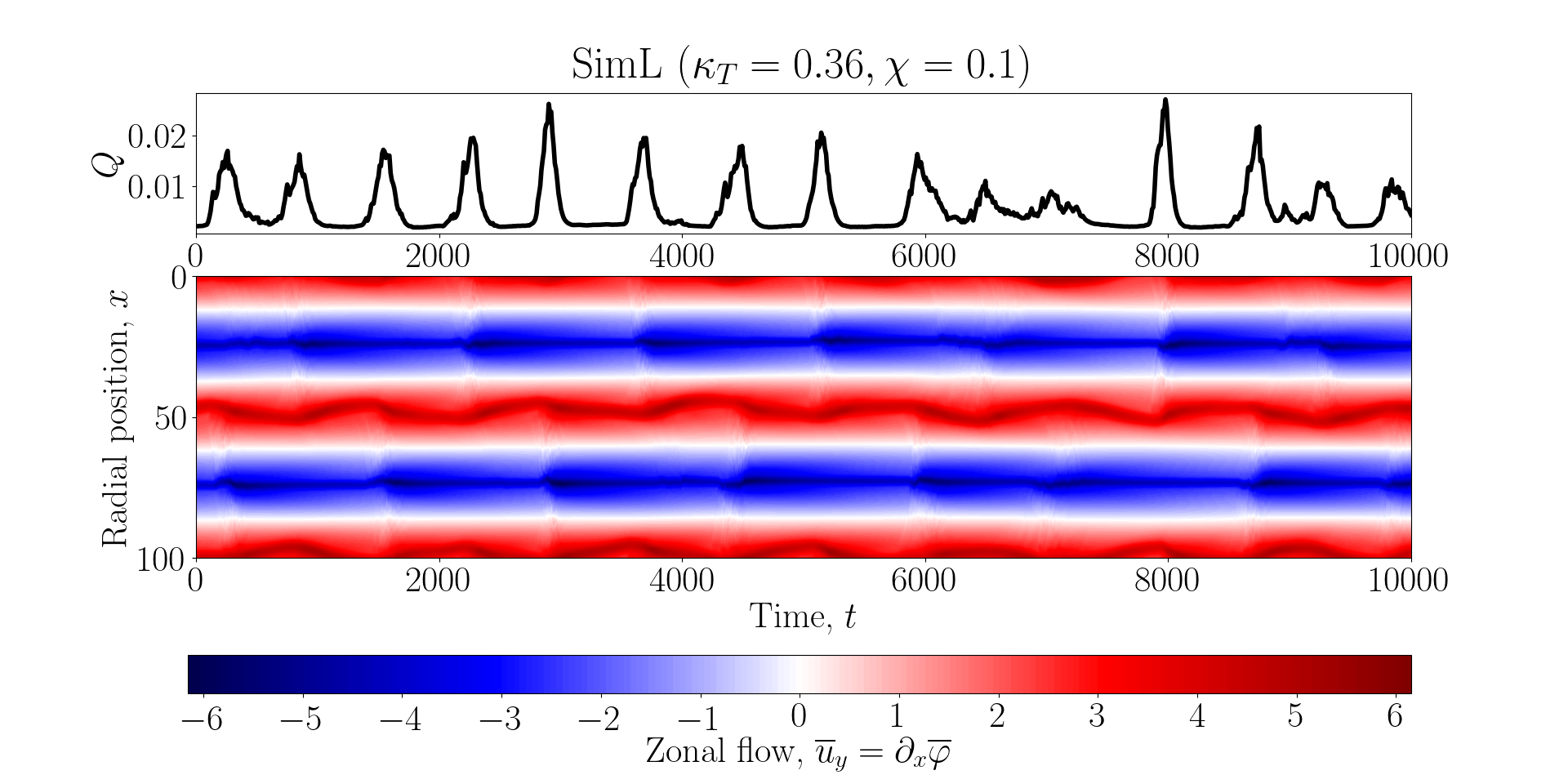}
	\caption{ Box-integrated radial heat flux \(Q\) and ZF velocity (\(\px \zf{\phinorm}\)) vs. time for SimL (\(\vt = 0.36\), \(\chi = 0.1\)). Each turbulent burst is accompanied by an order-of-magnitude increase in \(Q\) and a radial oscillation of the locations of the ZF maxima.}
	\label{fig_zf_r_t}
\end{figure}

We now proceed to investigate the saturated state of \eqref{curvy_phi} and \eqref{curvy_psi} numerically and semi-analytically. A well-defined saturated state is found only for temperature gradients below a critical gradient \(\vt < \vtcrit\), where \(\vtcrit\) is an increasing function of collisionality \(\chi\) (see Figure~\ref{fig_parspace}). The saturated state is always dominated by strong zonal flows and exhibits levels of turbulent transport that are low compared to the equilibrium diffusive transport (\(Q \lesssim \chi \vt\)). We will refer to this state as the Dimits state. The critical gradient \(\vtcrit\) is then the nonlinear critical gradient that marks the break up of the zonally dominated state and the onset of fully developed ITG turbulence. We will relate \(\vt^{c}\) to the resilience of the zonal profiles in the face of nonzonal perturbations, which is in turn determined by the behaviour of turbulence in the presence of strong (comparable to the ITG growth rate) zonal shear. For \(\vt < \vtcrit\), zonally sheared turbulence enhances the ZFs that are doing the shearing through a negative turbulent viscosity. Beyond the Dimits threshold (\(\vt > \vtcrit\)), the turbulent viscosity is positive, and strong, ITG-suppressing ZFs cannot be maintained. These results are presented in Section~\ref{sect_zfstab}, but first, in this section, we shall describe the saturated state near the Dimits threshold. 

Figure \ref{fig_parspace} shows the heat flux \(Q\) vs. \(\vt\) and \(\chi\). We have checked that all simulations have converged by inspection of their heat flux and ZF profiles, and by ensuring that they run for several box-scale diffusion times \(t_\text{box} = (L_x / 2\pi)^2 / a\chi\). The turbulent heat flux \(Q\) depends strongly on the temperature gradient and increases monotonically with increasing \(\vt\). In contrast, its dependence on the collisionality is much weaker and non-monotonic (see Figure~\ref{fig_parspace}, bottom panel). Close to the Dimits threshold, \(Q\) decreases with increasing \(\chi\) (which takes it away from the threshold), whereas farther away from the threshold, it increases and then plateaus with increasing \(\chi\). An increase of flux with collisionality for \(Z\)-pinch turbulence was noted by \citet{ricci2006}.

In what follows, a significant fraction of the detailed analysis is done using two simulations of the low-collisionality near-marginal state with parameters \(\vt = 0.36\), \(\chi=0.1\), \(L_x = 100\), \(L_y = 150\), one with higher (\(507\times337\)) and one with lower (\(167\times167\)) number of Fourier modes (the lower-resolution simulation is used for longer runs due to its lower computational cost). They have the same initial condition, taken from an already saturated simulation. Both the low- and high-resolution simulations show good convergence of their spectra (see Figure~\ref{fig_conv_spectra}). We shall refer to these two simulations as "SimL" and "SimH", respectively.

In the near-marginal Dimits state, turbulence is suppressed by a quasi-static "zonal staircase" arrangement of the ZFs and zonal temperature perturbations. This structure is reminiscent of the "\exb staircase" observed in global GK simulations \citep{pradalier2010, difpradalier2015, difpradalier2017, villard2013, villard2014}. The zonal staircase consists of interleaved regions of strong zonal shear that suppresses the ITG turbulence in those regions, and localised turbulent patches at the turning points of the ZF velocity. We shall refer to the former as the "shear zones" (Section~\ref{sect_shearzones}) and to the latter as the "convection zones" (Section~\ref{sect_convzones}). A typical near-marginal ZF configuration can be seen in Figure~\ref{fig_staircase_profiles} and corresponding snapshots of the perturbed temperature in Figure~\ref{fig_snapshot_t_dimits}. Turbulence is always present, in a highly localised form, in the convection zones, but not in the shear zones.

The ZF in the staircase is not steady, but subject to viscous decay. In the low-collisionality (\(\chi \lesssim 1\)), near-marginal regime, this decay is slow and turbulent bursts are triggered periodically when the zonal shear in the shear zones has decayed to a level that is insufficient for the suppression of turbulence. These bursts lead to a significant (order-of-magnitude) increase in the radial heat flux. Similar bursts were reported by \citet{kobayashi2012} in entropy-mode-driven \(Z\)-pinch turbulence. In our system, they are seeded by highly localised, coherent, turbulent structures, reminiscent of those reported by \citet{vanwyk2016} in gyrokinetic turbulence with an imposed equilibrium flow shear (see Section~\ref{sect_ferds}). A typical turbulent burst is illustrated in Figures~\ref{fig_staircase_profiles}, \ref{fig_snapshot_t_dimits} and \ref{fig_burst_heatflux} where we see the evolution of the quiescent state into a turbulent one and then back. Figure~\ref{fig_zf_r_t} shows a longer time evolution for the same parameters, illustrating the (quasi)periodic nature of the bursts. At higher collisionality, the ZFs decay faster, the turbulent bursts start to overlap, and it becomes difficult to isolate quiescent periods from turbulent ones. This state is more homogeneous in time and does not have well-defined oscillations, unlike the bursty state at low collisionality. However, we find that the mechanism that governs the stability of the Dimits state and the transition to strong turbulence is very similar for all values of collisionality that we have explored (see Section~\ref{sect_dimits_threshold}). 

We now proceed to describe the features of the zonal staircase in more detail.

\subsection{Shear Zones}
\label{sect_shearzones}
\begin{figure}
	\centering
	\includegraphics[scale=0.27]{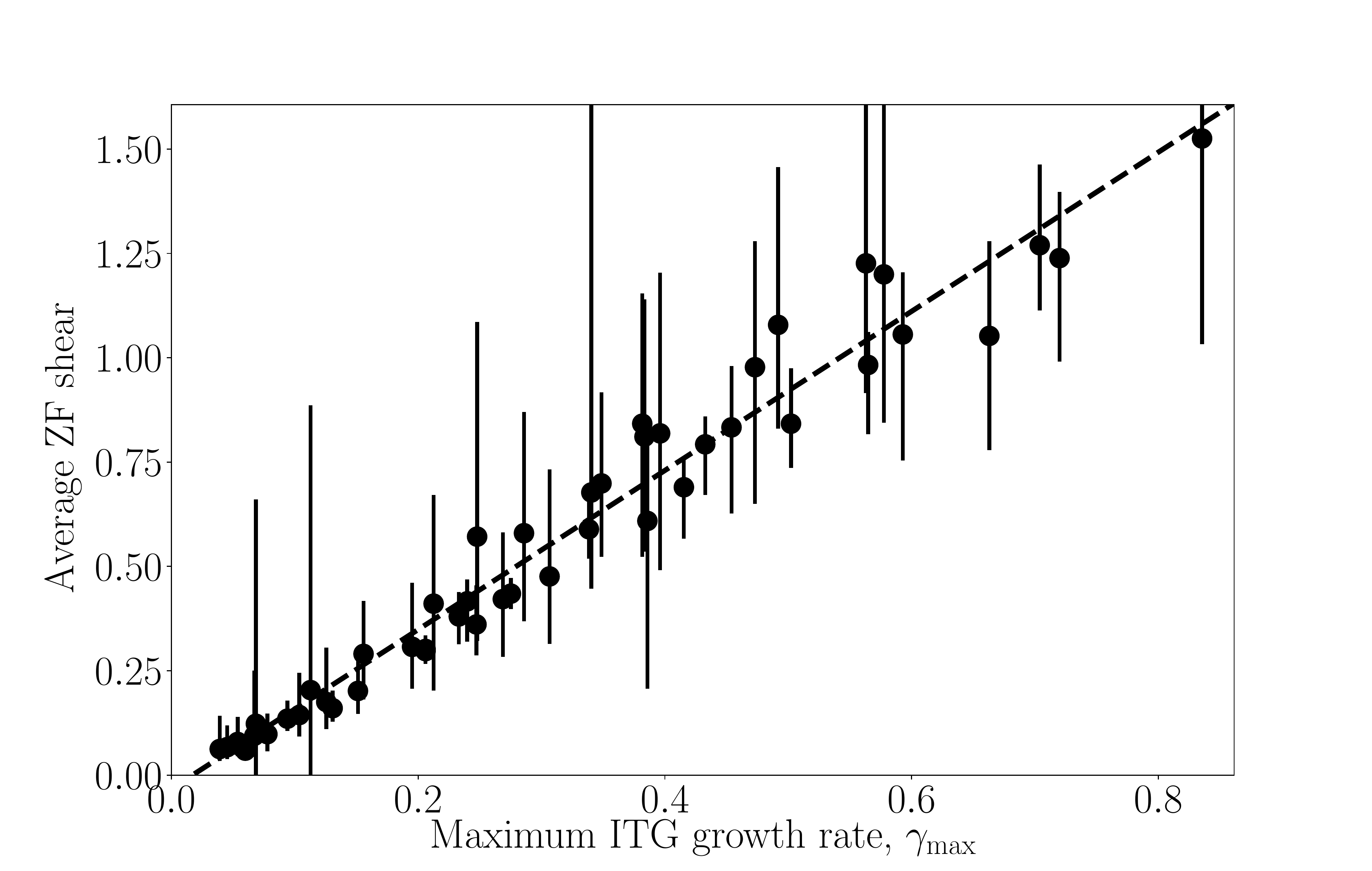}
	\caption{ZF shear \(S\) (time- and space-averaged over the shear zones) vs. maximum linear ITG growth rate \(\gamma_\text{max}\) [as given by \eqref{eq_disp_relation}]. The data is taken from a number of simulations over a range of parameters: \(\vt \in [0.16, 7.29]\) and \(\chi \in [0.1, 10]\). The error bars represent the smallest and the largest values of the spatially averaged ZF shear for each simulation. The best-fit line (dashed) is \(S \approx 2 \gamma_\text{max}\).}
	\label{fig_zfshear_growth}
\end{figure}
\subsubsection{Suppression of Turbulence}

The zonal staircase is arranged in such a way that it efficiently suppresses turbulence in the shear zones via strong ZF shear. We find that this shear, \(S \equiv \px^2 \zf{\phinorm}\), satisfies \(S \gtrsim \gamma_\text{max}\), where \(\gamma_\text{max}\) is the largest linear ITG growth rate determined from the dispersion relation \eqref{eq_disp_relation}. The notion that ITG turbulence requires comparable \(S\) and \(\gamma_\text{max}\) to be suppressed by shear is known as the "quench rule" \citep{waltz94, waltz98, kinsey2005, kobayashi2012}. Quantitatively, this is supported by Figure~\ref{fig_zfshear_growth}, which shows that the time- and space-averaged (over the shear zones) shear satisfies \(S \approx 2 \gamma_\text{max}\) over a range of simulation parameters\footnotemark. Note that the \textit{particular snapshots} of zonal profiles seen in Figure~\ref{fig_staircase_profiles} suggest \(S \approx \gamma_\text{max}\). However, the \textit{time-averaged} \(S\) is larger due to the variation of shear over time (see also Figure~\ref{fig_zfdecay_with_hf}). \footnotetext{The averaging is performed numerically over regions of near-uniform zonal shear, where, at every time step, we identify the radial locations of the uniform shear zones by applying the following conditions: \(\px^3 \zf{\phinorm} < 0.1 \text{max}\lbrace\px^3 \zf{\phinorm}\rbrace\) to isolate regions of near-uniform shear, and \(\px^2 \zf{\phinorm} < 0.5 \text{max}\lbrace\px^2 \zf{\phinorm}\rbrace\) to exclude the large variations of shear around the ZF extrema (see Figure~\ref{fig_staircase_profiles}).}

\FloatBarrier

\subsubsection{Decay of Zonal Flows}
\label{sect_zfdecay}

Let us study the viscous decay of the ZFs. The equation for the evolution of the ZFs is given by the zonal part of \eqref{curvy_phi}:
\begin{equation}
\label{eq_zf}
\pt \px^2 \zf{\phinorm} = \px^2 \zf{\px \phinorm \py \left( \phinorm + \deltaT \right)} + \chi \px^4 \left(a \zf{\phinorm} - b \zf{\deltaT}\right).
\end{equation}
Integrating \eqref{eq_zf} once yields
\begin{equation}
\label{eq_zfflux}
\pt \zf{u}_y = \px \left[ \zf{\left(\px \phinorm\right) \py \left( \phinorm + \deltaT \right)} + \chi \px^2 \left(a \zf{\phinorm} - b \zf{\deltaT}\right)\right] = -\px \left( \Pi_t + \Pi_d \right),
\end{equation}
where the zonal flow velocity is \(\zf{u}_y \equiv \px \zf{\phinorm}\) and we have identified the turbulent, \mbox{\( \Pi_t \equiv - \zf{\left(\px \phinorm\right) \py \left( \phinorm + \deltaT \right)}\)}, and diffusive, \(\Pi_d \equiv - \chi \px^2 \left(a \zf{\phinorm} - b \zf{\deltaT}\right)\), radial fluxes of poloidal momentum. The integration constant in \eqref{eq_zfflux} is zero because both sides of the equation are exact derivatives with respect to \(x\) and our domain is periodic. Integrating \eqref{eq_zfflux} once more yields a term that is not necessarily an exact derivative --- the turbulent momentum flux \(\Pi_t\):
\begin{equation}
\label{eq_zfflux_int}
\pt \zf{\phinorm} +  \Pi_t + \Pi_d = \Pi,
\end{equation}
where the integration constant \(\Pi = (1/L_x) \int_0^{L_x} dx \ \Pi_t\) is the total box-averaged poloidal momentum flux. However, \eqref{curvy_phi} and \eqref{curvy_psi} are invariant under the symmetry
\begin{equation}
\label{eq_symmetry}
x \mapsto -x, \ y \mapsto y, \ \phinorm \mapsto -\phinorm, \ \deltaT \mapsto - \deltaT.
\end{equation}
Under this symmetry, \(\Pi \mapsto -\Pi\), a property of our model inherited from gyrokinetics \citep{parra2011}. Therefore, assuming that the volume-averaged solutions to \eqref{curvy_phi} and \eqref{curvy_psi} respect \eqref{eq_symmetry}, we conclude that \(\Pi = 0\). This is confirmed by our numerical solutions. Thus, the right-hand side of \eqref{eq_zfflux_int} vanishes.

\begin{figure}
	\centering
	\includegraphics[scale=0.27]{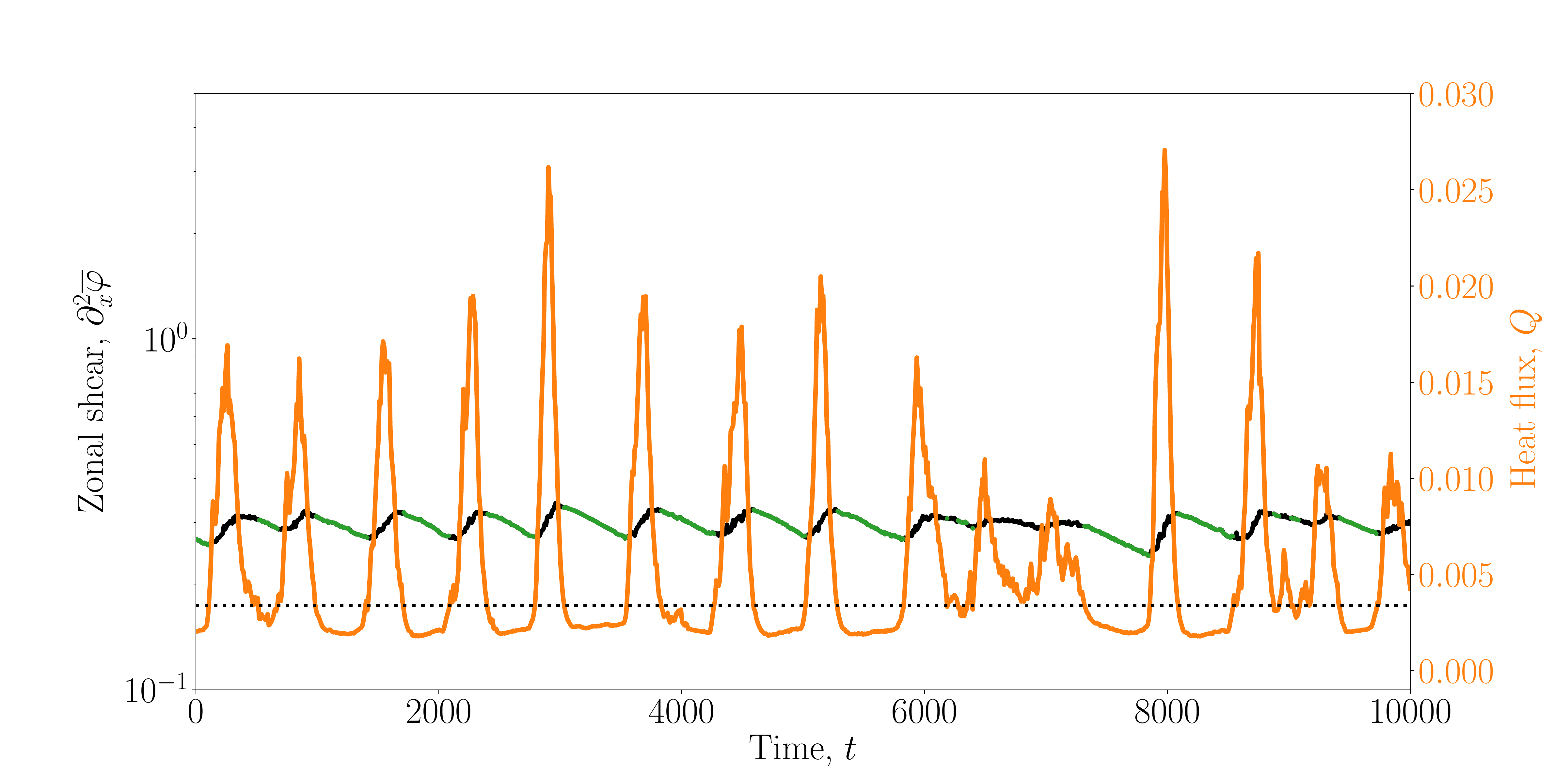}
	\caption{ Heat flux \(Q\) (orange) and zonal shear \(\px^2 \zf{\phinorm}\) (black) vs. time for SimL. The highlighted (in green) sections of the zonal shear correspond to the quiescent periods. They are identified as those in which \(Q\) is smaller than a threshold value (dashed black line), defined as \(60\%\) of the time-averaged \(Q\). The average decay rate of the zonal shear in the thus-identified quiescent periods is \(\gamma \approx -3.5\times10^{-4}\) and the decay rate given by \eqref{eq_def_gamma_box} is \(\gamma_\text{s} \approx - 3.6\times10^{-4}\). }
	\label{fig_zfdecay_with_hf}
\end{figure}

During the quiescent periods of the Dimits-state evolution (i.e., between turbulent bursts), the turbulent momentum flux in the shear zones is negligible compared to the diffusive momentum flux, \(\Pi_t \ll \Pi_d\). This is a consequence of the suppression of the \(k_y \neq 0\) ITG modes by the zonal shear\footnotemark.\footnotetext{Note that \(\Pi_t\) is not small if there is turbulence present in the shear zones (which happens in the run up to and during turbulent bursts) --- we shall investigate \(\Pi_t\) in Section~\ref{sect_zfstab}. } We also find that the zonal temperature gradient \(\px \zf{T}\) is approximately constant in the quiescent shear zones (see Figure~\ref{fig_staircase_profiles}), so \(\px^2 \zf{T} = 0\). Therefore, \eqref{eq_zfflux_int} becomes
\begin{equation}
	\label{eq_zfdiffusioneq}
	\pt \zf{\phinorm} = a\chi\px^2 \zf{\phinorm}.
\end{equation}
This is a diffusion equation governing the viscous decay of the ZFs with a collisional viscosity \(a\chi\). As Figure~\ref{fig_zfdecay_with_hf} shows, quiescent periods of low heat flux and, thus, low levels of nonzonal perturbations, are correlated with the periods of decay of the zonal shear. We find that, despite the ever-present turbulence in the convection zones, where \eqref{eq_zfdiffusioneq} does not hold, the decay rate of the zonal shear is closely approximated by the viscous decay rate of the longest-wavelength ZF that comprises the zonal staircase, viz., 
\begin{equation}
\label{eq_def_gamma_box}
\gamma_\text{s} = -a \chi \left(\frac{2 \pi n}{L_x}\right)^2,
\end{equation}
where \(n\) is the number of periods of the zonal staircase in the domain of radial size \(L_x\). 

Let us now discuss what the ZF periodicity is.

\begin{figure}
	\centering
	\includegraphics[scale=0.427]{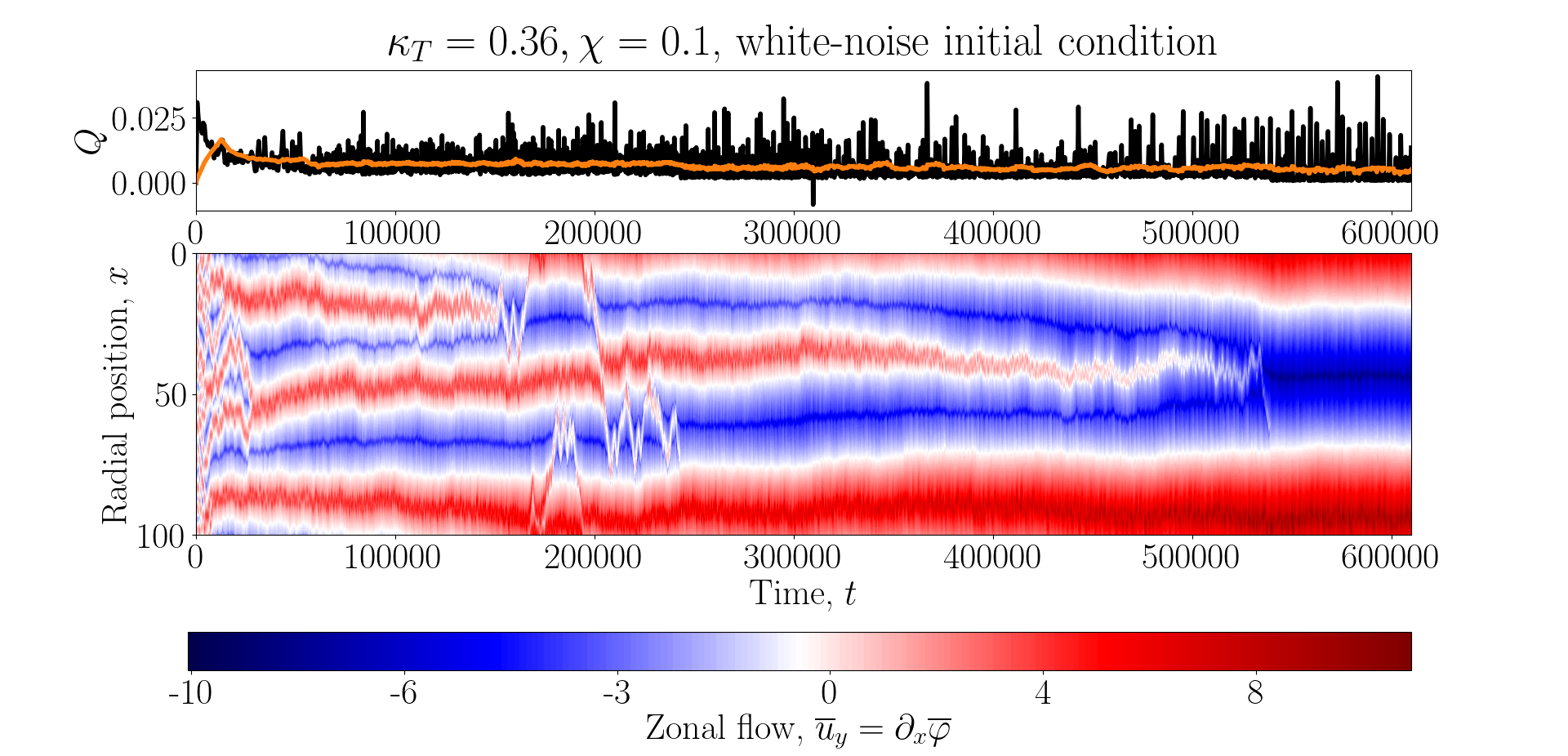}
	\includegraphics[scale=0.427]{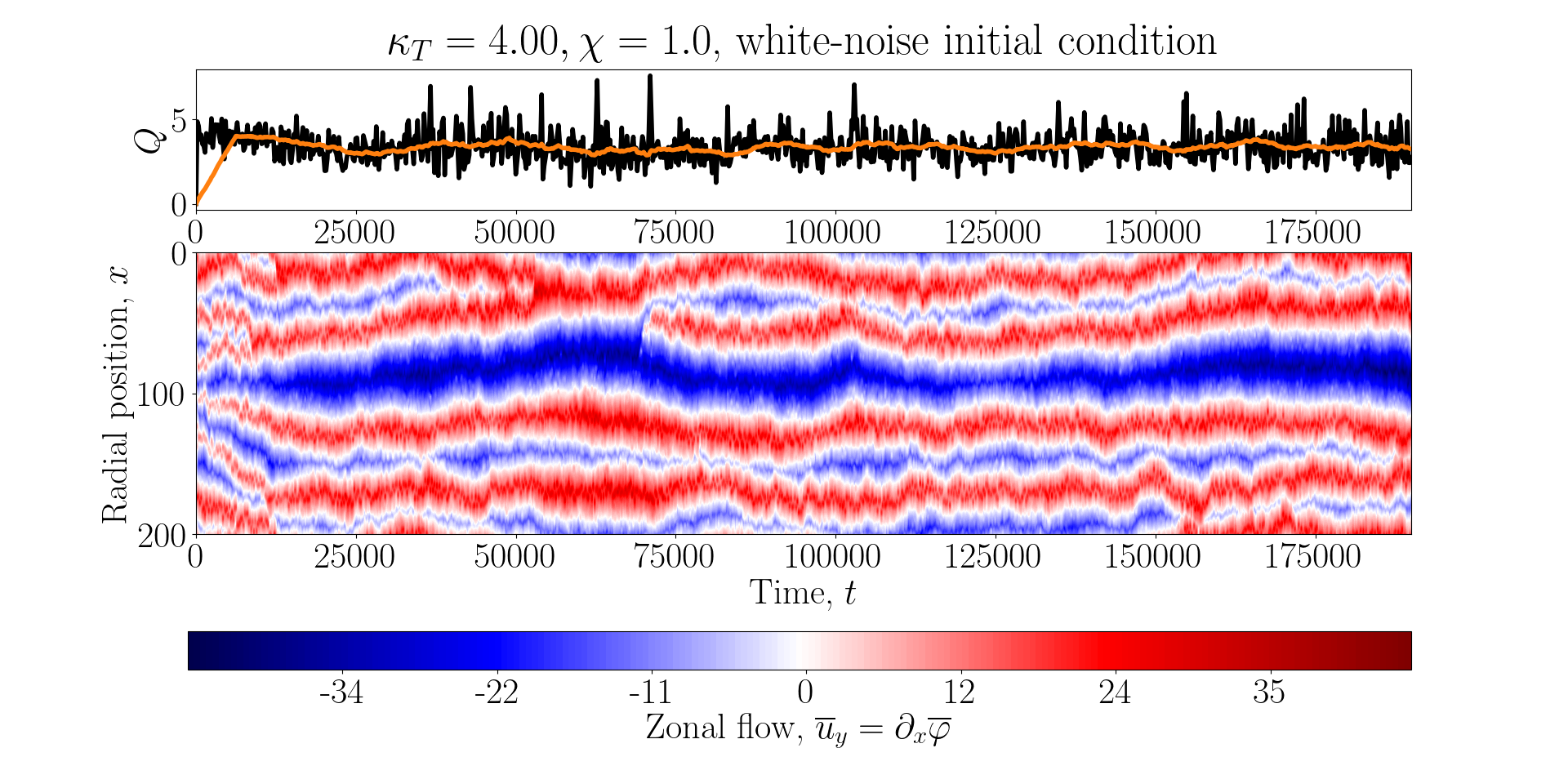}
	\includegraphics[scale=0.427]{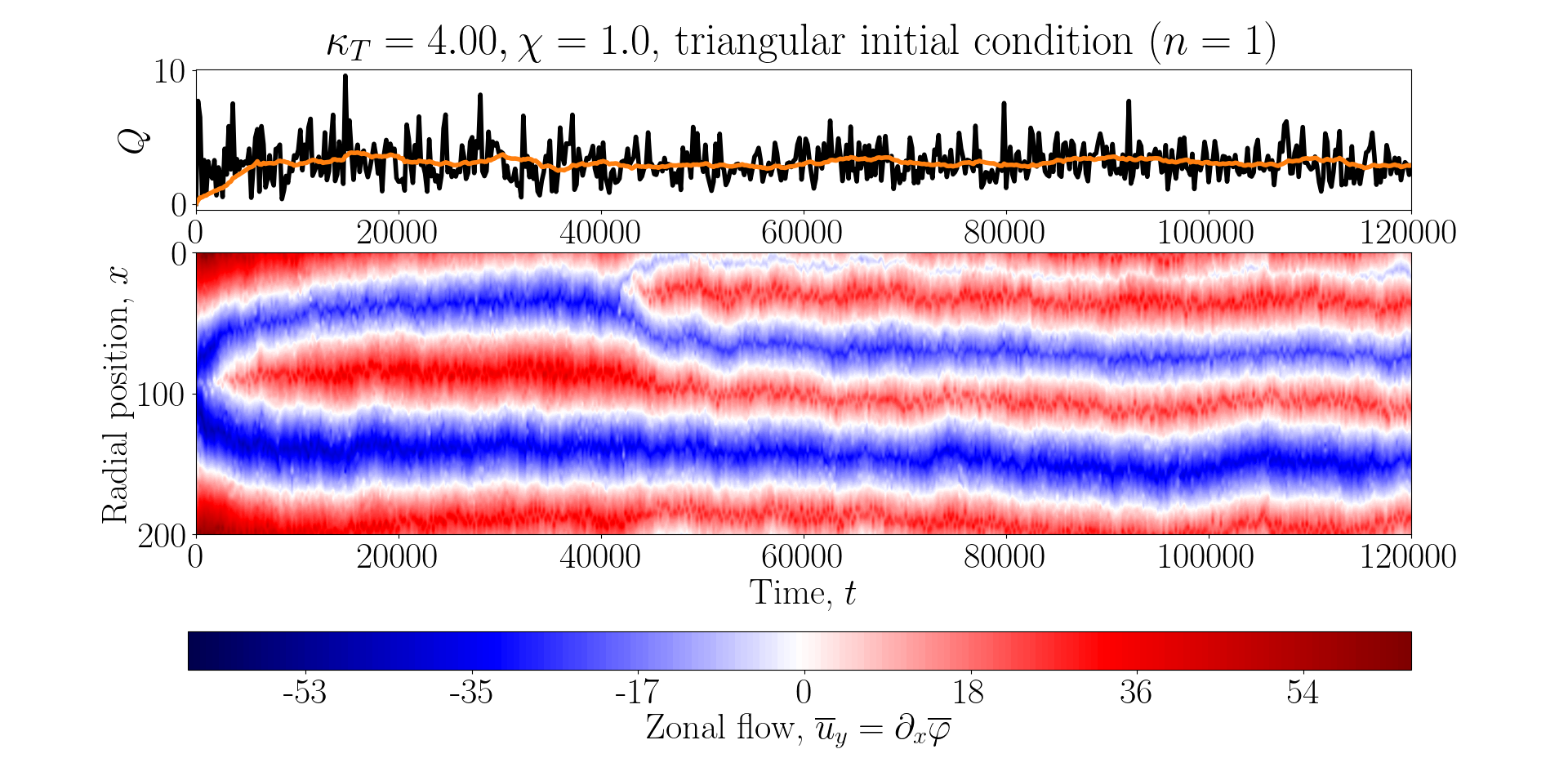}
	\caption{\textbf{Top:} Heat flux, raw (black) and rolling-averaged with a window of \(6000\) time units (orange) for \(\vt = 0.36\), \(\chi = 0.1\), \(L_x = 100\), \(L_y = 100\) and \(125\times125\) Fourier modes. This is the same as SimL, but we have used small-amplitude white noise as an initial condition. The diffusion time for the box-scale ZF is \(t_\text{box} = (L_x / 2\pi)^2 / a\chi \approx 11200\). Convergence to a box-sized ZF occurs on a very long timescale (\(>50t_\text{box}\)). \textbf{Middle:} Same as the top panel, but for \(\vt = 4\), \(\chi = 1\), \(L_x = 200\), and \(L_y = 200\). Here \(t_\text{box} = (L_x / 2\pi)^2 / a\chi \approx 4500\). \textbf{Bottom:} Same as the middle panel, but with a single-peak triangular ZF as an initial condition. It is evident that the ZF profile does not converge to a single peak.}
	\label{fig_joined_hf_zf_r_t}
	\label{fig_joined_hf_zf_r_t2}
	\label{fig_joined_hf_zf_r_t3}
\end{figure}

\begin{figure}
	\centering
	\includegraphics[scale=0.47]{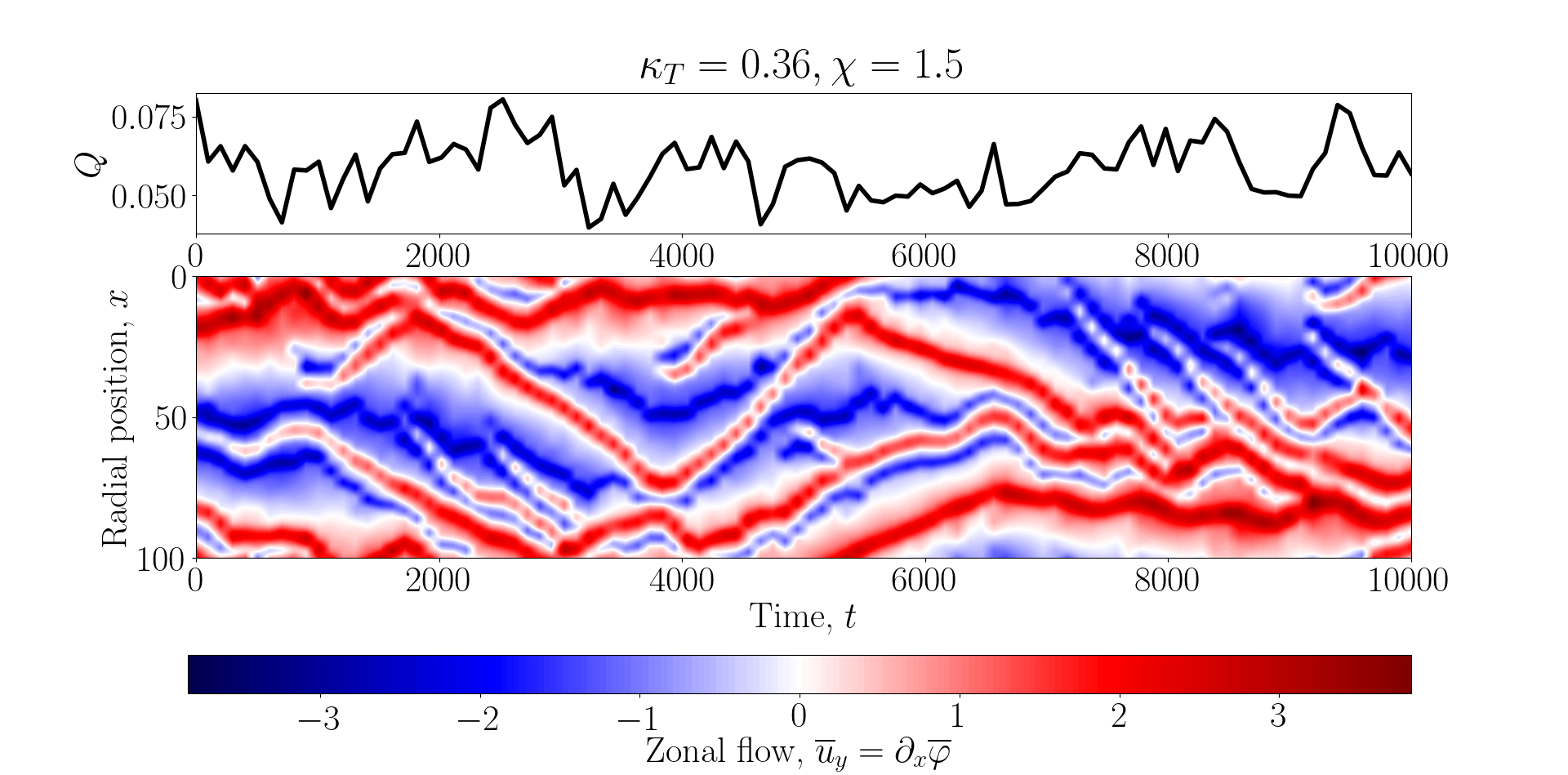}
	\caption{ Same as Figure~\ref{fig_zf_r_t}, but for \(\chi = 1.5\). The diffusion time for the box-scale ZF is \(t_\text{box} = (L_x / 2\pi)^2 / a\chi \approx 750\). The locations of the ZF extrema of the staircase drift significantly over times comparable to \(t_\text{box}\). }
	\label{fig_hf_and_zf_z06_chi1p5}
\end{figure}

\subsubsection{Scale of Zonal Flows}
\label{sect_zfscale}
In general, increasing/decreasing the radial extent of the integration domain by a factor increases/decreases the number \(n\) of shear zones by the same factor. This suggests that the characteristic length scale of the staircase, viz., the time-averaged radial separation of ZF extrema, is determined by a box-size-independent mechanism (see further discussion in Section~\ref{sect_discussion}). Ascertaining definitively whether this is the case is made difficult by the numerically observed time scales of convergence, which are at least an order of magnitude larger than the longest linear time scales, i.e., than the box-scale diffusion time \(t_\text{box} = (L_x / 2\pi)^2 / a\chi\) (see Figure~\ref{fig_joined_hf_zf_r_t}). Note that the long-time evolution of the zonal profile and its length scale is not accompanied by a significant change in the average heat flux. In fact, the latter appears to reach saturation on time scales comparable to the box-scale diffusion time. Therefore, it is reasonable to trust the numerical values of the box-averaged heat flux (e.g., those shown in Figure~\ref{fig_parspace}), even though we could not be certain that the zonal profiles have reached ultimate saturation.

As we increase collisionality and thus move away from the near-marginal regime and into the collisionality-independent regime (the plateau seen in the bottom panel of Figure~\ref{fig_parspace}), the ZFs become more dynamic --- they can merge, split and drift, as shown in Figure~\ref{fig_hf_and_zf_z06_chi1p5} (bottom panel). Here we focus on the near-marginal regime and the transition to strong turbulence, so this higher-collisionality regime will not be studied. 

Even though the zonal staircase arises naturally from white-noise initial conditions for both the zonal and the nonzonal fields, its shape suggests initialising the ZFs with a "triangular" pattern. We find that this helps achieve more quickly a "less noisy" and more symmetric final state, which is easier to handle both numerically and analytically. Of course, we do not know in advance how many "steps" the staircase will "choose" to have in the saturated state, so their number for the "triangular" initial condition is just an informed guess. Most results in this paper are from simulations that used such a triangular ZF initial condition, including SimL and SimH. Notable exceptions are Figures~\ref{fig_parspace}, \ref{fig_zfshear_growth}, and \ref{fig_crit_grad}, where we used data from many simulations, some with white-noise initial conditions and others with "triangular" ones.

\subsection{Convection Zones}
\label{sect_convzones}

\begin{figure}
	\centering
	\includegraphics[scale=0.27]{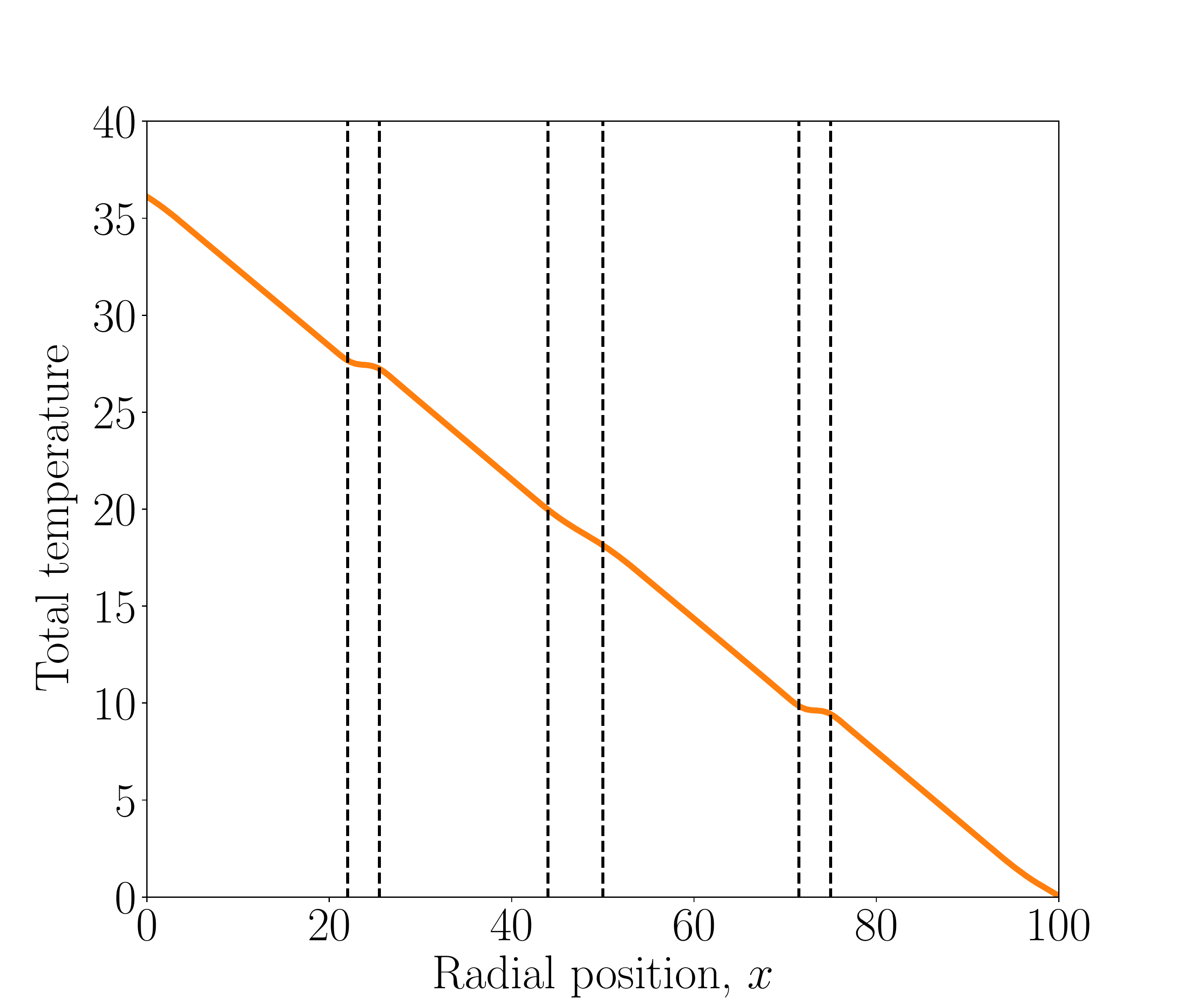}
	\caption{ Time-averaged normalised total temperature (SimL), relative to the absolute temperature \(T_R\) at the right edge of the domain: \(\left(\tau L_B / 2 \rho_s\right)\left(T_i - T_R + \zf{\delta T}\right) = (L_x-x)\vt + \zf{\deltaT} \). A strong flattening of the gradient is visible around the ZF minima at \(x \approx 25\) and \(75\), a weaker one around the ZF maximum at \(x \approx 50\). }
	\label{fig_temp_steps}
\end{figure}

The convection zones located at the extrema of the ZFs contain localised patches of ITG turbulence and have a high radial turbulent heat conductivity (see Figures~\ref{fig_staircase_profiles}~and~\ref{fig_snapshot_t_dimits}). The imposed equilibrium temperature gradient is flattened in the convection zones and slightly steepened in the shear zones. This results in a staircase-like radial temperature profile, shown in Figure~\ref{fig_temp_steps}. The turbulence in the convection zones is driven by a tertiary instability, localised by the zonal shear. In the low-collisionality, near-marginal regime, which we consider to be the most important (see the footnote on page \pageref{fn_cols}), there is a qualitative difference between the way in which the tertiary instability operates at the ZF maxima and minima. A similar difference exists in both the Hasegawa-Mima equation \citep{zhu2018} and gyrokinetics \citep{mcmillan2011}.

\subsubsection{Turbulence at ZF Minima}

At the ZF minima, we find both ITG and Kelvin-Helmholtz tertiary instabilities. The former is dominant (faster) and saturates by producing a zonal-temperature gradient that cancels the background temperature gradient. This effectively decouples the evolution of the temperature perturbations from that of the electrostatic potential and leaves a KH mode that seems to determine the poloidal wavenumber at the ZF minima (the peak at \(k_y \approx 0.26\) in Figure~\ref{fig_conv_spectra} is precisely the wavenumber of the fastest-growing KH mode at the ZF minima). Further details on the tertiary instability at the ZF minima and its saturation can be found in Appendix~\ref{appendix_zfminima}. 

\subsubsection{Turbulence at ZF Maxima}
\label{sect_convzones_maxima}

In contrast to the ZF minima, the regions around the ZF maxima cannot support a Kelvin-Helmholtz instability because the Rayleigh-Kuo criterion for instability is not satisfied there \citep{kuo49, zhu2018_tertiary}: see Appendix~\ref{appendix_khtertiary}. The ITG instability in these regions is significantly weaker than that at the ZF minima and does not appear to saturate in a similar fashion (by cancelling the equilibrium temperature gradient). The profile of the zonal-temperature gradient shown in Figure~\ref{fig_staircase_profiles} suggests that the instability might not even be localised to the ZF maximum itself: there are two peaks of the zonal-temperature gradient visible on either side of the ZF maximum at \(x \approx 47\) at \(t = 10\). The poloidal scale of the modes at the ZF maxima is significantly longer than that at the ZF minima, see Appendix~\ref{appendix_zfmaxima}. 

Additionally, an asymmetric flattening of the zonal shear develops on one side of the ZF maximum, accompanied by a drift of the location of this maximum in the opposite direction (such a flattening is seen to the right of the central ZF maximum at \(x \approx 47\) in Figure~\ref{fig_staircase_profiles}). Eventually, ferdinons are launched in the direction of the flattening (see also Section~\ref{sect_ferds}). This is likely due to the inability of the diminished zonal shear there to suppress the nonzonal perturbations. The burst of ferdinons causes the ZF maxima to change the direction of its drift and a flattening of the zonal shear develops on the opposite side. This causes an oscillation of the position of the ZF maximum, as seen in the bottom panel of Figure~\ref{fig_zf_r_t}.

Thus, while turbulence is suppressed by zonal shear in the shear zones and by the cancellation of the equilibrum temperature gradient by the zonal temperature around the ZF minima, the regions around the ZF maxima remain locally unstable. As long as the zonal shear in the shear zones is strong enough to suppress turbulence, this instability is tamed, with any perturbations launched from the unstable regions into the shear zones unable to survive. Once the zonal shear decays below a certain level, it is no longer able to suppress these perturbations ("ferdinons", see Section~\ref{sect_ferds}) and a turbulent burst is initiated. Thus, the quasi-stationary zonal staircase contains the seeds of its own destruction: the perilous combination of decaying ZFs and unstable convections zones around the ZF maxima.

\subsubsection{Scale of Convection Zones}

\begin{figure}
	\centering	
	\includegraphics[scale=0.27]{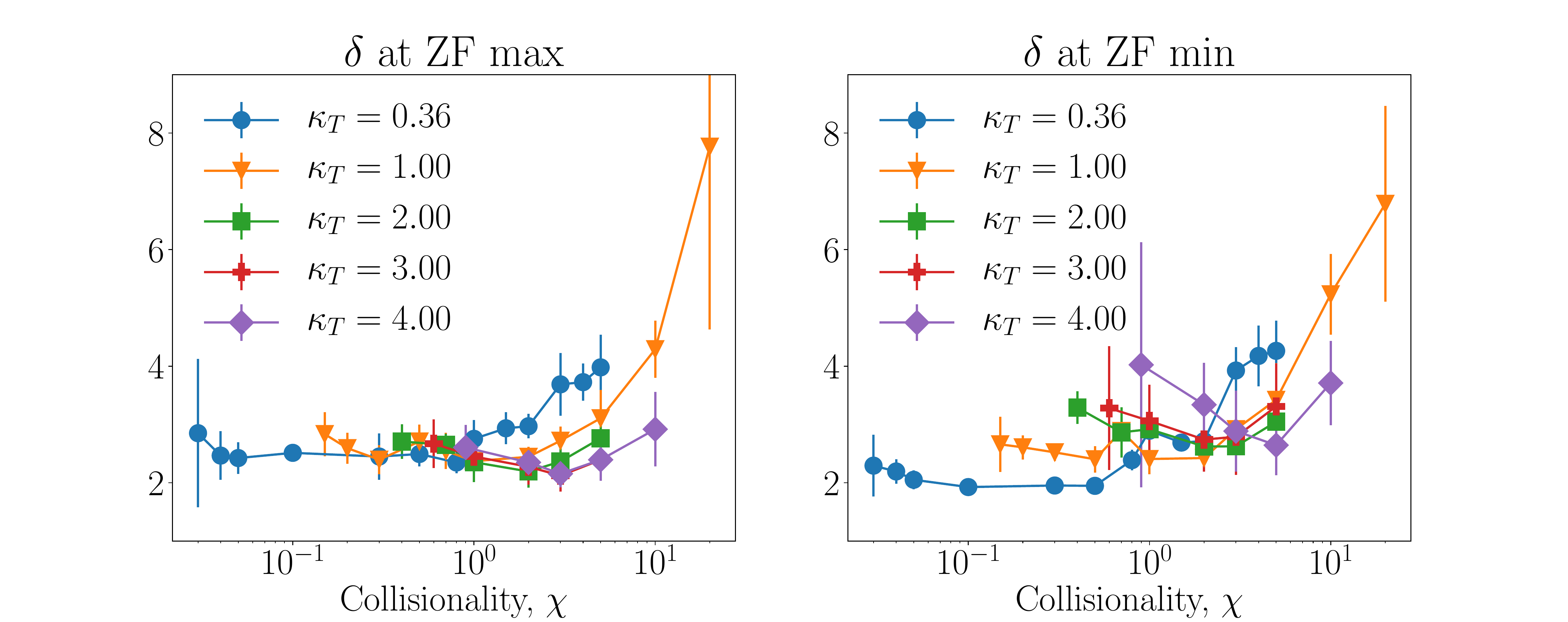} 
	\caption{ Numerically determined time-averaged values of \(\delta\) [see \eqref{eq_delta_def}] at the global maximum (left) and global minimum (right) of the ZFs in the simulations whose heat flux is shown in Figure~\ref{fig_parspace}. We find no significant variation of \(\delta\) with \(\chi\) or \(\vt\), except at large \(\chi\). }
	\label{fig_delta_parspace}
\end{figure}

The width of the convection zones can be characterised by the quantity
\begin{equation}
\label{eq_delta_def}
\delta \equiv \sqrt{\frac{\px \zf{\phinorm}}{\px^3 \zf{\phinorm}}}.
\end{equation}
Figure~\ref{fig_delta_parspace} shows that \(\delta\) does not depend very strongly on either \(\vt\) or \(\chi\), except far from the marginal state, where collisionality appears to smooth out the gradients in the convection zones and thus increase \(\delta\). This suggests that near the Dimits threshold, \(\delta\) is an \(\order{1}\) quantity in the normalised units of \eqref{curvy_phi} and \eqref{curvy_psi}, i.e., it is equal to a few times the sound radius \(\rho_s\).

\subsection{Ferdinons}
\label{sect_ferds}

After the ZF has decayed sufficiently to weaken its ability to suppress perturbations, vortex-like propagating structures are spawned from the ZF maxima and drift radially through the shear zones. Strikingly similar structures --- "ferdinons" --- have been observed in GK simulations with imposed background flow shear \citep{vanwyk2016,vanwyk2017}. Figures~\ref{fig_staircase_profiles} and \ref{fig_snapshot_t_dimits} (\(t = 136\)) show a particular instance of the launching of ferdinons (see also Figure~\ref{fig_burst_heatflux}). 

As the ferdinons smash into the turbulent modes in the convection zones at the ZF minima, more structures are produced and a burst of turbulence ensues (see Figure~\ref{fig_staircase_profiles} and \ref{fig_snapshot_t_dimits}, \mbox{\(t = 250\)}). These travelling structures, as well as the resulting turbulence, cause a significant spike in the box-averaged heat flux (see Figure~\ref{fig_burst_heatflux}). As Figure~\ref{fig_staircase_profiles} shows, they do not carry a significant ZF perturbation. They are created and propagate even if the ZF is held artificially constant in the numerical simulations, but the zonal temperature is left to evolve according to \eqref{curvy_psi}. In other words, a localised ZF perturbation is not an essential part of these structures.

\begin{figure}
	\centering
	\includegraphics[scale=0.45]{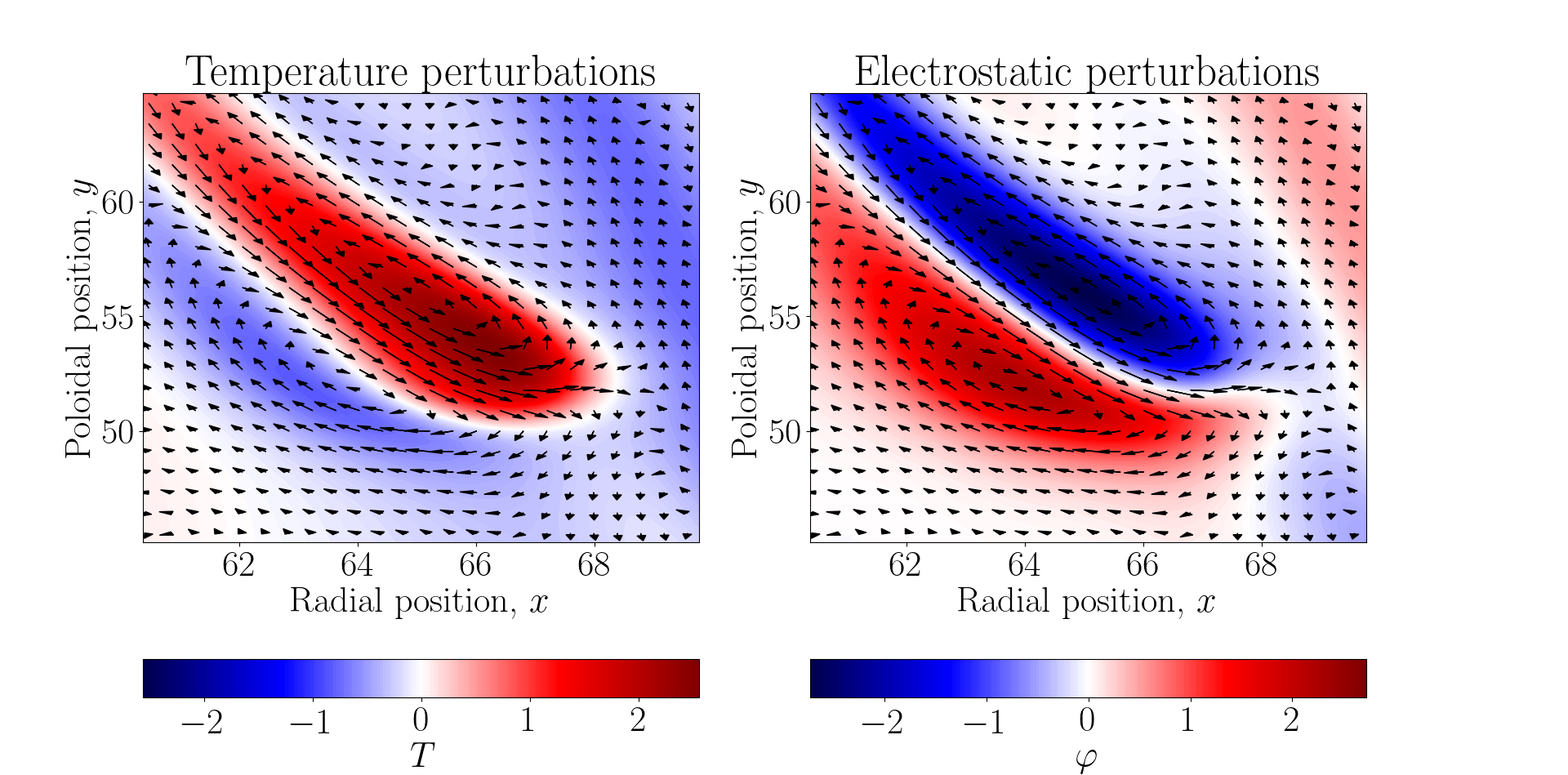}
	\includegraphics[scale=0.45]{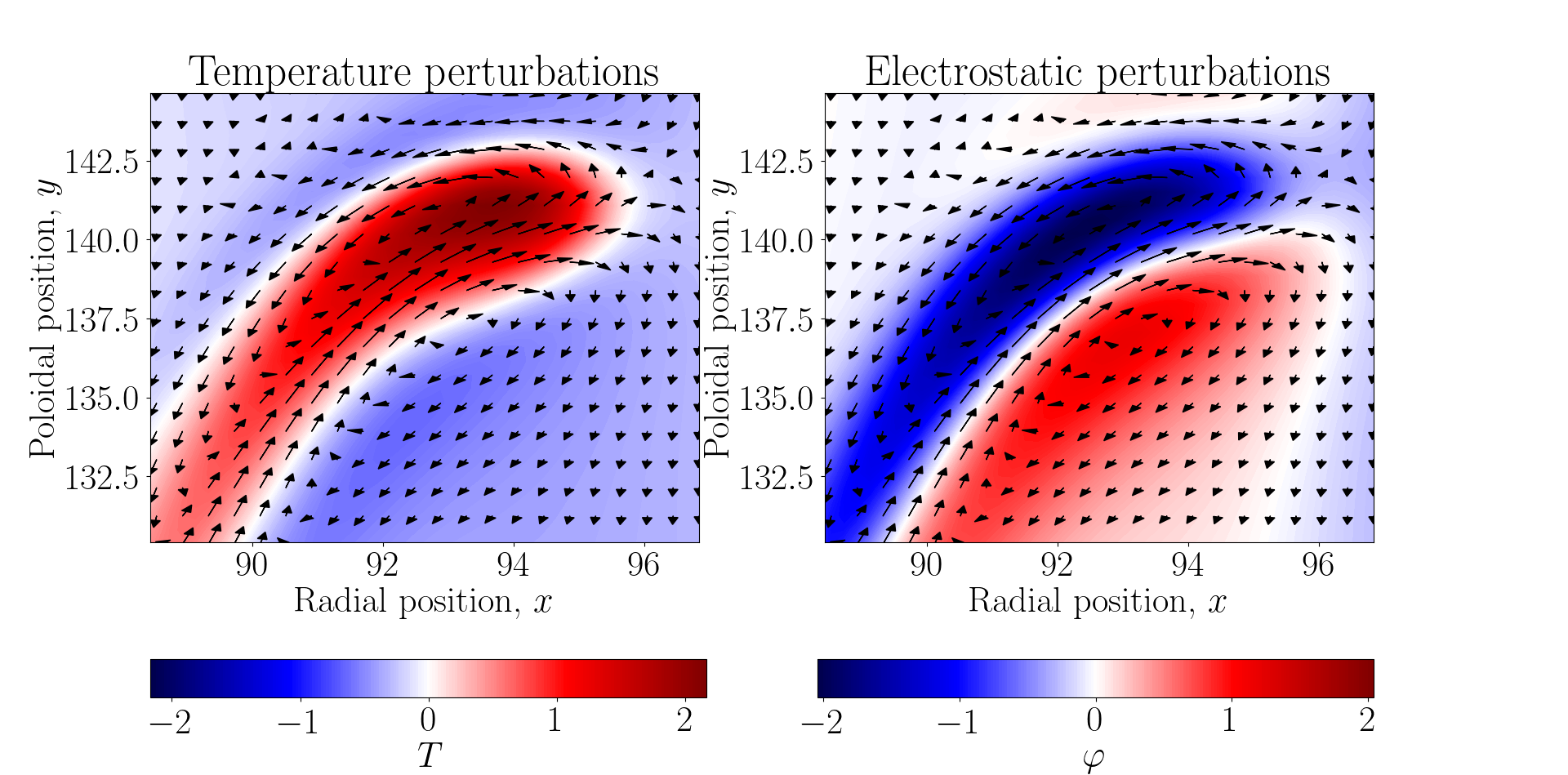}
	\includegraphics[scale=0.45]{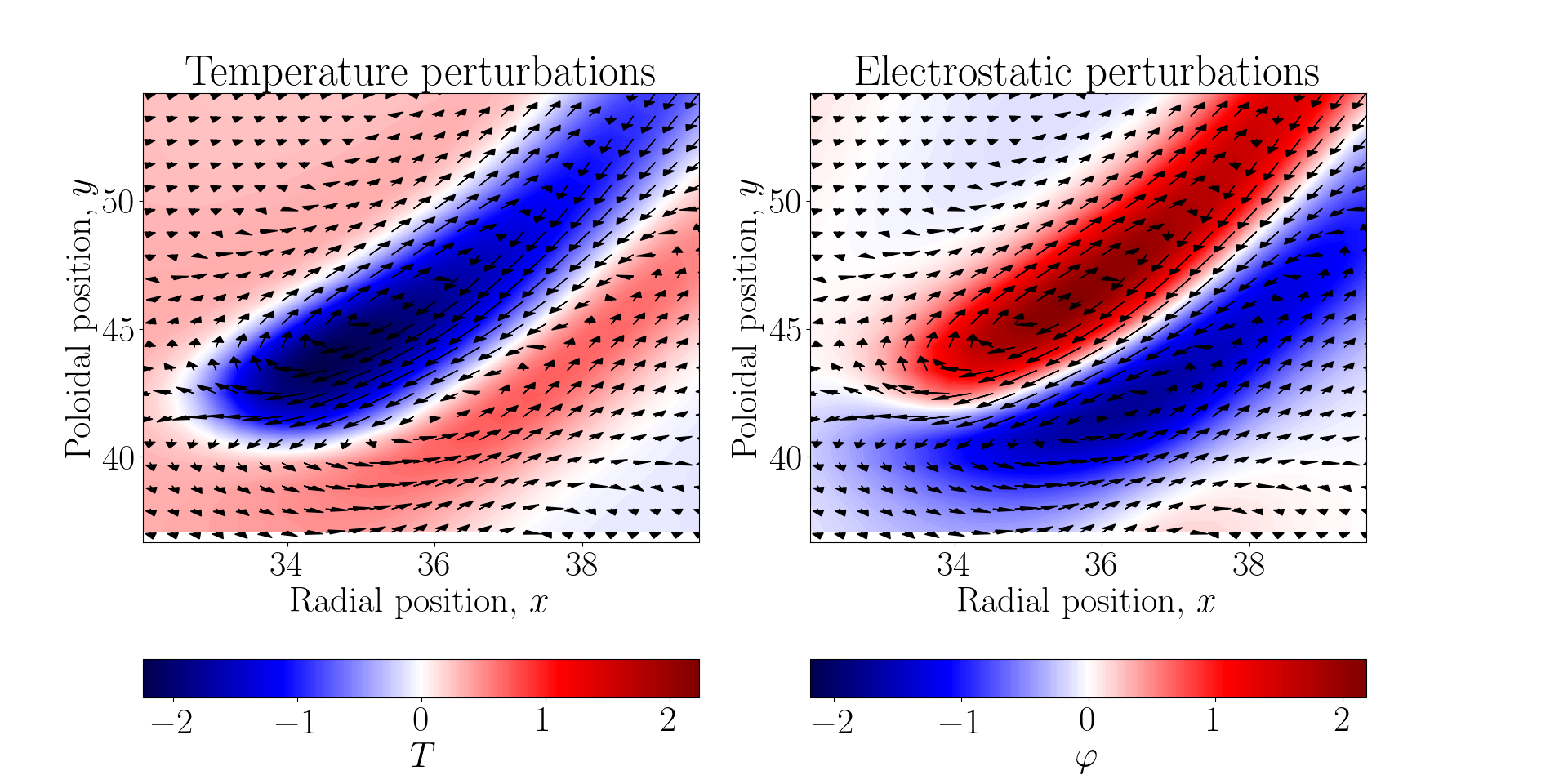}
	\caption{\textbf{Top:} A close-up view (from the top-right plot of Figure~\ref{fig_snapshot_t_dimits}) of the temperature and electrostatic-potential perturbations of a hot ferdinon in a region of negative zonal shear, \(S = \px^2 \zf{\phinorm} < 0\). The arrows represent the local nonzonal \exb velocity \( \dw{\vect{u}} = \uvect{z}\times\del\dw{\phinorm}\) (in arbitrary units). \textbf{Middle:} A hot ferdinon in a region of positive zonal shear \(S = \px^2 \zf{\phinorm} > 0\). \textbf{Bottom:} A cold ferdinon in a region of positive zonal shear \(S = \px^2 \zf{\phinorm} > 0\).}
	\label{fig_ferd_closeup}
	\label{fig_ferd_closeup_up}
	\label{fig_ferd_closeup_cold}
\end{figure}

Figure~\ref{fig_ferd_closeup} shows that ferdinons consist of a vortex dipole and a strong temperature perturbation trapped in one of the vortices of this dipole. There are ferdinons carrying both positive ("hot") and negative ("cold") temperature perturbations. Hot ferdinons drift towards the cooler (right) side of the domain, while the cold ones drift in the opposite direction, towards the hotter (left) side (see also Figure~\ref{fig_burst_heatflux}). The top and middle panels of Figure~\ref{fig_ferd_closeup} demonstrate that the direction of the drift does not depend on the sign of the zonal shear. Net flow circulation around the ferdinons is also independent of the sign of the shear --- it is always anticlockwise for hot and clockwise for cold ones (see bottom panels of Figure~\ref{fig_ferd_closeup_up}).

Note that the ferdinons that emerge in our simple ITG model bear a striking qualitative resemblance to the avalanches reported by \citet{villard2013} in global GK simulations, namely, they propagate both inwards and outwards, but always with a positive heat flux, and originate from the local maxima of the ZF. Simple soliton solutions have already been proposed as a model for GK avalanches \citep{mcmillan2009, mcmillan2018}. Vortex-dipole solitons called "modons" have been investigated in Hasegawa-Mima-like models of turbulence \citep{horton94}. We do not yet know how and whether any of these are related to the ferdinons that we observe.

Let us discuss what we expect the ferdinon solution to be. Numerically, we find that the existence and propagation of these structures depend crucially on the two ITG-drive terms in \eqref{curvy_phi} and \eqref{curvy_psi}, as well as on the nonlinear terms. In particular, the poloidal localisation of these structures is due to the nonzonal-nonzonal interactions. Indeed, we have found that \eqref{curvy_phi} and \eqref{curvy_psi} with the nonzonal-nonzonal nonlinear terms taken out \citep[in what is sometimes referred to as the "quasilinear approximation"; see][]{srinivasan2012} do not have ferdinon solutions. However, the quasilinear system does have soliton solutions that are \textit{not} localised poloidally, but rather appear to have a definite poloidal wavenumber \(k_y\). These solutions might be related to those described by \citet{mcmillan2009} and \citet{zhou2020}. Models have been proposed for structure formation in a sheared flow that rely on the tilting of turbulence by shear and a nonzero group velocity to produce moving structures \citep{mcmillan2018, zhou2020}. The radial group velocity is (at least in the Hasegawa-Mima-related models) proportional to the product \(k_x k_y\)\footnotemark, which acquires a definite sign in the presence of flow shear (Section~\ref{sect_stresses}). However, we observe ferdinons moving in both radial directions in regions of definite zonal shear and, thus, definite radial group velocity. Therefore, at the moment, we consider it unlikely that the propagation of ferdinons can be explained using such group-velocity arguments. We leave the detailed investigation of ferdinon generation and propagation for future work.

\footnotetext{The radial group velocity \(\partial \omega_\vk / \partial k_x\) is proportional to \(k_xk_y\) because \(\omega_\vk\propto k_y\) and \(\omega_\vk\) depends on \(k_x\) only through \(k^2\).}

Understanding ferdinons and their properties can also put an upper bound on the radial scale of the ZF. Indeed, our numerical simulations show that the ZFs can have a well-defined radial scale smaller than the box size (Section~\ref{sect_zfscale}). This scale could perhaps be estimated via a causality argument --- assuming that ferdinons, and, thus, turbulence, can only propagate a finite radial distance in a region of self-consistently evolving zonal shear, then an infinitely wide shear zone cannot be sustained for long. Note that finite-lifetime ferdinons over a dynamic ZF background, with which they can interact and gain or lose energy, are not in contradiction with the infinite-lifetime ferdinons seen by \citet{vanwyk2016}, where a constant flow shear was imposed, and thus the shear profile was unable to react to the presence of ferdinons.

Once ferdinons are generated and turbulence develops in the shear zones, our analysis of the viscous decay of the zonal staircase in Section~\ref{sect_zfdecay}, which ignored the turbulent momentum flux, is no longer valid. Instead, we must focus on the effect of the turbulence on the ZFs. We find that the turbulence in the shear zones has a restoring effect on the zonal staircase in the Dimits regime, whereas beyond the Dimits threshold, it inhibits staircase formation.

\section{Resilience of the Zonal State and the Dimits Threshold}
\label{sect_zfstab}

\subsection{Turbulent Momentum Flux}
\label{sect_turbmom}

In order to investigate the way in which ZF profiles are formed and maintained during turbulent periods, let us ask the following question: does ITG turbulence in the shear zones (produced in bursts) have a definite effect on the ZFs, and is it to oppose or to feed them? We shall find that this sheared turbulence enhances the ZFs in the Dimits regime and destroys them beyond the Dimits transition.

The ZF evolution equation \eqref{eq_zfflux_int} is
\begin{equation}
\label{eq_zfflux_int2}
\pt \zf{\phinorm} + \Pi_t + \Pi_d = 0.
\end{equation}
In Section~\ref{sect_zfdecay}, we discussed the effect of the diffusive momentum flux \(\Pi_d\), viz., the viscous decay of the ZFs. For the rest of this section, we focus on the effects of turbulence by examining the turbulent momentum flux \(\Pi_t\). 

It is evident from \eqref{eq_zfflux_int2} that ZF saturation requires 
\begin{align}
	\label{eq_zfsatcondition}
	\avgT{\Pi_t(t, x) + \Pi_d(t, x)} \approx 0
\end{align}
to be satisfied at every radial location \(x\), where \(\avgT{f(t)} \equiv (1 / \Delta t) \int_{\Delta t} dt f(t)\) is a time average in the saturated state over a time \(\Delta t\) longer than the typical evolution time of the ZF (e.g., longer than the duration of turbulent bursts if the saturated state is bursty). Recall that \mbox{\(\Pi_d \approx -a\chi\px^2\zf{\phinorm} = -a\chi S\)} in the shear zones (see Section~\ref{sect_zfdecay}). Therefore, within a shear zone with nearly constant (in time \textit{and} in space) zonal shear, \eqref{eq_zfsatcondition} tells us that the time-averaged turbulent momentum flux in that shear zone must have a definite value\footnotemark, determined by the local zonal shear. Thus, in the saturated state, \(\Pi_t\) will be correlated with \(S\). 

\footnotetext{Also, the spatial average over a shear zone of the turbulent momentum flux must be nonzero. This is not in contradiction with the argument in Section~\ref{sect_zfdecay} that the spatial average over the entire box is zero, viz., \(\Pi = 0\), because a definite uniform zonal shear breaks the symmetry \eqref{eq_symmetry} locally within each shear zone.}

\begin{figure}
	\centering
	\includegraphics[scale=0.27]{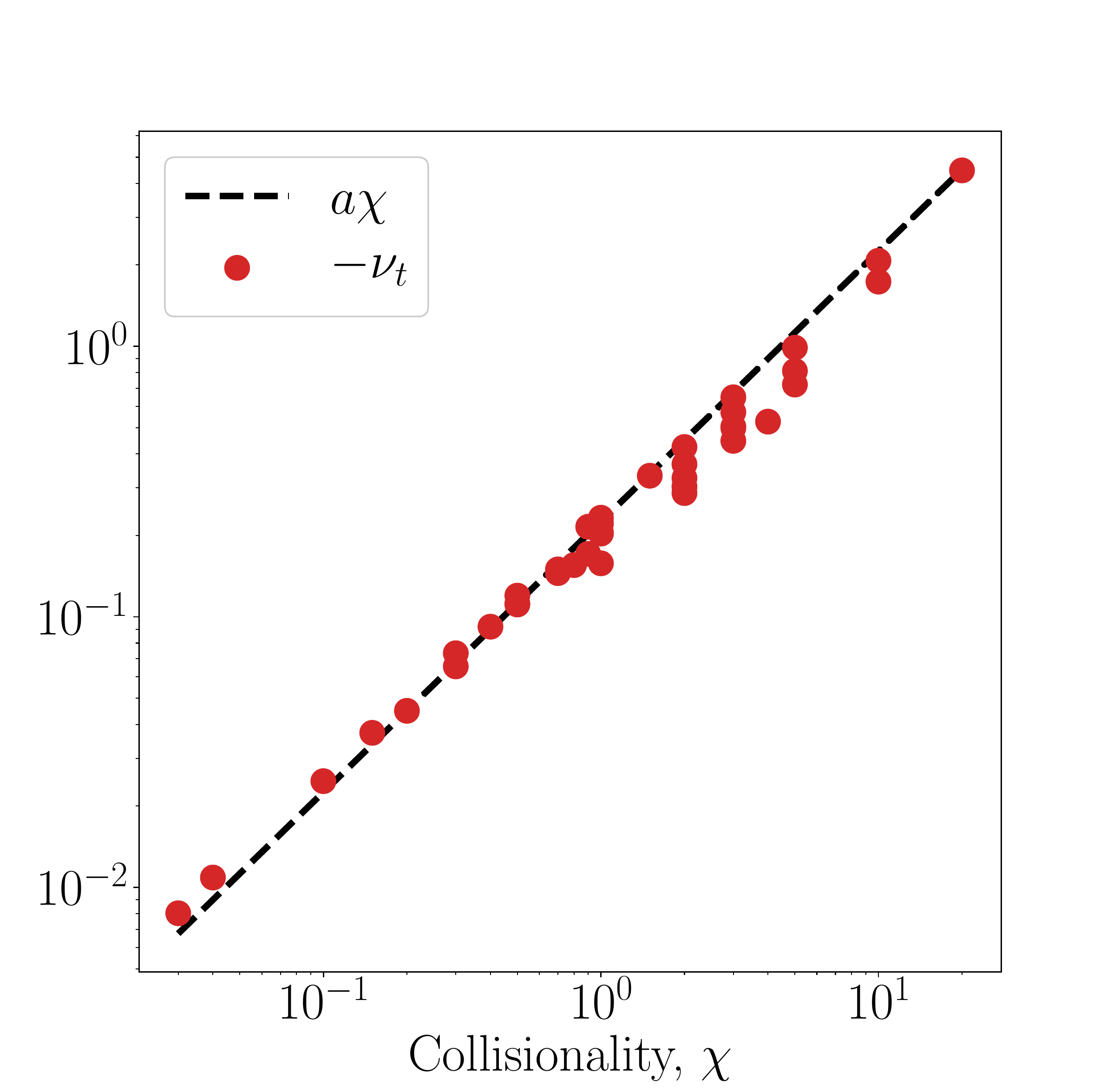}
	\caption{ Comparison of estimated turbulent viscosity \(\nu_t\) and the collisional viscosity \(a\chi\) for the simulations from Figure~\ref{fig_parspace}. We see that \(\nu_t \approx -a\chi\). }
	\label{fig_turbvisc_sat}
\end{figure}

To quantify this correlation, let us multiply both sides of \eqref{eq_zfflux_int2} by \(S = \px^2 \zf{\phinorm}\) and integrate across the radial extent of the domain. We find
\begin{align}
	\label{eq_flux_radial_int}
	\int_0^{L_x} dx \ \left(\px^2 \zf{\phinorm}\right) \pt \zf{\phinorm} + \int_0^{L_x} dx \ \Pi_t S + \int_0^{L_x} dx \ \Pi_d S = 0.
\end{align}
Since \(\Pi_d = -\chi \px^2 (a\zf{\phinorm} - b\zf{\deltaT}) = -a\chi S + b\chi\px^2\zf{\deltaT}\), we have
\begin{align}
	\int_0^{L_x} dx \ \Pi_d S = -a\chi \int_0^{L_x} dx \ S^2 + b\chi \int_0^{L_x} dx \ S \px^2\zf{\deltaT} \approx -a\chi \int_0^{L_x} dx \ S^2,
\end{align}
where we have assumed that the second term is negligible because the main contribution to \(S\) comes from the shear zones, where \(\px^2 \zf{\deltaT} \approx 0\) [see also the discussion leading to \eqref{eq_zfdiffusioneq}]. Therefore, after integrating by parts the first term in \eqref{eq_flux_radial_int} and time averaging the resulting equation, we find
\begin{align}
	\avgT{\int_0^{L_x} dx \ \left( \Pi_t S - a\chi S^2 \right)}
	= - \avgT{\frac{1}{2} \pt \int_0^{L_x} dx \ \left(\px \zf{\phinorm}\right)^2} \approx 0.
\end{align}
This gives a prediction for the effective "turbulent viscosity" in the shear zones:
\begin{align}
	\label{eq_momcorr_def}
	\nu_t \equiv -\frac{ \avgT{\int_0^{L_x} dx \ \Pi_t S}}{ \avgT{\int_0^{L_x} dx \ S^2}} \approx -a\chi.
\end{align}
Relation \eqref{eq_momcorr_def} is, of course, corroborated  by numerical simulations: see Figure~\ref{fig_turbvisc_sat}.

\subsection{Sign Reversal of the Turbulent Momentum Flux at the Dimits Threshold}
\label{sect_dimits_threshold}

An important consequence of \eqref{eq_momcorr_def} is that, in a shear zone, the sign of the turbulent momentum flux must coincide with the sign of the zonal shear. Therefore, if, for certain parameters, sheared turbulence has a momentum flux with a sign \textit{opposing} that of the local shear, saturation cannot be achieved. We shall see that this is exactly what happens beyond the Dimits threshold. 

\begin{figure}
	\centering
	\hspace*{0.4cm}\includegraphics[scale=0.27]{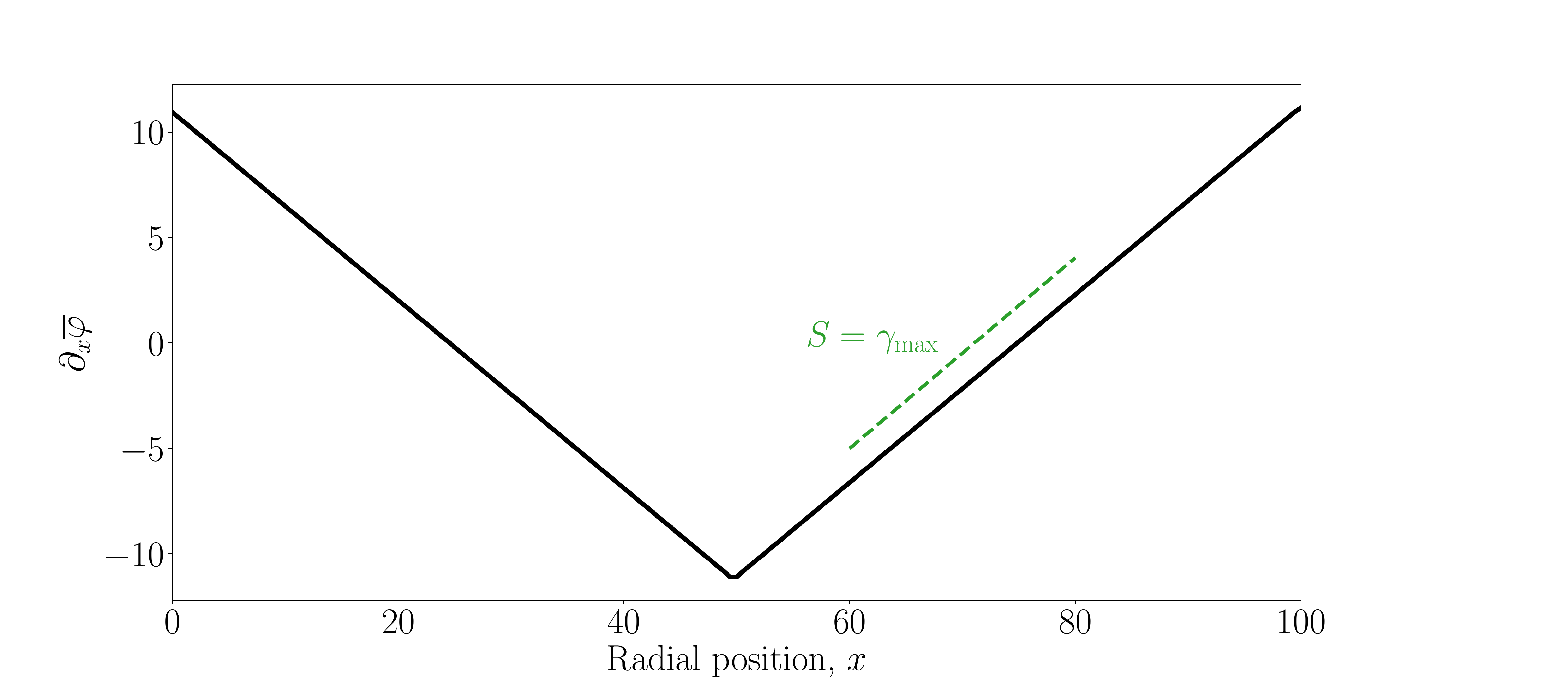}
	\includegraphics[scale=0.27]{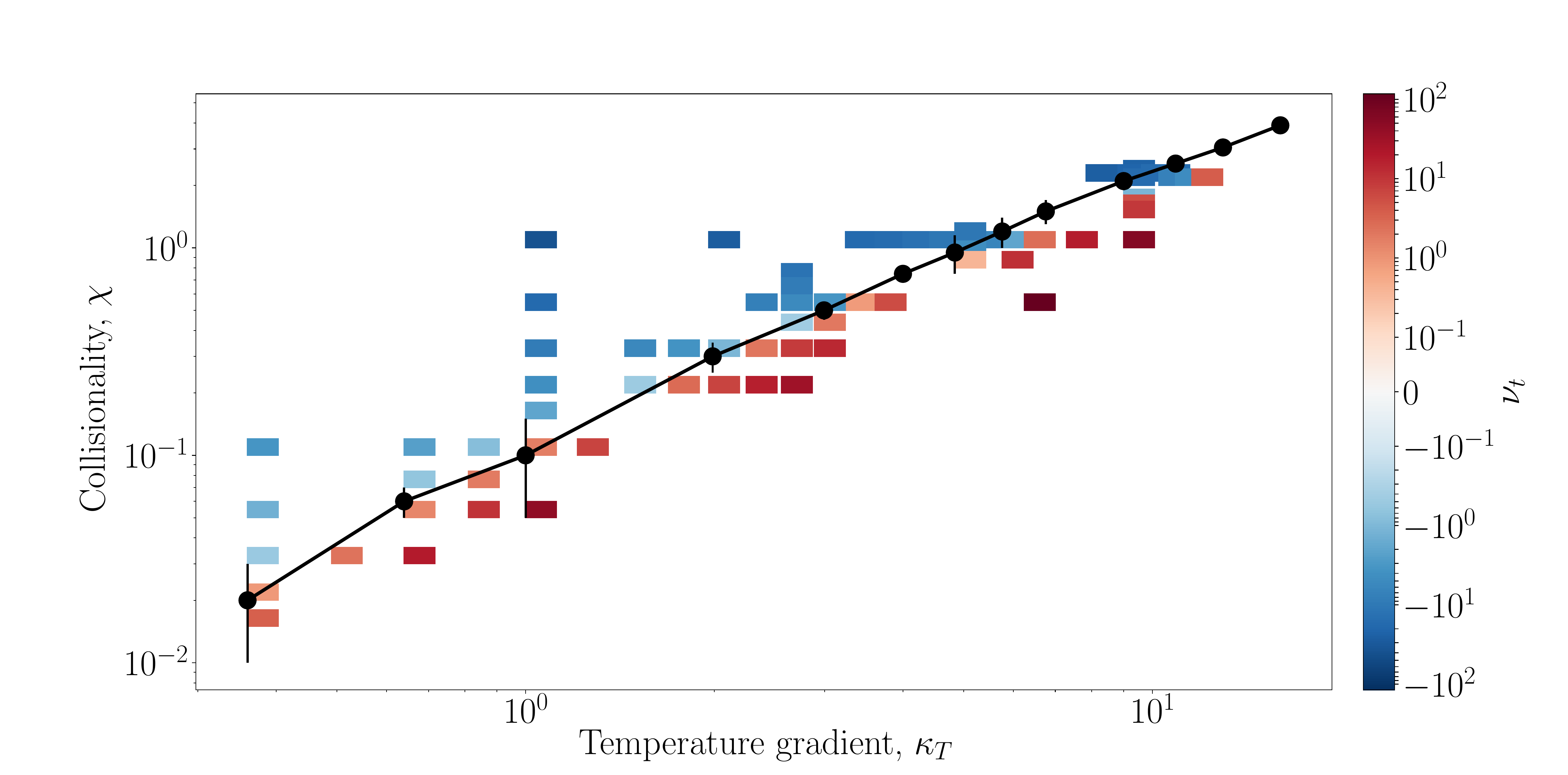}
	\caption{ \textbf{Top:} The artificial triangular zonal pattern used to generate the bottom panel of this figure. The zonal shear in the shear zones is chosen to be equal in absolute value to the largest ITG growth rate (represented by the dashed green line). \textbf{Bottom:} The effective turbulent viscosity \(\nu_t\), as defined by \eqref{eq_momcorr_def}, for the static triangular ZF profile given in the top panel of this figure (coloured data points). The black line represents the numerically established Dimits threshold. }
	\label{fig_momcorr}
\end{figure}

Let us investigate how turbulence responds to an imposed static zonal profile. We solve \eqref{curvy_phi} and \eqref{curvy_psi} numerically with an imposed static triangular ZF pattern (i.e., we do not evolve the ZFs at all), in a box of size \(L_x = L_y = 100\) and \(169\times169\) Fourier modes, for a range of parameters around the Dimits transition. The chosen ZF pattern is shown in Figure~\ref{fig_momcorr} (top panel) and is adjusted for every simulation so that the value of the zonal shear in the shear zones matches the largest ITG growth rate for that simulation. The chosen radial scale of the ZF (\(= 100\)) is inspired by the typical ZF scale that we observe in the low-collisionality regime, and is held fixed as we vary \(\chi\) and \(\vt\). Then we calculate the effective turbulent viscosity \(\nu_t\) associated with the turbulent momentum flux. 

As the bottom panel of Figure~\ref{fig_momcorr} shows, we find a negative turbulent viscosity \(\nu_t\) (and, thus, a positive correlation between local zonal shear and turbulent momentum flux) in the Dimits regime and a positive \(\nu_t\) beyond it (thus, a negative correlation). Let us denote by \(\vtstatic\) the temperature gradient at which \(\nu_t\) reverses its sign. The designation "\textit{static}" reflects the fact that this is a numerical result for ITG turbulence with an artificially imposed \textit{static} ZF profile. We find that the value of \(\vtstatic\) is insensitive to the exact shape of the ZF profile, and, most importantly, it nearly perfectly coincides with the Dimits threshold, i.e., \(\vtstatic \approx \vtcrit\).

Thus, in the Dimits regime, shear zones are resilient because, when the zonal shear there decays due to viscosity and turbulence is thus unleashed, this turbulence acts to reinforce the ZFs and the zonal shear in the shear zones is restored to its turbulence-suppressing level. Beyond the Dimits regime, the zonal staircase cannot be sustained because both turbulence and collisional viscosity act to flatten out the ZFs.

\subsection{Reynolds Stress and Diamagnetic Stress}
\label{sect_stresses}

\begin{figure}
	\centering
	\includegraphics[scale=0.3]{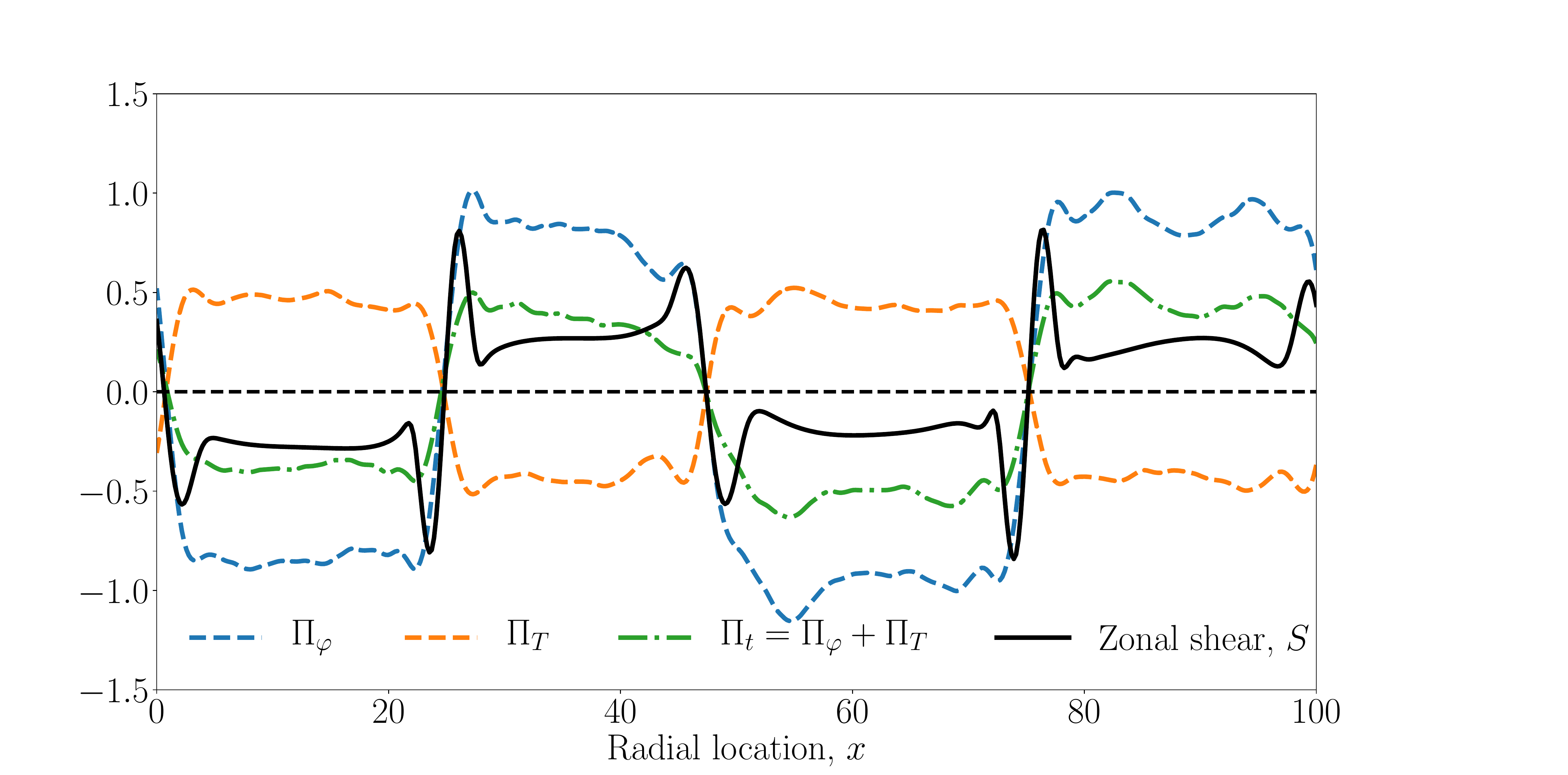}
	\includegraphics[scale=0.3]{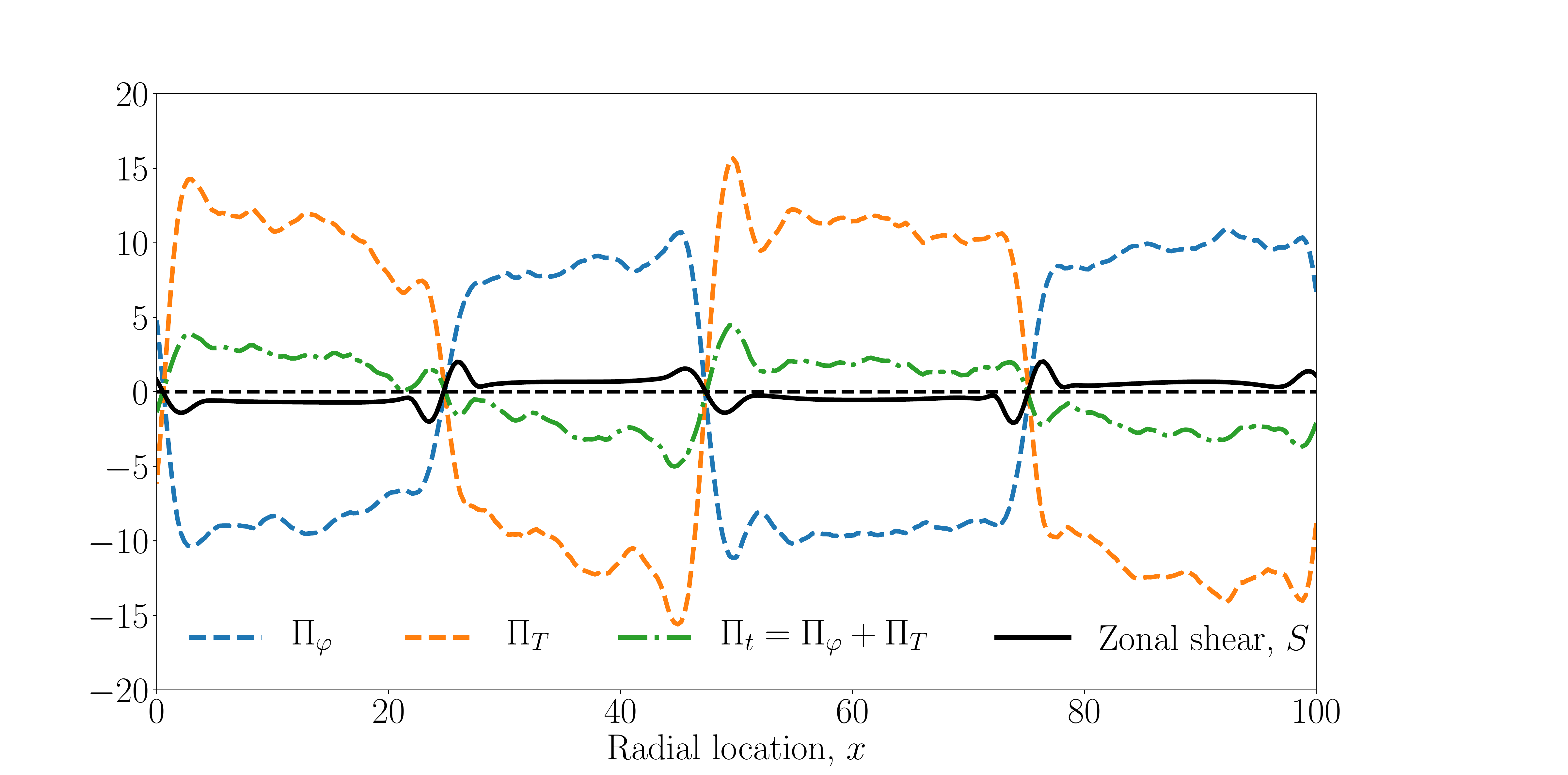}
	\caption{\textbf{Top:} Time-averaged momentum fluxes \(\Pi_\phinorm\), \(\Pi_\deltaT\) and \(\Pi_t = \Pi_\phinorm + \Pi_\deltaT\) for saturated ITG turbulence over a fixed zonal background. Note the correlation between the signs of the various fluxes and the zonal shear \(S\): the sign of \(\Pi_t\) coincides with that of \(\Pi_\phinorm\) and \(S\) and opposes the sign of \(\Pi_\deltaT\). This reflects that the temperature gradient is lower than the Dimits threshold, \(\vt = 0.36 < \vtcrit \approx 1\). The ZF profile used here was extracted from SimH at \(t = 10\), but reduced by a factor of \(0.8\) in order to allow ITG turbulence to develop in the shear zones. \textbf{Bottom:} Same as the top panel, but with \(\vt = 1.21\). The sign of \(\Pi_t\) now opposes the sign of \(\Pi_\phinorm\) and \(S\) and coincides with the sign of \(\Pi_\deltaT\). This reflects that \(\vt = 1.21 > \vtcrit \approx 1\). The extracted ZF is augmented by a factor of \(2\) to account for the increased ITG growth rate (due to the larger \(\vt\)). This is necessary for the turbulence to saturate at numerically feasible amplitudes. Note that saturation is possible only because we have fixed the ZF profile. If the ZF is left to evolve according to \eqref{eq_zfflux}, the poloidal momentum generated by the nonzonal perturbations flattens it and the system fails to reach a finite-amplitude saturated state (see Section~\ref{sect_blowup}). }
	\label{fig_momflux_z06chi01}
	\label{fig_momflux_z1p10chi01}
\end{figure}

\begin{figure}	
	\centering
	\includegraphics[scale=0.27]{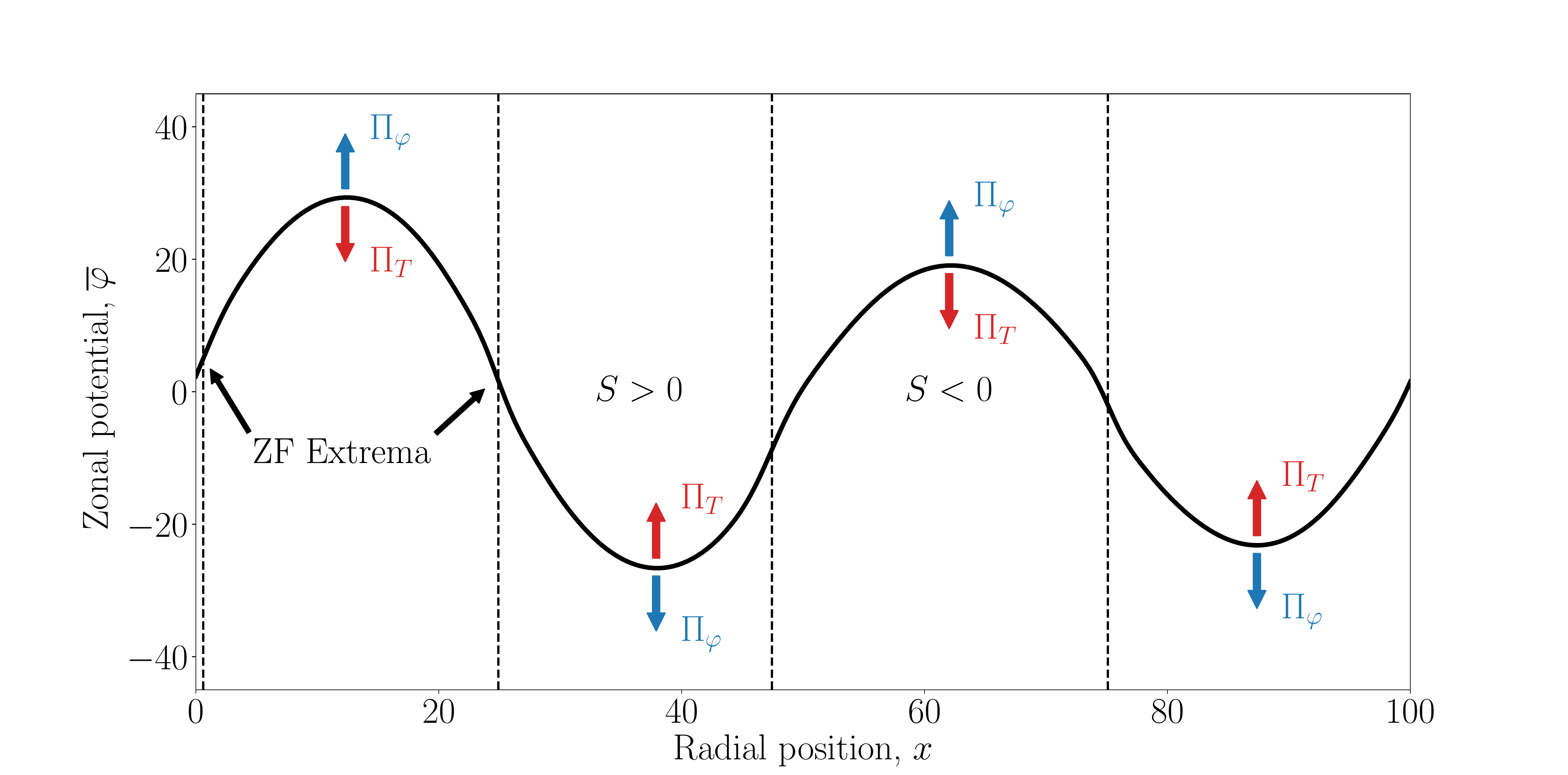}
	\caption{An illustration of the enhancing and suppressing effects of \(\Pi_\phinorm\) and \(\Pi_\deltaT\) on the ZF. The black curve shows the \(\zf{\phinorm}(x)\) profile taken from SimH at \(t = 10\). The ZF extrema are the locations where \(S = \px^2 \zf{\phinorm} = 0\) (marked by dashed lines). }
	\label{fig_zfstab_illustration}
\end{figure}

Let us analyse what causes the turbulent momentum flux \(\Pi_t\) to reverse its sign at the Dimits transition. We split \(\Pi_t = \Pi_\phinorm + \Pi_\deltaT\) and define
\begin{equation}
\label{eq_piphi_and_pit_def}
\Pi_\phinorm \equiv - \zf{(\px \phinorm) (\py \phinorm)} = \zf{u_y u_x}, \qquad \Pi_\deltaT \equiv - \zf{(\px \phinorm) (\py \deltaT)} = \zf{u_y w_x},
\end{equation}
where \(\vect{u} = (-\py \phinorm, \px \phinorm)\) is the \exb flow and \(\vect{w} = (-\py \deltaT, \px \deltaT)\) is the diamagnetic flow. Here \(\Pi_\phinorm\) is the radial flux of the poloidal momentum due to the Reynolds stress of the \exb flow and the "diamagnetic stress" \(\Pi_\deltaT\) is a contribution to the momentum flux that physically arises due to the advection of the poloidal diamagnetic flow \(w_y = \px T\) by the radial \exb flow \(u_x = -\py \phinorm\)\footnotemark. To see this, let us take the zonal average of \eqref{curvy_psi} and differentiate once with respect to \(x\), to obtain an equation for the zonal diamagnetic~flow:
\begin{align}
	\label{eq_zonal_diamagnetic}
	\pt \zf{w}_y + \px \zf{u_x w_y} - \px \Pi_T = \chi \px^2 \zf{w}_y.
\end{align}
The evolution of the zonal poloidal \exb flow is described by \eqref{eq_zfflux}, which can be recast~as
\begin{align}
	\label{eq_zfflow}
	\pt \zf{u}_y + \px \zf{u_x u_y} + \px \Pi_\deltaT = -\px \Pi_d.
\end{align}
Added together, \eqref{eq_zonal_diamagnetic} and \eqref{eq_zfflow} describe the advection of the total poloidal flow \mbox{(\exb + diamagnetic)}, \(v_y \equiv u_y + w_y\), by the radial \exb flow:
\begin{align}
	\label{eq_totalpoloidalflow}
	\pt \zf{v}_y + \px \zf{u_x v_y} = \text{dissipative terms}.
\end{align}
This makes physical sense because the diamagnetic flow is not a real flow and thus cannot advect anything.

\footnotetext{Similar terms in the momentum flux play an important role in the GK theory of momentum transport \citep{parra2009, parra2010, abiteboul2012, calvo2015}.}

The numerical solutions of \eqref{curvy_phi} and \eqref{curvy_psi} reveal that \(\Pi_\phinorm\) and \(\Pi_\deltaT\) are in competition: on average, \(\Pi_\phinorm\) has the same sign as the zonal shear \(S = \px^2\zf{\phinorm}\), while \(\Pi_\deltaT\) has the opposite sign. This is evident in Figure~\ref{fig_momflux_z06chi01}. Equation \eqref{eq_zfflux_int2} then tells us that \(\Pi_\phinorm\) feeds the ZFs by increasing the zonal potential \(\zf{\phinorm}\) in the shear zones of negative zonal shear (where \(\zf{\phinorm}\) is concave) and decreasing it in the shear zones of positive zonal shear (where \(\zf{\phinorm}\) is convex), whereas \(\Pi_\deltaT\) relaxes the ZFs by opposing \(\Pi_\phinorm\). Their combined effect either steepens or relaxes the ZF velocity \(\zf{u}_y = \px\zf{\phinorm}\) at the turning points \(\px^2 \zf{\phinorm} = 0\), depending on which stress is larger. Figure~\ref{fig_zfstab_illustration} is an illustration of this. This competition is crucial for the ability of the ZF to reconstitute itself after a turbulent burst and thus sets the threshold for the Dimits regime.

In order to assess what decides the outcome of this competition (i.e., the relative size of \(\Pi_\phinorm\) and \(\Pi_\deltaT\)), let us consider how zonal shear affects ITG turbulence. For this purpose, consider a shear zone of radial extent \(d\) with a constant zonal shear \(S = \px^2 \zf{\phinorm}\) throughout it. We can then perform the usual shearing-box change of variables \((t, x, y) \mapsto (\shf{t}, \shf{x}, \shf{y})\), where
\begin{equation}
\shf{t} = t, \ \shf{x} = x, \ \shf{y} = y - Stx.
\end{equation}
This coordinate transformation eliminates the spatially inhomogeneous zonal-advection terms \((\px \zf{\phinorm})\py = Sx\py\) in \eqref{curvy_phi} and \eqref{curvy_psi}. Consider a Fourier mode \(\phinorm, \deltaT \propto \exp(i \shf{k}_x \shf{x} + i \shf{k}_y \shf{y})\) in this shearing frame. In the laboratory frame \((t, x, y)\), this mode has the form \(\phinorm, \deltaT \propto \exp(i k_x x + i k_y y)\), where
\begin{equation}
\label{eq_labframe_ks}
k_x = \shf{k}_x - St \shf{k}_y, \ k_y = \shf{k}_y.
\end{equation}
Thus, the ZF shear introduces an effective drift of the laboratory-frame radial wavenumber\footnotemark.\footnotetext{It is certainly true that an equilibrium shear would have such an effect on the turbulence. However, this is not guaranteed for ZFs. Their influence on the turbulence depends crucially on the modified electron response \eqref{eq_eresponse}. This is a distinguishing feature of ion-scale physics that does not exist in, e.g., the electron-scale version of the model presented here. } The direction of this drift is given by the sign of \(S\), viz., \(S > 0\) gives rise to an anticorrelation of \(k_x\) and \(k_y\), i.e., \(k_xk_y < 0\), whereas for \(S < 0\), \(k_xk_y > 0\). Integrating the effect of \(\Pi_\phinorm\) over the sheared region, we obtain
\begin{equation}
\frac{1}{d}\int dx \ \Pi_\phinorm =  -\frac{1}{dL_y}\int dxdy \ \left(\px \phinorm\right) \left(\py \phinorm\right) = -\sum_\vect{k} k_xk_y |\phinorm_\vk|^2.
\end{equation}
Therefore, on average, \(\Pi_\phinorm\) has the same sign as \(S\), and, thus, feeds the ZFs that generate the shear zones\footnotemark.

\footnotetext{This is a well-known result in the context of Rossby-wave turbulence \citep[see][chapter 15.1.2]{vallis2017}}

We can write a similar expression for the diamagnetic stress:
\begin{equation}
\frac{1}{d}\int dx \ \Pi_\deltaT = -\sum_\vect{k} k_xk_y |\phinorm_\vk|^2 \re{\frac{\deltaT_\vk}{\phinorm_\vk}}.
\end{equation}
Then the total turbulent momentum flux integrated over a shear region is
\begin{equation}
\label{eq_bothstresses}
\frac{1}{d}\int dx \ \Pi_t = -\sum_\vect{k} k_xk_y |\phinorm_\vk|^2 \left(1 + \re{\frac{\deltaT_\vk}{\phinorm_\vk}} \right).
\end{equation}
Recall that we already encountered the quantity \(\re{(\deltaT_\vk/\phinorm_\vk)}\) when dealing with the secondary instability in Section~\ref{sect_sec}. There we found that \(\re{(\deltaT_\vk/\phinorm_\vk)} < 0\) for all linearly unstable modes. Thus, linear theory predicts that \(\Pi_\deltaT\) and \(\Pi_\phinorm\) are anti-correlated due to the negative sign of \(\re{(\deltaT_\vk/\phinorm_\vk)}\). Now let us perform a more detailed analysis of the linear modes and attempt to construct a model for the Dimits threshold based on it.

\subsection{Dimits Threshold from Linear Physics}
\label{sect_dimits_linearapprox}

\begin{figure}
	\centering
	\includegraphics[scale=0.27]{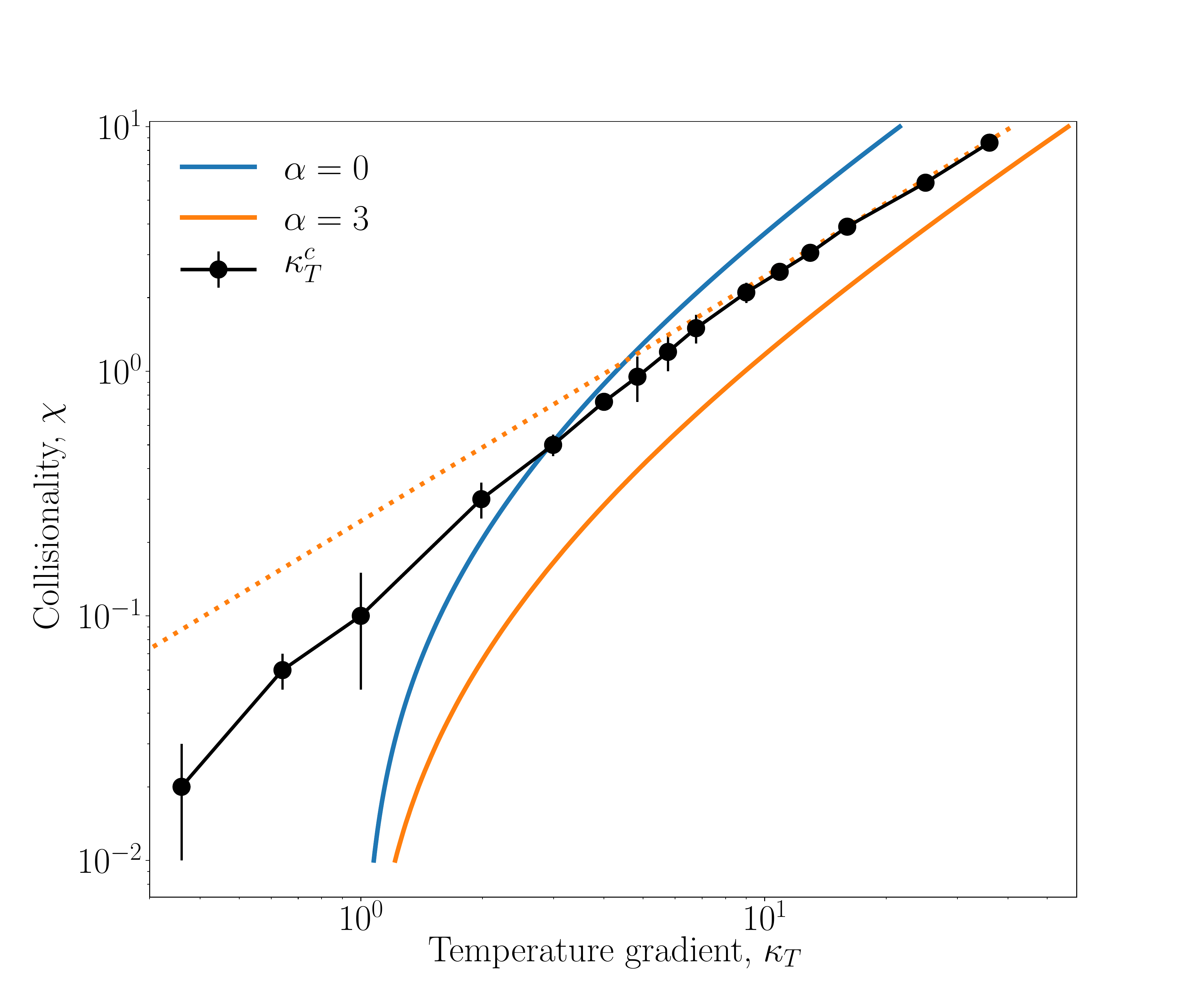}
	\caption{Comparison of numerical data to the analytical estimate for the threshold of the Dimits regime (Section~\ref{sect_dimits_linearapprox}). The black points represent the numerically observed \(\vtcrit\). The other two curves correspond to the parameters for which the fastest-growing mode with \(k_x = \alpha k_y\) satisfies \(\re{(\deltaT_\vk/\phinorm_\vk)} = -1\). The case \(\alpha = 0\) corresponds to the threshold for the suppression of the secondary instability of the fastest streamer (\(k_x = 0\)) mode (see Section~\ref{sect_sec}), and the curve with \(\alpha = 3\) asymptotes to the numerically determined slope of the Dimits threshold as \(\vt \to \infty\). Its asymptote is represented here by the dotted line and is given by \eqref{eq_vtcrit_linear_approx}.}
	\label{fig_crit_grad}
\end{figure}

Using our knowledge of ITG perturbations in a region of uniform ZF shear, and of the turbulent momentum flux produced by them, we can make a heuristic linear-physics-based estimate for the Dimits threshold \(\vtcrit\). In view of \eqref{eq_bothstresses}, it is given by the temperature gradient at which the \textit{relevant} ITG modes have \(\re{(\deltaT_\vk/\phinorm_\vk)} = -1\). By \textit{relevant} we mean those ITG modes that dominate the turbulence in the shear zones. It is tempting to assume that these modes would be the most unstable modes in the system. This, however, cannot be the case because the most unstable modes are the radial streamers with \(k_x = 0\), but the zonal shear that we find is comparable in magnitude to the largest ITG growth rate (\(S \sim \gamma_\text{max}\)), and, therefore, is bound to break these streamers. Following the discussion in Section~\ref{sect_stresses}, we may assume that the typical ITG modes in sheared turbulence satisfy \(k_x \sim \alpha k_y\), where \(\alpha \sim S\tau_\text{nl}\) characterises how tilted the mode is, \(\tau_\text{nl}\) being the nonlinear correlation time of the turbulence in the shear zones. \mbox{If \(\tau_\text{nl}^{-1} \sim \gamma_\text{max} \sim S \), then \(\alpha \sim 1\)}. 

Thus, we assume that the relevant modes are tilted with \(k_x =\alpha k_y\), where \(\alpha \sim 1\) is an unknown tilt parameter that depends on the structure of the turbulence. We then look for the temperature gradient \(\vt\) at which the fastest-growing ITG mode with \(k_x = \alpha k_y\) satisfies \(\re{(\deltaT_\vk/\phinorm_\vk)} = -1\). This yields a prediction for the Dimits threshold in the \((\vt, \chi)\) plane that we refer to as the "fastest-mode approximation". Note that there is no \textit{a priori} reason to assume that \(\alpha\) is itself not a function of \(\vt\). 

\subsubsection{High-Collisionality Limit}

We can take the \(\chi \to \infty\) limit of the fastest-mode approximation analytically. We formally order \(\chi \sim \vt\), as suggested by Figure~\ref{fig_crit_grad}. We then use the dispersion relation \eqref{eq_disp_relation} to find the growth rate \(\gamma_\vk\) and real frequency \(\omega_\vk\) of the fastest mode with \(k_x =\alpha k_y\). In Section~\ref{sect_unstable_region_col}, we showed that a mode of wavenumber \(\vk\) is unstable if and only if
\begin{equation}
	\vt k_y^2 > a\chi^2k^6 = a\chi^2(1 + \alpha^2)^3k_y^6,
\end{equation}
so all unstable modes with \(k_x = \alpha k_y\), where \(\alpha \sim \order{1}\), satisfy
\begin{equation}
	k_y^4 < \frac{\vt}{a\chi^2(1 + \alpha^2)^3} \sim \order{\vt^{-1}}.
\end{equation}
Similarly, using the results in Section~\ref{sect_unstable_region_cless} for the FLR bounds on the region of unstable wavenumbers, we find \(k_y^4 \sim \order{\vt^{-1}}\) in the limit \(\vt \to \infty\). Thus, both mechanisms that bound the region of instability (and hence restrict the largest ITG growth rate), lead to the same scaling for the unstable wavenumbers. Therefore, the wavenumber of the most unstable mode must also satisfy \(k_y \sim \vt^{-1/4}\). Applying the ordering \(\chi \sim \vt \gg 1\) and \(k_x = \alpha k_y \sim \vt^{-1/4}\) to \eqref{eq_disp_short_defs}, we find
\begin{equation}
\label{eq_shortdefs_ordering}
A \sim \order{\vt^{1/2}}, \qquad B \sim \order{1}, \qquad C \sim \order{\vt^{1/4}}, \qquad f \sim \order{1}, \qquad g \sim \order{1}.
\end{equation}
The unstable solution of \eqref{eq_disp_short} is
\begin{equation}
	\gamma_\vk - i \omega_\vk = \frac{-A-B+iC + \sqrt{(A-B+iC)^2 + 4fAB - 4igAC}}{2}.
\end{equation}
After expanding it using \eqref{eq_shortdefs_ordering}, we find
\begin{align}
	\label{eq_mostunstable_orderings}
	&\gamma_\vk \sim \order{1} \ll \omega_\vk \sim \order{\vt^{1/4}} \ll \chi k^2 \sim \order{\sqrt{\vt}}, \\
	&\omega_\vk = -(1-g)C + \order{\vt^{-1/4}} = -\vt (1-b)(1+\alpha^2)k_y^3 + \order{\vt^{-1/4}}.
\end{align}
Therefore, \eqref{eq_re_tphi} gives
\begin{align}
	\label{eq_dimits_highcol}
	\re{\frac{\deltaT_\vk}{\phinorm_\vk}} = \frac{k_y\vt\omega_\vk}{|\gamma_\vk -i\omega_\vk + \chi k^2|^2} \approx \frac{k_y\vt\omega_\vk}{\chi^2k^4} \approx -\frac{1-b}{1+\alpha^2} \frac{\vt^2}{\chi^2}.
\end{align}
Thus, the large-temperature-gradient fastest-mode approximation of the Dimits threshold is a straight line in the \((\vt, \chi)\) plane, given by
\begin{align}
	\label{eq_vtcrit_linear_approx}
	\vtcrit \approx \chi \sqrt{\frac{1 + \alpha^2}{1 - b}},
\end{align}
a posteriori confirming the ordering \(\chi \sim \vt\). The numerically determined Dimits threshold is indeed close to a straight line. Fitting the slope of that line to \eqref{eq_vtcrit_linear_approx} yields \(\alpha \approx 3\). Comparison of the prediction for the Dimits threshold for this value of \(\alpha\), as well as \(\alpha = 0\), which corresponds to the threshold for the secondary instability of a primary streamer (as discussed in Section~\ref{sect_sec}), can be found in Figure~\ref{fig_crit_grad}. The convergence is slow (\(\propto \vt^{-1/4}\)), hence the sizeable discrepancy for the values of \(\vt\) shown there, but we consider the asymptotic result to be sound.

\subsubsection{Low-Collisionality Limit}

Using a calculation that is nearly identical to the one in Section~\ref{sect_sec_longwave}, we can analytically take the limit \(\chi \to 0\) of the fastest-mode approximation using the collisionless dispersion relation \eqref{eq_disp_relation_nodamping} and inserting it into \eqref{eq_collisionless_re}. We obtain that \(\vtcrit \to 1\) as \(\chi \to 0\) for the fastest mode with \(k_x =\alpha k_y\), regardless of the value of \(\alpha\). This is a weakness of our "fastest-mode approximation" because the numerical data suggests instead that \(\vtcrit \to 0\) as \(\chi \to 0\). Thus, the assumptions that we made above about the relevance of the fastest-growing modes appear to be inadequate at low collisionality. \linebreak

To summarise, the assumption that the momentum flux and \(\re{(\deltaT_\vk/\phinorm_\vk)}\) are dominated by the most unstable mode with some tilt given by \mbox{\(\alpha = k_x / k_y\)} allows us to predict the Dimits threshold at high collisionality, but fails at low collisionality. This partial success is likely due to the fact that \(\re{(\deltaT_\vk/\phinorm_\vk)}\) for the most unstable mode is independent of \(\vk\) for \(\vt \sim \chi \gg 1\) [see \eqref{eq_dimits_highcol}]. So, not only the most unstable, but in fact all modes in its vicinity will have the same value of \(\re{(\deltaT_\vk/\phinorm_\vk)}\). On the other hand, the failure of these assumptions at low collisionality suggests that we cannot use linear theory to predict the threshold there, but must rather focus on the nonlinear structure of the ITG turbulence seeded by ferdinons during bursts. This will be further discussed in Section~\ref{sect_discussion}.

\subsection{Beyond the Dimits Regime}
\label{sect_blowup}

\begin{figure}
	\centering
	\includegraphics[scale=0.46]{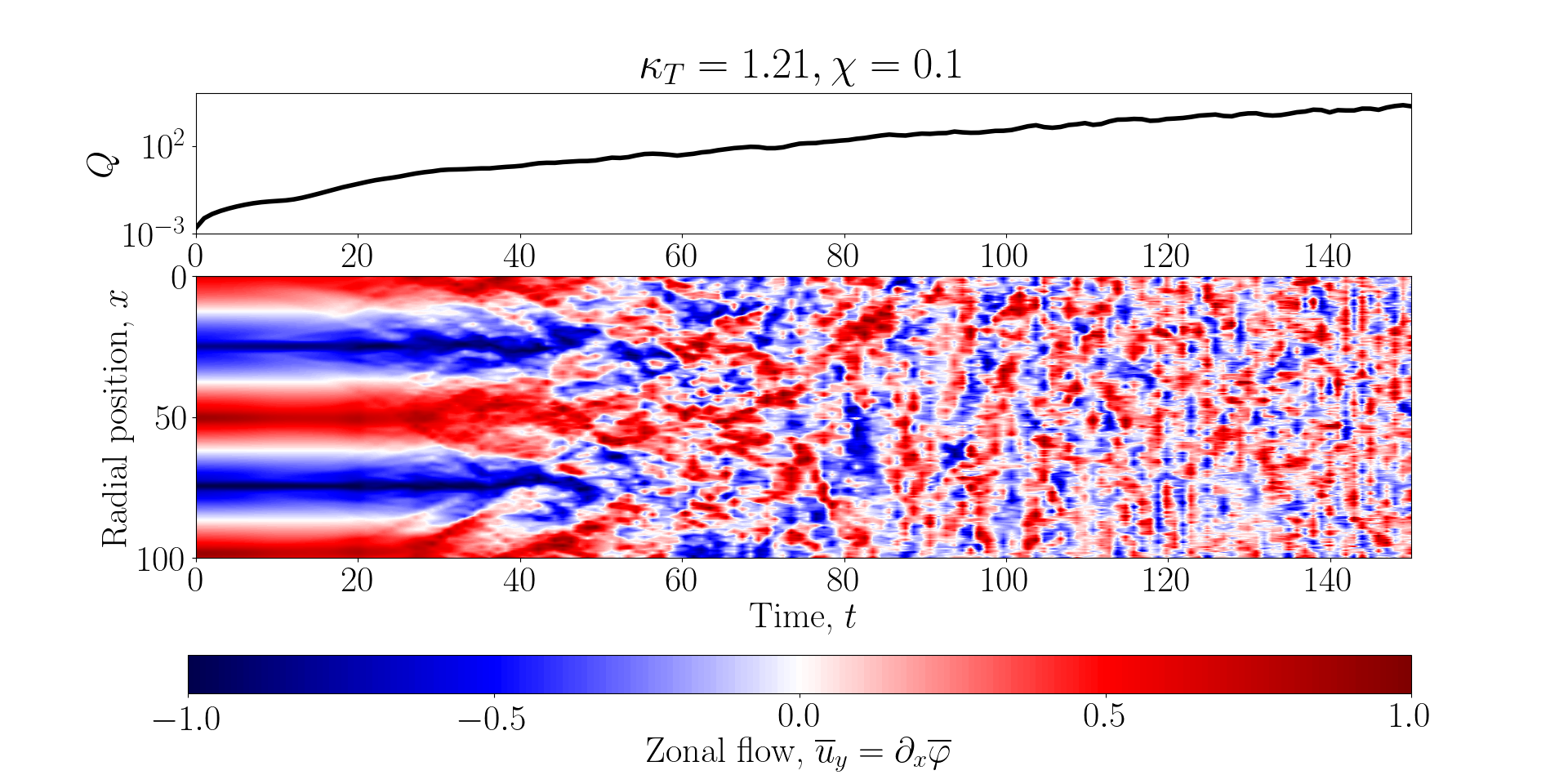}
	\caption{Time evolution of the heat flux (top) and ZF velocity \(\zf{u}_y\) (bottom) beyond the Dimits regime. The ZF amplitude in the lower panel is normalised to a maximum of \(1\) at each time. The initial conditions are the same as those for SimH, but with an augmented temperature gradient \(\vt = 1.21 > \vtcrit \approx 1\) and a lower resolution of \(337\times167\) Fourier modes due to the numerical cost of simulating the blow-up regime. The ZFs are quickly destroyed and large-scale ZFs never reappear, while the box-averaged heat flux \(Q\) grows exponentially. This state is eventually dominated by a streamer with a poloidal scale equal to that of the integration domain (see Figure~\ref{fig_temp_snapshots_blowup}). }
	\label{fig_hf_blowup}
\end{figure}

\begin{figure}
	\centering
	\includegraphics[scale=0.46]{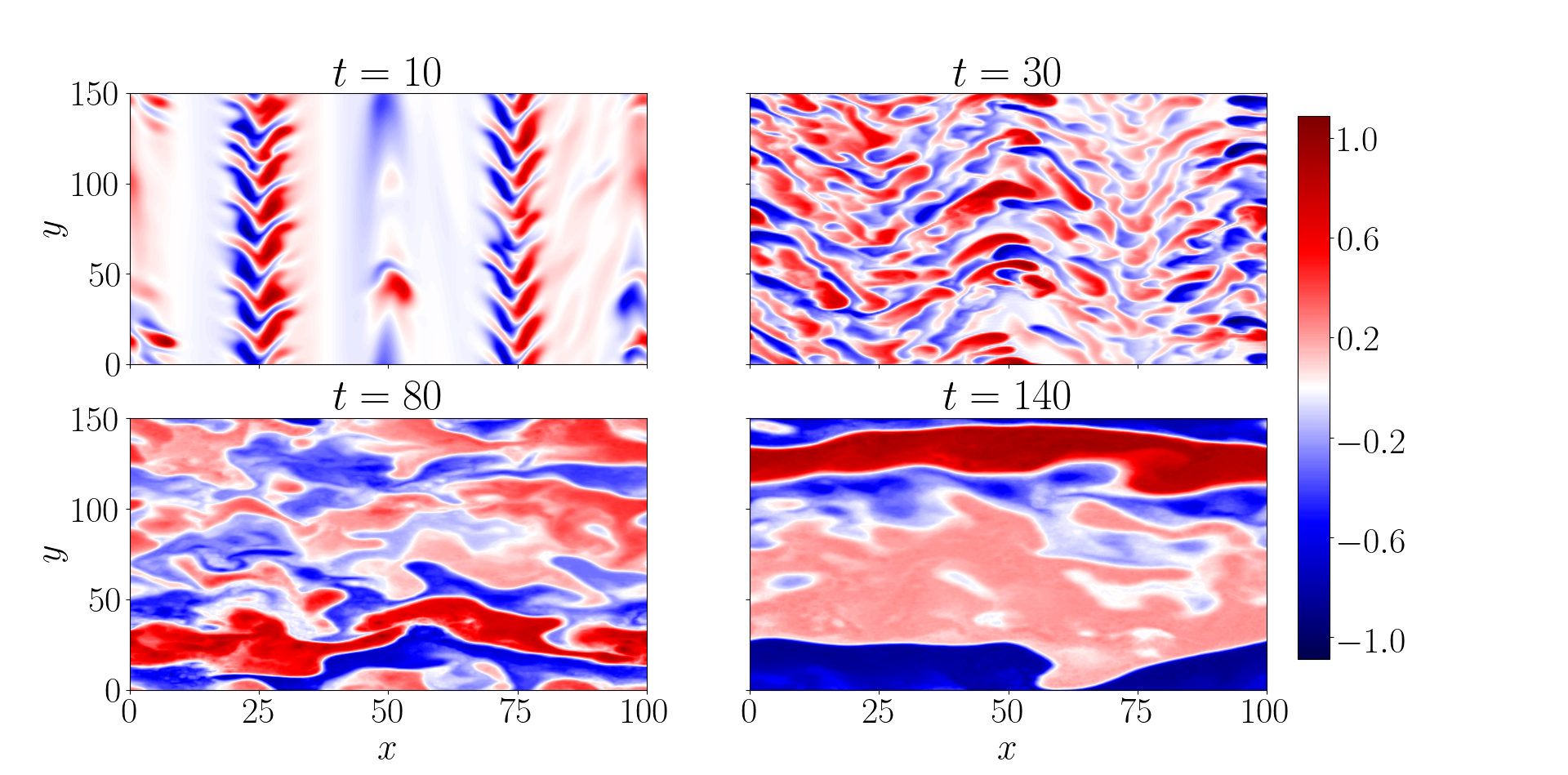}
	\caption{Snapshots of temperature perturbations in the blow-up state beyond the Dimits threshold (a movie is available in the supplementary material). The amplitudes are normalised to a maximum of \(1\) at each time. The initial conditions are the same as those for SimH, but with an augmented temperature gradient \(\vt = 1.21 > \vtcrit \approx 1\) and a lower resolution of \(337\times167\) Fourier modes due to the numerical cost of simulating the blow-up regime.  }
	\label{fig_temp_snapshots_blowup}
\end{figure}

Beyond the Dimits threshold (\(\vt > \vtcrit\)), our 2D system fails to reach saturation on a scale smaller than the domain size --- perturbations grow exponentially and the box-sized streamer (\(k_x = 0, k_y = 2\pi/L_y\)) eventually dominates the spectrum. Figures \ref{fig_hf_blowup} and \ref{fig_temp_snapshots_blowup} show that the large-scale, coherent ZFs that comprise the zonal staircase are quickly destroyed and never reappear. This is consistent with the illustration in Figure~\ref{fig_zfstab_illustration} and the discussion in Section~\ref{sect_stresses}. For \(\vt > \vtcrit\), if a shear zone of coherent zonal shear (like the ones we observe in the Dimits regime) were formed, the turbulent stress \(\Pi_t\) would flatten out the ZF profile. Any coherent zonal shear is thus the harbinger of its own demise due to the momentum flux of the tilted turbulent eddies. The nonzonal perturbations grow exponentially, and so do the ZFs, but the latter are now dominated by small-scale time-incoherent zonal modes that are unable to quench the instability. 

The lack of saturation beyond the Dimits regime in 2D is not surprising. GK simulations have shown that, beyond the Dimits regime in saturated 3D ITG turbulence, the ITG frequency at the injection ("outer") scale of the perpendicular plane is balanced by the parallel propagation time --- the turbulence is in "critical balance" \citep{barnes2011}:
\begin{equation}
\label{eq_critbalance}
\omega^\text{o} \sim v_\parallel k_\parallel,
\end{equation}
where \(v_\parallel\) is some appropriate speed of parallel propagation (e.g., the ion thermal speed \(\vti\)), and \(\omega^\text{o} \propto k_y\) is the ITG frequency at the outer scale, proportional to \(k_y\) by \eqref{eq_disp_relation}. In a tokamak, the smallest allowed value of \(k_\parallel\) is \(k_\parallel \sim L_\parallel^{-1}\), where \(L_\parallel\) is the parallel connection length of the device. Thus, a parallel length scale is enforced by the magnetic geometry. The poloidal outer scale (\(k_y\)) then follows by \eqref{eq_critbalance} and the radial outer scale is enforced by zonal shearing (\(k_x \sim k_y\)). The 2D approximation can be obtained as the \(k_\parallel \to 0\) limit of the 3D system. In this case, \eqref{eq_critbalance} implies that \(k_y \to 0\), in agreement with the blow up dominated by the box-sized streamer that we observe beyond the Dimits threshold. Thus, the 2D approximation is fundamentally inadequate as a description of fully developed ITG turbulence\footnotemark. \footnotetext{Also because of the presence of 2D invariants (see Section~\ref{sect_cons}), which can lead to an inverse cascade and energy pile-up at the largest available (box) scale, as they do in 2D hydrodynamic turbulence \citep{frisch1995}.}However, we have shown that ZF-mediated saturation and the Dimits transition are captured by a 2D model. Of course, it is an outstanding task (left for future work) to confirm that the physics of the 2D Dimits transition remains (qualitatively) valid in 3D.

\section{Discussion}
\label{sect_discussion}

We have found that the saturation of 2D ITG turbulence in \(Z\)-pinch geometry is mediated by strong quasi-static ZFs with patchwise-constant zonal shear (Section~\ref{sect_nl}). There is a clear transition between a ZF-dominated Dimits regime and a strongly turbulent state, which in 2D fails to saturate at a finite amplitude (Section~\ref{sect_blowup}). The mechanism that sustains the ZFs in the Dimits regime (\(\vt < \vtcrit\)) and undermines them beyond it (\(\vt > \vtcrit\)) is linked to the turbulent momentum flux of ITG modes in the presence of a coherent zonal shear. Namely, in the Dimits regime, the response of ITG turbulence to strong (comparable to the ITG-instability growth rate), coherent zonal shear can be described in terms of a negative turbulent viscosity that reinforces the ZFs. This turbulent viscosity vanishes at the Dimits threshold and becomes positive beyond it, thus impeding any strong zonal shear that could suppress turbulence (Section~\ref{sect_dimits_threshold}). Viewed this way, the Dimits transition is caused by a change in the properties of sheared ITG turbulence. In the model considered here, the turbulent momentum flux consists of the usual Reynolds stress, familiar from hydrodynamics, and a diamagnetic contribution. We find that the former acts to reinforce the ZFs, while the latter opposes the ZFs (Section~\ref{sect_stresses}). 

In general, therefore, determining whether a set of equilibrium parameters lies within the Dimits regime, requires one to make a statement about the combined momentum flux of all turbulent modes. In Section~\ref{sect_dimits_linearapprox}, we employed the heuristic assumption that the momentum flux is determined predominantly by the most unstable modes with a finite tilt (\(k_x = \alpha k_y\)). We found that \(\alpha \approx 3\) models reasonably well the Dimits threshold for large temperature gradients and collisionalities\footnotemark. At low collisionalities, such simple considerations do not produce quantitatively satisfactory results. \footnotetext{Note that \(\alpha \sim 1\) is consistent with a balance of zonal shear and turbulent turnover time (\(S \sim \tau_\text{nl}^{-1}\)).}

The mechanism for the Dimits transition described above is not directly tied to the onset of the radially localised tertiary instability found at the extrema of the ZF (Section~\ref{sect_tert} and Appendix~\ref{appendix_tert}). It is more appropriate to view it as the quenching of some nonlinear version of the secondary instability (Section~\ref{sect_sec}). This appears to be in contrast with the transition seen in the Hasegawa-Wakatani equations, where the tertiary instability was argued to determine the threshold for the strong-turbulence regime \citep{zhu2019_dimits}. The Hasegawa-Wakatani system does not contain the diamagnetic stress discussed in Section~\ref{sect_stresses} or, indeed, any other form of poloidal momentum flux apart from the Reynolds stress. Thus, by construction, it lacks the effects discussed in this paper. However, it is worth mentioning that \citet{zhu2019_dimits} have observed turbulent bursts triggered by travelling structures in the Hasegawa-Wakatani system. Due to the lack of diamagnetic stress, these turbulent bursts are bound to restore the ZFs.

As concluded in Section~\ref{sect_dimits_linearapprox}, an accurate prediction of the Dimits threshold requires a detailed understanding of the properties of sheared ITG turbulence. The nature of this turbulence is likely to be closely related to the properties of the localised structures ("ferdinons") that are seeded by the tertiary-unstable regions around the ZF maxima (see Section~\ref{sect_convzones_maxima}) and then drift through the shear zones (see Section~\ref{sect_ferds}). If the averaged properties of sheared turbulence correspond to those of a collection of (independent) ferdinons, we might be able to use the momentum flux of a ferdinon in order to make predictions about sheared turbulence. Developing an analytical approximation of a simple ferdinon is therefore an important outstanding task for future research.

The insight into the significance of the turbulent momentum flux of sheared turbulence provides us with a natural starting point for the investigation of the Dimits threshold in 3D gyrokinetics. To reiterate, we have showed that, within the Dimits regime, the time-averaged poloidal momentum flux of turbulence sheared by a region of constant zonal shear acts to reinforce the ZFs and, thus, zonal shear. On the other hand, beyond the Dimits regime, the overall sign of the momentum flux reverses and the ZFs are destroyed by the turbulence. The validity of this statement is a well-posed question that can be answered by 3D gyrokinetic numerical simulations, regardless of the number of parameters (which, in a realistic scenario, greatly outnumbers two). 

\section*{Acknowledgements}

The authors would like to thank I.~Y.~Dodin, M.~R.~Hardman, B.~F.~McMillan, \mbox{D.~A.~St-Onge}, and the Oxford Plasma Theory group for useful discussions.

This work has been carried out within the framework of the EUROfusion Consortium and has received funding from the Euratom research and training programme 2014-2018 and 2019-2020 under grant agreement No. 633053. The views and opinions expressed herein do not necessarily reflect those of the European Commission. The work of A.A.S. and F.I.P. was funded in part by the Engineering and Physical Sciences Research Council (EPSRC) [EP/R034737/1]. W.D. was supported by the US Department of Energy Grants DEFG0293ER54197 and DE-SC0018429. 


\appendix

\section{Derivation of the Model Equations}
\label{appendix_curvy_derivation}

\subsection{Gyrokinetic Equation}
\label{appendix_gk}

The derivation of our fluid model is similar to the one by \citet{newton2010}. We start from the 2D ion gyrokinetic (GK) equation in a \(Z\)-pinch-like equilibrium, as discussed in Section~\ref{sect_magn_geometry}, omitting the parallel streaming term \citep[for a review of gyrokinetics, see][]{abelgk1}. The ion distribution function is \(f_i = F_i + \delta f_i\), where \(F_i = n_i/\left(\pi^{3/2} \vti^3\right) \exp\left(-v^2 / \vti^2\right)\) is the equilibrium ion distribution function and
\begin{equation}
\label{eq_f_def}
\delta f_i = -\frac{Ze\phinonnorm}{T_i}F_i + h\left(t, \vect{R}, v_{\parallel}, v_{\perp}\right).
\end{equation}
The gyrocentre distribution \(h\) satisfies
\begin{equation}
\label{gk_eq}
\frac{\partial}{\partial t} \left(h  - \frac{ZeF_i}{T_i} \avgR{\phinonnorm}\right) + \vd \bcdot \frac{\partial h}{\partial \vect{R}} + \langle \ve \rangle_R \bcdot \bigg[\frac{\partial h}{\partial \vect{R}} - \uvect{x} \left( \frac{v^2}{v^2_{ti}} - \frac{3}{2}\right) \frac{F_i}{L_T} \bigg] =\langle C_{l}[h] \rangle_R.
\end{equation}
Here \(T_i = m_i \vti^2 / 2\) and \(L_T^{-1} = -\px \ln T_i\) are the ion equilibrium temperature and temperature gradient, respectively. The ion mass is \(m_i\), the ion charge is \(Ze\), and \(\Omega_i = ZeB/m_ic\) is the ion gyrofrequency. \(\vect{B}\) is the equilibrium magnetic field and \(\avgr{.}\) and \(\avgR{.}\) denote, respectively, the gyroaverages at fixed position and fixed ion guiding centre \(\vect{R} = (X, Y, Z) = \vect{r} - \uvect{b} \times \vect{v} / \Omega_i\), where \(\uvect{b} = \vect{B} / B\) is the unit vector parallel to the magnetic field. These gyroaverages are defined as
\begin{align}
\label{avgrdef}
&\avgr{f(\vect{R})} \equiv \avgTheta{f(\vect{r} - \vect{\rho}(\theta))} = \int_0^{2\pi} \frac{d\theta}{2\pi} \ f(\vect{r} - \vect{\rho}(\theta)), \\
\label{avgRdef}
&\avgR{f(\vect{r})} \equiv \avgTheta{f(\vect{R} + \vect{\rho}(\theta))} = \int_0^{2\pi} \frac{d\theta}{2\pi} \ f(\vect{R} + \vect{\rho}(\theta)),
\end{align}
where \(\avgTheta{.}\) denotes the average with respect to the gyroangle \(\theta\), \(\vect{\rho}(\theta) \equiv \vect{b} \times \vect{v} / \Omega_i\), \(\vect{v} = v_\parallel \uvect{b} + v_\perp \left(\cos\theta \uvect{y} - \sin\theta\uvect{x}\right)\), and the unit vectors \(\lbrace \uvect{x}, \uvect{y}, \uvect{b} \rbrace\) form a right-handed orthonormal basis, as shown in Figure~\ref{fig_zpinch}. We will require the following properties of the gyroaverage:
\begin{align}
	\label{eq_rho_averages}
	\avgTheta{\vect{\rho}} = 0, \qquad \avgTheta{\vect{\rho}\vect{\rho}} = \frac{1}{2}\frac{v_\perp^2}{\vti^2} \rho_i^2 \mathds{1}_\perp,
\end{align}
where \(\mathds{1}_\perp\) is the identity matrix in the \((x, y)\) plane. Using \eqref{eq_rho_averages} and Taylor expanding \eqref{avgrdef} and \eqref{avgRdef}, we obtain
\begin{alignat}{2}
\label{eq_avgr_lowestorder}
&\avgr{h} &&= \left[1 + \frac{1}{4} \frac{v_\perp^2}{\vti^2} \rhoidelperpsq + \order{\kperprhoisqq} \right]h(\vect{r}), \\
\label{eq_avgR_lowestorder}
&\avgR{\phinorm} &&= \left[1 + \frac{1}{4} \frac{v_\perp^2}{\vti^2} \rhoidelperpsq  + \order{\kperprhoisqq}\right]\phinorm(\vect{R}).
\end{alignat}

The velocities \(\vd\) and \(\avgR{\ve}\) in \eqref{gk_eq} are the magnetic and (gyroaveraged) \exb drifts, respectively, given by 
\begin{align}
\label{eq_drifts_def}
\vd &= \frac{1}{\Omega} \left[ v^2_\parallel \hat{\vect{b}} \times (\vect{b} . \nabla \vect{b}) + \frac{v_\perp^2}{2} \vect{b} \times \nabla \ln B \right]= \frac{1}{\Omega_i B} \left( v_\parallel^2 + \frac{v_\perp^2}{2} \right) \hat{\vect{b}} \times \del B,
\\\avgR{\ve} &= \frac{c}{B} \hat{\vect{b}} \times \partd{\avgR{\phinonnorm}}{\vect{R}}.
\end{align}
The second equality in \eqref{eq_drifts_def} is obtained by assuming that the magnetic field is created by currents external to the spatial domain, so \(\del \times \vect{B} = 0\). For the \(Z\)-pinch geometry and coordinates discussed in Section~\ref{sect_model} and shown in Figure~\ref{fig_zpinch}, we obtain
\begin{equation}
    \vd = -\frac{\rhoivti}{L_B} \left( \frac{v_\parallel^2}{\vti^2} + \frac{v_\perp^2}{2\vti^2} \right) \uvect{y}.
\end{equation}

The term on the right-hand side of \eqref{gk_eq} is the gyroaveraged linearised Landau collision operator. To lowest order in the mass-ratio expansion ($m_e / m_i \ll 1$), only ion-ion collisions contribute and the linearised operator is given by
\begin{equation}
\label{eq_collision_op}
C_{l}[h] = \frac{\nu_i\vti^3}{n_i} \partdat{}{\vect{v}}{\vect{r}} \bcdot \left\{ F_i(\vect{v}) \int d^3 \ \vect{v'} F_i(\vect{v'}) \tens{U} \bcdot \left[ \partdat{}{\vect{v}}{\vect{r}}  \frac{h(\vect{v})}{F_i(\vect{v})} - \partdat{}{\vect{v'}}{\vect{r}}  \frac{h(\vect{v'})}{F_i(\vect{v'})}  \right] \right\},
\end{equation}
where \(\tens{U} = \left(\vect{u}^2 \tens{I} - \vect{u} \vect{u}\right) / u^3 \),  \(\vect{u} = \vect{v} - \vect{v'}\), and
\begin{equation}
	\label{eq_nu_i_def}
	\nu_i = \frac{2\pi Z^4e^4n_i \ln\Lambda_{i}}{m_i^2\vti^3}
\end{equation}
is the ion-ion collision frequency, where \(\ln\Lambda_{i}\) is the Coulomb logarithm \citep{helander2002}.

The electrostatic gyrokinetic equation \eqref{gk_eq} is closed by the quasineutrality condition \(\delta n_e = Z \delta n_i\) and the modified adiabatic electron response \citep{abelgk2}
\begin{equation}
\label{eq_fe}
\delta f_e = \frac{e\dw{\phinonnorm}}{T_e} F_e.
\end{equation}
Therefore,
\begin{equation}
\label{eq_qn}
\frac{1}{n_i}\int  d^3 \vect{v}  \  \avgr{h} = \frac{Z e }{T_i} \phinonnorm + \frac{e}{T_e} \phinonnorm'.
\end{equation}

Putting all of this together, we arrive at the 2D electrostatic gyrokinetic system in our \(Z\)-pinch equilibrium:
\begin{align}
\label{eq_curvy_gk}
&\partd{}{t} \left(h - \avgR{\phinorm}F_i\right)  + \frac{\rho_i \vti}{2L_T}\left( \frac{v^2}{v^2_{ti}} - \frac{3}{2} \right)F_i \partd{\avgR{\phinorm}}{Y} - \frac{\rho_i \vti}{L_B} \left( \frac{v^2_\parallel}{\vti^2} + \frac{v^2_\perp}{2 \vti^2} \right)\partd{h}{Y}
 \nonumber \\ & \quad + \frac{1}{2} \rho_i \vti \left\{ \avgR{\phinorm}, h \right\}  = \avgR{C_{l}[h]},
\end{align}
\begin{equation}
\label{eq_curvy_qn}
\frac{1}{n_i}\int d^3\vect{v}  \ \avgr{h} = \phinorm + \tau \phinorm',
\end{equation}
where the normalised (to ion units) electric potential is \(\phinorm = Ze\phinonnorm / T_i\) and the temperature ratio is \(\tau = T_i / ZT_e\).

\subsection{Lowest-Order Solution}

We now apply the high-collisionality, long-wavelength, cold-ion ordering \eqref{eq_ordering}. In this expansion, we write \(h = h^{(0)} + h^{(1)}\), where \(\order{h^{(1)}} \sim  \order{\kperprhoisq h^{(0)}}\). Then, to lowest order, equation \eqref{eq_curvy_gk} gives $C_l [h^{(0)}] = 0$, whose solution is a perturbed Maxwellian
\begin{equation}
\label{eq_h0}
h^{(0)}(\vect{R}) = \left[\frac{\delta N(\vect{R})}{n_i} + \frac{\delta T (\vect{R})}{T_i} \left(\frac{v^2}{\vti ^2} - \frac{3}{2}\right)\right]F_i.
\end{equation}
Here the abstract quantities $\delta N (\vect{R})$ and $\delta T (\vect{R})$ are taken to be functions of the guiding centre $\vect{R}$. Substituting the expansion for \(h\) into the quasineutrality equation \eqref{eq_curvy_qn} and using \eqref{eq_avgr_lowestorder}, we obtain
\begin{equation}
\label{eq_lowestorderstuffinappendix}
\frac{1}{n_i} \intv \left(1 + \frac{1}{4} \frac{v_\perp^2}{\vti^2} \rhoidelperpsq\right) \left[\frac{\delta N}{n_i} + \frac{\delta T}{T_i} \tempvel\right]F_i = \phinorm + \tau \dw{\phinorm},
\end{equation}
where we have absorbed the density and temperature moments of \(h\) into \(h^{(0)}\) by imposing
\begin{equation}
\intv h^{(1)} = \intv v^2 h^{(1)} = 0.
\end{equation}
Formally, in writing down \eqref{eq_lowestorderstuffinappendix}, we have only assumed that the density moment of \(h^{(1)}\) vanishes. We will use the condition that \(h^{(1)}\) has a zero temperature moment in Appendix~\ref{appendix_cons}. 

Under the ordering \(\tau \sim \kperprhoisq \ll 1\), \eqref{eq_lowestorderstuffinappendix} yields
\begin{equation}
\label{eq_deltaN}
\frac{\delta N}{n_i} = \phinorm + \tau \dw{\phinorm} - \frac{1}{4} \rhoidelperpsq \left(\phinorm + \frac{\delta T}{T_i}\right) + \order{\kperprhoisqq \phinorm}.
\end{equation}
We now proceed to take density and temperature moments of the GK equation \eqref{eq_curvy_gk} at fixed particle position \(\vect{r}\) and retain only the lowest-order terms in the ordering \eqref{eq_ordering}. We will find that the density moment of \eqref{eq_curvy_gk} vanishes to lowest order, hence we are required to expand that moment to order \(\order{\kperprhoisq h}\). We shall only require terms up to \(\order{h}\) for the temperature moment.

\subsection{Density Moment}
\label{appendix_phi}

Let us consider one-by-one the density moments at fixed particle position, \((1/n_i)\int \dv \ \avgr{.}\), of the terms in \eqref{eq_curvy_gk}. The first term is
\begin{align}
	\label{eq_density_term1}
	\frac{1}{n_i}\int \dv \ \avgr{h - \avgR{\phinorm}F_i} = \phinorm + \tau \dw{\phinorm} - \left(1 + \frac{1}{2}\rhoidelperpsq\right)\phinorm  = \tau \dw{\phinorm} - \frac{1}{2}\rhoidelperpsq \phinorm,
\end{align}
where we have used quasineutrality \eqref{eq_curvy_qn} and the lowest-nontrivial-order expressions for the gyroaverages \eqref{eq_avgr_lowestorder} and \eqref{eq_avgR_lowestorder}. 

The next term is
\begin{align}
	&\frac{1}{n_i}\int \dv \ \frac{\rho_i \vti}{2L_T}\left( \frac{v^2}{v^2_{ti}} - \frac{3}{2} \right)F_i \avgr{\partd{\avgR{\phinorm}}{Y}} \nonumber \\
	&\approx\frac{1}{n_i}\int \dv \ \frac{\rho_i \vti}{2L_T}\left( \frac{v^2}{v^2_{ti}} - \frac{3}{2} \right)F_i \left(1 + \frac{1}{2}\frac{v_\perp^2}{\vti^2}\rhoidelperpsq\right)\frac{\partial \phinorm}{\partial y} \nonumber \\
	&=\frac{\rho_i \vti}{2L_T} \frac{1}{2}\rhoidelperpsq\frac{\partial \phinorm}{\partial y}.
\end{align}

To lowest order in \(\kperprhoisq \ll 1\), the magnetic-drift term gives
\begin{align}
	&\frac{1}{n_i}\int \dv \ \frac{\rho_i \vti}{L_B} \left( \frac{v^2_\parallel}{\vti^2} + \frac{v^2_\perp}{2 \vti^2} \right)\avgr{\partd{h(\vect{R})}{Y}}\nonumber \\
	&\approx \frac{1}{n_i}\int \dv \ \frac{\rho_i \vti}{L_B} \left( \frac{v^2_\parallel}{\vti^2} + \frac{v^2_\perp}{2 \vti^2} \right)\partd{h^{(0)}(\vect{r})}{y} \nonumber \\
	&=\frac{1}{n_i}\int \dv \ \frac{\rho_i \vti}{L_B} \frac{2}{3} \frac{v^2}{\vti^2} \partd{h^{(0)}(\vect{r})}{y}
	= \frac{\rho_i \vti}{L_B} \partd{}{y} \left(\phinorm + \frac{\delta T}{T_i}\right),
\end{align}
where we have used 
\begin{align}
	\int dv \ v_\parallel^2 h^{(0)} = \int dv \ \frac{v_\perp^2}{2} h^{(0)} = \int dv \ \frac{v^2}{3} h^{(0)},
\end{align}
which is a consequence of the isotropic form \eqref{eq_h0} of \(h^{(0)}\). 

The density moment of the nonlinear term in \eqref{eq_curvy_gk} is
\begin{alignat}{2}
	\label{eq_density_nl1}
	&\frac{1}{n_i}\int \dv \ \avgr{\pbra{\avgR{\phinorm}(\vect{R})}{h(\vect{R})}} \span\span \nonumber \\
	&\approx\frac{1}{n_i}\int \dv \ &&\avgr{\pbra{\phinorm(\vect{R}) + \frac{1}{4} \frac{v_\perp^2}{\vti^2}\rhoidelperpsq\phinorm(\vect{R})}{h(\vect{R})}} \nonumber \\
	&\approx\frac{1}{n_i}\int \dv \ && \Bigg\langle\pbra{\phinorm(\vect{r}) - \vect{\rho}\cdot\del \phinorm(\vect{r}) + \frac{1}{2}\vect{\rho}\vect{\rho}:\del\del\phinorm(\vect{r})+ \frac{1}{4} \frac{v_\perp^2}{\vti^2}\rhoidelperpsq\phinorm(\vect{r})}{h(\vect{r} - \vect{\rho})}\Bigg\rangle_r  \nonumber \\
	&=\frac{1}{n_i}\int \dv \ &&\Bigg\langle\pbra{\phinorm(\vect{r})}{h(\vect{r} - \vect{\rho})} + \vect{\rho}\vect{\rho}:\pbra{\del\phinorm(\vect{r})}{\del h(\vect{r})} \nonumber \\
	& &&\quad + \frac{1}{2}\vect{\rho}\vect{\rho}:\pbra{\del\del\phinorm(\vect{r})}{h(\vect{r})} + \frac{1}{4} \frac{v_\perp^2}{\vti^2}\pbra{\rhoidelperpsq\phinorm(\vect{r})}{h(\vect{r})} \Bigg\rangle_r \nonumber \\
	&=\frac{1}{n_i}\int \dv \ && \Bigg[\pbra{\phinorm(\vect{r})}{\avgr{h}(\vect{r})} + \avgr{\vect{\rho}\vect{\rho}}:\pbra{\del\phinorm(\vect{r})}{\del h(\vect{r})} \nonumber \\
	& &&\quad + \frac{1}{2}\avgr{\vect{\rho}\vect{\rho}}:\pbra{\del\del\phinorm(\vect{r})}{h(\vect{r})} + \frac{1}{4} \frac{v_\perp^2}{\vti^2}\pbra{\rhoidelperpsq\phinorm(\vect{r})}{h(\vect{r})} \Bigg] \nonumber \\
	&= \frac{1}{n_i}\int \dv \ && \Bigg[\pbra{\phinorm(\vect{r})}{\avgr{h}(\vect{r})} + \frac{1}{2}\frac{v_\perp^2}{\vti^2} \rho_i^2 \mathds{1}_\perp:\pbra{\del\phinorm(\vect{r})}{\del h(\vect{r})} \nonumber \\
	& &&\quad + \frac{1}{2} \frac{v_\perp^2}{\vti^2}\pbra{\rhoidelperpsq\phinorm(\vect{r})}{h(\vect{r})} \Bigg],
\end{alignat}
where we have used \eqref{eq_rho_averages}. Using the lowest-order contribution \eqref{eq_h0} to \(h\) and the fact that \(\pbra{g}{g} = 0\) for any \(g\), we find that \eqref{eq_density_nl1} becomes
\begin{align}
	\label{eq_density_nl2}
	&\pbra{\phinorm}{\tau\dw{\phinorm}} + \frac{1}{2}\rho_i^2\mathds{1}_\perp:\pbra{\del\phinorm}{\del\frac{\delta T}{T_i}} + \frac{1}{2}\rho_i^2 \pbra{\nabla_\perp^2\phinorm}{\phinorm + \frac{\delta T}{T_i}} \nonumber \\
	&=\pbra{\phinorm}{\tau\dw{\phinorm} - \frac{1}{2}\rhoidelperpsq\phinorm} + \frac{1}{2}\rho_i^2\del_\perp \cdot \pbra{\del_\perp\phinorm}{\frac{\delta T}{T_i}}.
\end{align}

Finally, collecting terms, dividing by \(\tau\) and introducing the ion sound radius \mbox{\(\rho_s = \rho_i / \sqrt{2\tau}\)}, we obtain 
\begin{align}
\label{eq_phi_nonnorm}
	&\frac{\partial}{\partial t} \left( \dw{\phinorm} - \rho_s^2 \nabla_\perp^2 \phinorm\right) - \frac{\rho_i \vti}{\tau L_B} \frac{\partial}{\partial y} \left( \phinorm + \frac{\delta T}{T_i} \right) + \frac{\rho_i \vti}{2L_T} \frac{\partial}{\partial y} \left(\rho_s^2\nabla_\perp^2 \phinorm\right)
	\nonumber \\&\quad + \frac{1}{2} \rho_i \vti \left( \pbra{\phinorm}{\dw{\phinorm} - \rho_s^2 \nabla_\perp^2 \phinorm} + \rho_s^2\del_\perp \bcdot \pbra{\del_\perp \phinorm}{\frac{\delta T}{T_i}}\right) = \frac{1}{\tau n_i} \int d^3\vect{v}  \  \avgr{\avgR{C_{l}[h]}}. 
\end{align}
This will become \eqref{phiEq} after we calculate the collisional term in Appendix~\ref{appendix_col_phi}.
\subsection{Temperature Moment}
\label{appendix_temp}

In a similar way, let us consider the temperature moments, \((1/n_i)\int \dv v^2/\vti^2 \avgr{.}\), of the terms in \eqref{eq_curvy_gk} to lowest order in \(\kperprhoisq \ll 1\). The first term is
\begin{align}
\frac{1}{n_i}\int \dv \ \frac{v^2}{\vti^2} \avgr{h - \avgR{\phinorm}F_i} \approx \frac{1}{n_i}\int \dv \ \frac{v^2}{\vti^2} \left(h^{(0)} - \phinorm F_i\right) =\frac{3}{2}\frac{\delta T}{T_i},
\end{align}
where we have used \eqref{eq_h0}.

The temperature-gradient term is
\begin{align}
\label{eq_tempmom_tempgrad}
&\frac{1}{n_i}\int \dv \ \frac{\rho_i \vti}{2L_T}\frac{v^2}{\vti^2}\left( \frac{v^2}{v^2_{ti}} - \frac{3}{2} \right)F_i \avgr{\partd{\avgR{\phinorm}}{Y}}\nonumber \\ 
&\approx \frac{1}{n_i}\int \dv \ \frac{\rho_i \vti}{2L_T}\frac{v^2}{\vti^2}\left( \frac{v^2}{v^2_{ti}} - \frac{3}{2} \right)F_i\frac{\partial \phinorm}{\partial y} 
=\frac{3}{2} \frac{\rho_i \vti}{2L_T} \partd{\phinorm}{y}.
\end{align}

The magnetic-drift term is
\begin{align}
&\frac{1}{n_i}\int \dv \ \frac{\rho_i \vti}{L_B} \frac{v^2}{\vti^2}  \left( \frac{v^2_\parallel}{\vti^2} + \frac{v^2_\perp}{2 \vti^2} \right)\avgr{\partd{h(\vect{R})}{Y}} \nonumber \\
&\approx \frac{1}{n_i}\int \dv \ \frac{\rho_i \vti}{L_B} \frac{v^2}{\vti^2} \left( \frac{v^2_\parallel}{\vti^2} + \frac{v^2_\perp}{2 \vti^2} \right)\partd{h^{(0)}(\vect{r})}{y} \nonumber \\
&=\frac{1}{n_i}\int \dv \ \frac{\rho_i \vti}{L_B} \frac{2}{3} \frac{v^4}{\vti^4}  \partd{h^{(0)}(\vect{r})}{y}
= \frac{5}{2} \frac{\rho_i \vti}{L_B} \partd{}{y} \left(\phinorm + 2\frac{\delta T}{T_i}\right),
\end{align}
where we have used the isotropy of \(h^{(0)}\) again. By the ordering \eqref{eq_ordering}, this term is an order \(L_T / L_B \sim \order{\kperprhoisq} \ll 1\) smaller than the temperature-gradient term \eqref{eq_tempmom_tempgrad}. Hence it will not contribute to the final expression for the temperature moment of \eqref{eq_curvy_gk}.

The nonlinear term is
\begin{align}
	\frac{1}{n_i}\int \dv \ \frac{v^2}{\vti^2} \avgr{\pbra{\avgR{\phinorm}}{h}} \approx \frac{1}{n_i}\int \dv \ \frac{v^2}{\vti^2} \pbra{\phinorm}{h^{(0)}} = \frac{3}{2}\pbra{\phinorm}{\frac{\delta T}{T_i}}.
\end{align}

Collecting terms, we find that the temperature moment of \eqref{eq_curvy_gk} is
\begin{equation}
\label{eq_temp_nonnorm}
\frac{\partial}{\partial t} \frac{\delta T}{T_i} + \frac{\rho_i \vti}{2L_T} \frac{\partial\phinorm}{\partial y} + \frac{1}{2}\rho_i \vti \pbra{\phinorm}{\frac{\delta T}{T_i}} = \frac{2}{3 n_i }\int d^3\vect{v}  \  \frac{v^2}{\vti^2} \avgr{\avgR{C_{l}[h]}}.
\end{equation}
This will become \eqref{psiEq} after we calculate the collisional term in Appendix~\ref{appendix_col_psi}.
\subsection{Moments of the Collision Operator}
\label{appendix_col} 

\subsubsection{Gyroaveraged Collision Operator}

The gyroaveraged collision term in \eqref{gk_eq}, expanded in a Fourier basis, \(h(\vect{R}) = \sum_\vect{k} h_\vect{k} e^{i \vect{k}.\vect{R}}\), is
\begin{align}
\label{eq_col_exp}
\avgR{C_{l}\left[h\right]} &= \sum_{\vect{k}}\avgR{C_{l}\left[ h_\vect{k} e^{i \vect{k}\bcdot\vect{R}}\right]}
= \sum_{\vect{k}}\avgR{C_{l}\left[ h_\vect{k} e^{-i \vect{k}\bcdot\vect{\rho}}\right] e^{i \vect{k}\bcdot\vect{r}} } 
\nonumber\\&=\sum_{\vect{k}} \avgTheta{C_{l}\left[ h_\vect{k} e^{-i \vect{k}\bcdot\vect{\rho}}\right] e^{i \vect{k}\bcdot\vect{\rho}}} e^{i \vect{k}\bcdot\vect{R}},
\end{align}
where all derivatives and integrals with respect to \(\vect{v}\) in the collision operator are taken at fixed \(\vect{r}\), and the \(\avgTheta{.}\) operation is the gyroangle average [see \eqref{avgrdef} and \eqref{avgRdef}]. In order to obtain collisional terms in \eqref{eq_phi_nonnorm} and \eqref{eq_temp_nonnorm}, we expand the exponential factors \(e^{\pm i \vect{k}\bcdot\vect{\rho}}\) in the small quantity \(\vect{k}\bcdot\vect{\rho}\). Recall that \eqref{eq_phi_nonnorm} and \eqref{eq_temp_nonnorm} are contained at different orders in the GK equation. We need \eqref{eq_col_exp} to order \(\mathcal{O}(k_\perp^2\rho_i^2 \nu_ih)\) for the temperature moment and to order \(\mathcal{O}(k_\perp^4\rho_i^4 \nu_ih)\) for the density moment. We first consider the collisional term in the temperature equation \eqref{eq_temp_nonnorm}, which represents thermal diffusion, before turning to the more involved calculation for the viscosity in \eqref{eq_phi_nonnorm}.

\subsubsection{Collisional Thermal Diffusion}
\label{appendix_col_psi}
The collisional term in the temperature equation \eqref{eq_temp_nonnorm} is
\begin{align}
\label{eq_tempcol_1}
&\frac{2}{3n_i}\avgr{\sum_{\vect{k}} e^{i \vect{k}\bcdot\vect{R}} \int \dv \ \frac{v^2}{\vti^2} \avgTheta{ C_{l}\left[ h_\vect{k} e^{-i \vect{k}\bcdot\vect{\rho}}\right] e^{i \vect{k}\bcdot\vect{\rho}}}} \nonumber \\
&\approx \frac{2}{3n_i} \sum_{\vect{k}} e^{i \vect{k}\bcdot\vect{r}} \int \dv \ \frac{v^2}{\vti^2} \avgTheta{ C_{l}\left[ h_\vect{k} e^{-i \vect{k}\bcdot\vect{\rho}}\right] e^{i \vect{k}\bcdot\vect{\rho}}},
\end{align}
where, to order \(\mathcal{O}(k_\perp^2\rho_i^2\nu_i h)\), we can expand
\begin{align}
\label{eq_colop_firstorder}
&\avgTheta{C_{l}\left[ h_\vect{k} e^{-i \vect{k}\bcdot\vect{\rho}}\right] e^{i \vect{k}\bcdot\vect{\rho}}} = C_{l}\left[h^{(1)}_\vk\right] +\avgTheta{i \krhonb C_{l}\left[\left(-i  \vect{k}\bcdot\vect{\rho}\right) \hok\right]}
\nonumber\\& \ - \frac{1}{2} \avgTheta{\left(\vect{k}\bcdot\vect{\rho}\right)^2} C_{l}\left[h^{(0)}_\vk\right] 
 +\avgTheta{C_{l}\left[ - \frac{1}{2} \left(\vect{k}\bcdot\vect{\rho}\right)^2\hok\right]} + \mathcal{O}( \kperprhoisqq\nu_i h).
\end{align}
Taking a temperature moment of \eqref{eq_colop_firstorder} annihilates the first and fourth terms due to the conservation-of-energy property of the collision operator (\(\intv v^2 \ C_{l}\left[f\right] = 0\) for any \(f\)) and the third term vanishes because \(C_{l}\left[h^{(0)}\right] = 0\). Just as in \cite{newton2010}, after performing the integration in \eqref{eq_tempcol_1}, we find that the second term of \eqref{eq_colop_firstorder} gives
\begin{equation}
\intv \frac{v^2}{\vti^2} \left(i \vect{k}\bcdot\vect{\rho}\right) C_{l}[(-i \vect{k}\bcdot\vect{\rho})\hok] = -\frac{3}{2} n_i \chi k_\perp^2 \frac{\delta T_\vect{k}}{T_i},
\end{equation}
where
\begin{equation}
\chi \equiv \frac{8}{9} \sqrt{\frac{2}{\pi}} \nu_{i}\rho_i^2
\end{equation}
and the ion-ion collision frequency \(\nu_i\) is defined in \eqref{eq_nu_i_def}. Thus, \eqref{eq_tempcol_1} is
\begin{align}
	\label{eq_tempmom_final}
	-\chi \sum_{\vect{k}} e^{i \vect{k}\bcdot\vect{r}}k_\perp^2 \frac{\delta T_\vect{k}}{T_i} 
	= \chi \nabla_\perp^2 \frac{\delta T(\vect{r})}{T_i}.
\end{align}
The right-hand side of \eqref{eq_temp_nonnorm} is \eqref{eq_tempmom_final}, hence we arrive at the temperature equation \eqref{psiEq}. 

\subsubsection{Collisional Viscous Damping}
\label{appendix_col_phi}
As we discussed before, under the ordering \eqref{eq_ordering}, the density moment of \eqref{eq_curvy_gk} is obtained at an order higher in \(\kperprhoisq \ll 1\) than the temperature moment. As a sanity check, note that taking a density moment (\(\intv \avgr{.} \)) of \eqref{eq_colop_firstorder} annihilates all terms: terms one and four vanish because the collision operator conserves particle number (\(\intv C_{l}\left[f\right] = 0\) for any \(f\)), term two vanishes because of conservation of momentum (\(\intv \vect{v} C_{l}\left[f\right] = 0\) for any \(f\)), and, as before, term three is identically zero due to the form of \(h^{(0)}\). We thus need to expand \eqref{eq_col_exp} to order \(\order{\kperprhoisqq \nu_ih}\).

The collisional term of the density moment in \eqref{eq_phi_nonnorm} is
\begin{align}
\label{eq_colvisc1}
&\sum_\vect{k}\int d^3\vect{v} \  \avgr{\avgTheta{C_{l}\left[ h_\vect{k} e^{-i \vect{k}\bcdot\vect{\rho}}\right] e^{i \vect{k}\bcdot\vect{\rho}}} e^{i \vect{k}\bcdot\vect{R}}} = \sum_\vect{k}\int d^3\vect{v} \  J_0\left(\frac{k_\perp v_\perp}{\Omega}\right) \avgTheta{C_{l}\left[ h_\vect{k} e^{-i \vect{k}\bcdot\vect{\rho}}\right] e^{i \vect{k}\bcdot\vect{\rho}}}  e^{i \vect{k}\bcdot\vect{r}} \nonumber 
\\& = \sum_\vect{k}\int d^3\vect{v} \  J_0\left(\frac{k_\perp v_\perp}{\Omega}\right)C_{l}\left[ h_\vect{k} e^{-i \vect{k}\bcdot\vect{\rho}}\right] e^{i \vect{k}\bcdot\vect{\rho}}  e^{i \vect{k}\bcdot\vect{r}} \equiv \sum_\vect{k} C_{D\vect{k}}e^{i \vect{k}\bcdot\vect{r}}.
\end{align}
Note that any term with an odd power of $\vect{k}\cdot\vect{\rho}$ is annihilated by gyroaveraging. Expanding \eqref{eq_colvisc1} to order \(\mathcal{O}(\kperprhoisqq\nu_ih)\), we find
\begin{align}
\label{eq_cdk}
C_{D\vect{k}}&= \int d^3\vect{v}  \ 
\Bigg\lbrace -\frac{1}{2}\krho^2 C_l\left[ -\frac{1}{2}  \krho^2\hok\right] 
\nonumber\\
&\quad\quad -\frac{1}{2}\krho^2 C_l\left[h_\vect{k}^{(1)}\right]-\frac{1}{6}i\krho^3 C_l\left[(-i\vect{k}\bcdot\vect{\rho})\hok\right]\nonumber\\
&\quad\quad -\frac{k_\perp^2v_\perp^2}{4\Omega^2}\left(\krhonb C_l\left[ \krhonb\hok\right] -\frac{1}{2} C_l\left[ \krho^2\hok\right] + C_l\left[h_\vect{k}^{(1)}\right]\right)\Bigg\rbrace.
\end{align}
Since the linearised collision operator is isotropic in velocity space and the gyrokinetic distribution \(h\) is gyroangle independent, it follows that \(C[h^{(1)}]\) is also gyroangle independent. Therefore, we can write 
\begin{equation}
\label{eq_h1_sub}
\int d^3\vect{v}  \ \frac{1}{2} \krho^2 C_l\left[h_\vect{k}^{(1)}\right] = \int d^3\vect{v}  \ \frac{1}{2}  \avgTheta{\krho^2} C_l\left[h_\vect{k}^{(1)}\right]
= \int d^3\vect{v}  \ \frac{k_\perp^2v_\perp^2}{4\Omega^2} C_l\left[h_\vect{k}^{(1)}\right].
\end{equation}
Substituting \eqref{eq_h1_sub} into \eqref{eq_cdk}, we obtain
\begin{align}
\label{denscol1}
C_{D\vect{k}} = \int d^3\vect{v}  \ 
\Bigg\lbrace &\frac{1}{4}\krho^2 C_l\left[  \krho^2\hok\right] - \frac{k_\perp^2v_\perp^2}{2\Omega^2} C_l\left[h_\vect{k}^{(1)}\right]
-\frac{1}{6}\krho^3 C_l\left[\krhonb\hok\right]
\nonumber\\&-\frac{k_\perp^2v_\perp^2}{4\Omega^2} \krhonb C_l\left[\krhonb\hok \right] + \frac{k_\perp^2v_\perp^2}{8\Omega^2} C_l\left[ \krho^2\hok\right] \Bigg\rbrace.
\end{align}

The high-collisionality limit allowed us to obtain the form of \(h^{(0)}\), but \(h^{(1)}\) is unknown without inverting the collision operator. We can take advantage of the form of the GK equation \eqref{eq_curvy_gk} to lowest non-trivial order and express directly the required moment of \(C[h^{(1)}]\) using moments of \(C[h^{(0)}]\). Consider the \(\int d^3\vect{v} (v_\perp^2 - 2v^2/3)\) moment of the GK equation up to order \(\mathcal{O}(\kperprhoisq \nu_i  h)\). In Appendix~\ref{appendix_temp}, where we took a \(v^2\) moment, we found that all non-collisional terms in the temperature moment of \eqref{eq_curvy_gk} are isotropic in velocity space to order \(\order{\kperprhoisq \nu_i  h}\), hence they are annihilated by the operator \(\int d^3\vect{v} \left(v_\perp^2 - 2v^2/3\right)\). Using the expansion \eqref{eq_colop_firstorder} of the collision operator, we find
\begin{alignat}{2}
\label{eq_tempcol_nonisoterms}
0&=\int d^3\vect{v}  \ \bigg(&&v_\perp^2 - \frac{2}{3}v^2\bigg) \Bigg\lbrace \krhonb C_l\left[ \krhonb\hok\right] -\frac{1}{2} C_l\left[ \krho^2\hok\right] + C_l\left[h_\vect{k}^{(1)}\right] \Bigg\rbrace
\nonumber\\&= \int d^3\vect{v} \ \Bigg\lbrace &&v_\perp^2 \krhonb C_l\left[ \krhonb\hok\right] -\frac{1}{2} v_\perp^2 C_l\left[ \krho^2\hok\right] 
\nonumber\\& &&+ v_\perp^2 C_l\left[h_\vect{k}^{(1)}\right]
- \frac{2}{3}v^2 \krhonb C_l\left[ \krhonb\hok\right] \Bigg\rbrace.
\end{alignat}
Here a few terms have dropped out due to the energy-conservation properties of the collision operator. Extracting from \eqref{eq_tempcol_nonisoterms} an expression for \(v_\perp^2 C_l\left[h_\vect{k}^{(1)}\right]\) and substituting this expression into \eqref{denscol1}, we obtain an expression for \(C_{D\vect{k}}\) involving only \(h^{(0)}\):
\begin{align}
C_{D\vect{k}} = \int d^3\vect{v} \ 
\Bigg\lbrace &\frac{1}{4}\krho^2 C_l\left[  \krho^2\hok\right] -\frac{1}{6}\krho^3 C_l\left[\krhonb\hok\right]
\nonumber\\&+\frac{k_\perp^2v_\perp^2}{4\Omega^2} \krhonb C_l\left[ \krhonb\hok\right] -\frac{k_\perp^2v^2}{3\Omega^2} \krhonb C_l\left[ \krhonb\hok\right] 
\nonumber\\&- \frac{k_\perp^2v_\perp^2}{8\Omega^2} C_l\left[ \krho^2\hok\right] \Bigg\rbrace.
\end{align}
We proceed to evaluate these integrals:

\begin{alignat}{2}
&\int d^3\vect{v}  \ \frac{1}{4}\krho^2 C_l\left[  \krho^2\hok\right] &&= -n_i \chi \rho_i^2 k_\perp^4 \left(\frac{3}{20}\phinorm_\vk + \frac{3}{16} \frac{\delta T_\vk}{T_i} \right),
\\& \int d^3\vect{v}  \ \frac{1}{6}\krho^3 C_l\left[\krhonb\hok\right] &&= -n_i \chi \rho_i^2k_\perp^4 \frac{3}{20} \frac{\delta T_\vk}{T_i},
\\& \int d^3\vect{v}  \ \frac{k_\perp^2v_\perp^2}{4\Omega^2} \krho C_l\left[ \krhonb\hok\right] &&= -n_i \chi \rho_i^2 k_\perp^4 \frac{3}{10} \frac{\delta T_\vk}{T_i},
\\& \int d^3\vect{v}  \ \frac{k_\perp^2v^2}{3\Omega^2} \krho C_l\left[ \krhonb\hok\right] &&= -n_i \chi \rho_i^2 k_\perp^4 \frac{1}{2} \frac{\delta T_\vk}{T_i},
\\& \int d^3\vect{v}  \ \frac{k_\perp^2v_\perp^2}{8\Omega^2} C_l\left[ \krho^2\hok\right] &&= -n_i \chi \rho_i^2 k_\perp^4 \left( \frac{3}{80} \phinorm_\vk + \frac{3}{64} \frac{\delta T_\vk}{T_i} \right).
\end{alignat}
Combining all of these, we get the following expression for the collision term on the right-hand side of \eqref{eq_phi_nonnorm}:
\begin{equation}
\label{eq_densmom_final}
\sum_\vect{k} C_{D\vect{k}}e^{i \vect{k}\bcdot\vect{r}} =  -\sum_\vect{k} n_i \chi \frac{\rho_i^2}{2} k_\perp^4 \left( \frac{9}{40} \phinorm_\vk - \frac{67}{160} \frac{\delta T_\vk}{T_i} \right) e^{i \vect{k}\bcdot\vect{r}}
= - \frac{1}{2} n_i \chi \rho_i^2 \nabla_\perp^4 \left( a \phinorm - b \frac{\delta T}{T_i} \right),
\end{equation}
where \(a = 9/40\) and \(b = 67/160\). The right-hand side of \eqref{eq_phi_nonnorm} is \eqref{eq_densmom_final}\(/\tau n_i\), hence follows the ion-density equation \eqref{phiEq}. 

\section{Conservation Laws}
\label{appendix_cons}
All calculations in this section are done to order \(\order{\kperprhoisqq}\). 

The nonlinearly conserved free energy in electrostatic GK is given by
\begin{equation}
	\label{eq_cons_app_w}
	W =\sum_s \intrv \frac{T_s \delta f_s^2}{2 F_s},
\end{equation}
where \(s\) labels the particle species and \(F_s\) is the corresponding Maxwellian equilibrium distribution. 

Using the modified adiabatic electron response \eqref{eq_fe}, the electron contribution to \eqref{eq_cons_app_w} is
\begin{equation}
	\label{eq_consapp_e}
	\intrv \frac{T_e \delta f_e^2}{2 F_e} = \frac{1}{2} T_i n_i \intr \tau \dw{\phinorm}^2.
\end{equation}

The ion contribution to \(W\) requires some work. Using \eqref{eq_f_def}, viz., \(\delta f_i = h - \phinorm F_i\), we obtain
\begin{equation}
	\label{eq_consapp_i}
	\intrv \frac{T_i \delta f_i^2}{2F_i} = \intrv \frac{1}{2} T_i \left(\frac{\avgr{h^2}}{F_i} + \phinorm^2 F_i - 2 \phinorm \avgr{h}\right).
\end{equation}
Using the quasineutrality condition \eqref{eq_curvy_qn} and the gyroaverage expansion \eqref{eq_avgr_lowestorder}, we find
\begin{align}
	&\intrv \frac{T_i\avgr{h^2}}{2F_i} = \frac{1}{2} T_i n_i \intr \left[\frac{3}{2} \frac{\delta T}{T_i} + \phinorm^2 + 2\tau\dw{\phinorm}^2 - \frac{1}{2} \phinorm \rhoidelperpsq\left(\phinorm + \frac{\delta T}{T_i}\right)\right], \\
	&\intrv \phinorm \avgr{h} = \phinorm^2 + \tau \dw{\phinorm}^2.
\end{align}
Finally, substituting these expressions into \eqref{eq_consapp_i} and combining with \eqref{eq_consapp_e}, we get
\begin{equation}
	\label{eq_free_e_appendix_full}
	W = \frac{1}{2} T_i n_i \intr \left[\frac{3}{2} \left(\frac{\delta T}{T_i}\right)^2 + \tau \dw{\phinorm}^2 - \frac{1}{2}\phinorm\rhoidelperpsq\left(\phinorm + \frac{\delta T}{T_i}\right)\right].
\end{equation}
To lowest order in \(\kperprhoisq \), the free energy is, therefore, 
\begin{equation}
	\label{eq_free_e_appendix}
	W = \frac{3}{2} T_i n_i \intr \frac{1}{2} \left(\frac{\delta T}{T_i}\right)^2,
\end{equation}
as promised in Section \ref{sect_cons}. Using \eqref{curvy_phi} and \eqref{curvy_psi}, it is straightforward to show that, up to multiplicative constants related to the normalisations \eqref{eq_normalisations}, \eqref{eq_free_e_appendix} satisfies \eqref{eq_free_e_cons}. 

The 2D (\(k_\parallel = 0\)) GK equation for species \(s\) has an additional conserved quantity \citep{schekochihingk2009}, given by
\begin{equation}
	I_s = \frac{T_s}{2F_s} \intR \avgR{\delta f_s}^2 = \frac{T_s}{2F_s} \intR \left(h_s - \frac{Z_s e}{T_s} \avgR{\phinonnorm} F_s\right)^2.
\end{equation}
In the model that we consider in this paper, only the ions are assumed to have 2D dynamics --- indeed, the modified adiabatic electron response \eqref{eq_eresponse} arises as a consequence precisely of the fast parallel streaming of the electrons. Thus, for the ions, 
\begin{align}
	\label{eq_i}
	I &\equiv I_i = \frac{T_i}{2 F_i} \intR \left(h - \avgR{\phinorm} F_i\right)^2  \nonumber \\
	 &= \frac{1}{2} T_i \intR \Bigg\lbrace \left(\frac{\delta T}{T_i}\right)^2 \tempvel^2 F_i + \frac{\delta T}{T_i} \tempvel h^{(1)} \nonumber  \\ 
	 & \quad + 2 \frac{\delta T}{T_i} \tempvel \left[\tau \dw{\phinorm} - \frac{1}{4} \rhoidelperpsq \left(\phinorm + \frac{\delta T}{T_i}\right) - \frac{1}{4} \frac{v_\perp^2}{\vti^2} \rhoidelperpsq \phinorm\right]F_i \Bigg\rbrace.
\end{align}
Note that \(I\) is a function of velocity \(\vect{v}\). In order to eliminate the unknown \(h^{(1)}\), we can integrate \(I\):
\begin{equation}
	\intv I = \frac{1}{2} T_i n_i \intr \left[\frac{3}{2} \left(\frac{\delta T}{T_i}\right)^2 - \frac{1}{2} \frac{\delta T}{T_i} \rhoidelperpsq \phinorm \right].
\end{equation}
Subtracting this from the free energy \eqref{eq_free_e_appendix_full}, we obtain
\begin{equation}
	\label{eq_phicons_appendix}
	W - \intv I = \frac{1}{2} T_i n_i \intr \left(\tau \dw{\phinorm}^2 - \frac{1}{2} \phinorm \rhoidelperpsq\phinorm\right) = \frac{1}{2} T_i n_i \tau \intr  \left[\dw{\phinorm}^2 + \rho_s^2 \left(\del_\perp\phinorm\right)^2\right],
\end{equation}
which is the conserved quantity in \eqref{eq_phi_cons}. This can be viewed as a version of the electrostatic GK invariant \citep{schekochihingk2009}
\begin{equation}
	Y \equiv W - \sum_s \intv I_s,
\end{equation}
but without the electron contribution \(I_e\), which is not conserved because the electrons do not obey \(k_\parallel = 0\). 

In order to obtain the third conserved quantity, we go back to \eqref{eq_i} and consider
\begin{equation}
	\intv \left(\frac{v^2}{\vti^2} - \frac{3}{2}\right)^{-1}I =  T_i n_i \tau \intr \left(\dw{\phinorm}\frac{\delta T}{T_i} - \frac{1}{2} \frac{\delta T}{T_i} \rho_s^2\nabla_\perp^2 \frac{\delta T}{T_i} - \frac{\delta T}{T_i} \rho_s^2\nabla_\perp^2 \phinorm \right).
\end{equation}
Adding this to \eqref{eq_phicons_appendix}, we obtain
\begin{equation}
	W - \intv \left[1 - \left(\frac{v^2}{\vti^2} - \frac{3}{2}\right)^{-1}\right]I = T_i n_i \tau \intr \left[ \frac{1}{2} \dw{\phinorm}^2 + \frac{\delta T}{T_i} \phinorm' + \frac{1}{2} \rho_s^2 \left(\del \phinorm + \del \frac{\delta T}{T_i}\right)^2 \right].
\end{equation}
This is the quantity that satisfies \eqref{eq_phi_weird_cons}. 

\section{Tertiary Instability}
\label{appendix_tert}

Using the decomposition \eqref{eq_tert_ordering} and dropping the nonzonal-nonzonal interaction terms in \eqref{curvy_phi} and \eqref{curvy_psi}, we obtain the linearised tertiary-mode equations:
\begin{align}
\label{eq_curvy_phi_zf}
&(\partial_t + \underbrace{\px \zf{\phinorm} \py}_\text{\circled{1}} ) \left( 1 - \delsq\right) \dw{\phinorm} - (1 \underbrace{-\px^3 \zf{\phinorm}}_\text{\circled{2}}) \partial_y \left( \dw{\phinorm} + \dw{\deltaT} \right) + (\vt \underbrace{-\px \zf{\deltaT}}_\text{\circled{3}}) \partial_y  \delsq \dw{\phinorm}  \nonumber \\ 
&\qquad + \underbrace{\left(\px^2\zf{\phinorm}\right)\px\py\dw{\deltaT} - \left(\px^2\zf{\deltaT}\right)\px\py\dw{\phinorm}}_\text{\circled{4}}=-\chi \nabla^4 (a\dw{\phinorm} - b\dw{\deltaT}), \\
\label{eq_curvy_psi_zf}
&(\partial_t + \underbrace{\px \zf{\phinorm} \py}_\text{\circled{5}} )\dw{\deltaT} +(\vt \underbrace{-\px \zf{\deltaT}}_\text{\circled{6}})\partial_y \dw{\phinorm} = \chi \nabla^2 \dw{\deltaT}.
\end{align}
Let us examine \eqref{eq_curvy_phi_zf} and \eqref{eq_curvy_psi_zf} to gain some insight into the way in which zonal fields might affect the ITG instability. Terms "1" and "5" represent the advection of density and temperature perturbations by the ZF. In the locations of nonzero ZF shear (\(\px^2 \zf{\phinorm} \neq 0\)), the zonal advection is responsible for shearing the turbulent eddies and thus suppressing turbulence. We do indeed find that the growing tertiary modes are localised where the zonal shear vanishes, \(\px^2 \zf{\phinorm} = 0\). Terms "3" and "6" reflect the modification of the background temperature gradient by the zonal temperature gradient \(\px \zf{\deltaT}\). Their presence suggests that a possible mechanism for ITG saturation (or mitigation) is to excite a zonal temperature gradient that cancels the background gradient, thus effectively eliminating the turbulent drive (see Appendix~\ref{sect_zonaltempsat}). Term "2" shows that the derivative of the zonal shear, \(\px^3 \zf{\phinorm}\), modifies the effective background magnetic field gradient. Terms "4" do not have an obvious simple interpretation that we know of.

Even though we can always embark on a 4-mode calculation similar to the one done in Section~\ref{sect_sec}, it is, in fact, not useful in this case. In contrast to the linear regime, where there is always a well-defined fastest-growing primary mode, neither a monochromatic zonal profile nor a monochromatic tertiary mode are ever observed in simulations. Instead, we find ITG modes that are localised around the points of vanishing zonal shear. 

\subsection{ITG Tertiary Instability}
\label{appendix_itgtert}

Let us derive a heuristic approximation for the growth rate of the ITG instability localised around a local extremum of the ZF and of the zonal temperature gradient, i.e., around \(x_0\) such that \(\px^2\zf{\phinorm}(x_0) = \px^2\zf{\deltaT}(x_0) = 0\). Note that terms "4" in \eqref{eq_curvy_phi_zf} then vanish at \(x_0\). Assuming that the tertiary modes are sufficiently localised, we can take \(\px^3\zf{\phinorm}\) and \(\px\zf{\deltaT}\) to be constant \citep{rogers2005}. We can then repeat the linear calculation of Section~\ref{sect_linear}, but now including the zonally modified gradients of the equilibrium magnetic field and background temperature, and obtain a dispersion relation for the tertiary modes. For simplicity, let us consider the collisionless (\(\chi = 0\)) modes. The resulting growth rate for a mode with poloidal wavenumber \(k_y\) is
\begin{equation}
\label{eq_itgtert}
\gamma_3 = \frac{k_y\sqrt{4(1+k^2)(1 - \px^3 \zf{\phinorm})(\vt - \px \zf{\deltaT}) - \left[1 - \px^3 \zf{\phinorm} + (\vt - \px \zf{\deltaT})k^2\right]^2}}{2(1+k^2)},
\end{equation}
which is a modified version of \eqref{eq_gammak_nodamping}.

Under the same assumptions that we used before to obtain the simplified collisionless ITG growth rate \eqref{eq_itg_verysimple}, viz., \(\vt - \px \zf{\deltaT} \gg 1\) and \(k \ll (\vt - \px \zf{\deltaT})^{-1/4}\), we find that the tertiary growth rate is
\begin{equation}
\label{eq_rogers_tert}
\gamma_3 \approx k_y \sqrt{(1-\px^3\zf{\phinorm})(\vt - \px \zf{\deltaT})}.
\end{equation}
If there are no background gradients, then \(\gamma_3 \approx k_y \sqrt{\px^3\zf{\phinorm}(x_0)\px \zf{\deltaT} (x_0)}\), as found by \citet{rogersdorland2000}. To work out how wide this mode can be, note that we have ignored the effects of the nonzero zonal shear away from the point \(x_0\). Taylor-expanding terms "1" and "5" in \eqref{eq_curvy_phi_zf} and \eqref{eq_curvy_psi_zf} around \(x_0\), we have 
\begin{equation}
i k_y\px \zf{\phinorm}(x) = i k_y\px \zf{\phinorm}(x_0) + \frac{1}{2} i k_y\px^3 \zf{\phinorm}(x_0) (x-x_0)^2 + \order{(x-x_0)^3},
\end{equation}
where the first term represents a Doppler shift in the frequency and the second captures the effect of the nonzero zonal shear. Dropping the latter is, therefore, valid only in an interval around \(x_0\) such that
\begin{align}
&\gamma_3 \gg \frac{1}{2} k_y \px^3 \zf{\phinorm}(x_0) (x-x_0)^2\nonumber \\
\label{eq_deltatert}
&\implies (x-x_0)^2 \ll \Delta^2 \equiv \frac{2 \gamma_3}{k_y\px^3 \zf{\phinorm}(x_0)} \approx 2\sqrt{\frac{\left[1-\px^3\zf{\phinorm}(x_0)\right]\left[\vt - \px \zf{\deltaT} (x_0)\right]}{\left[\px^3 \zf{\phinorm}(x_0)\right]^2}},
\end{align}
where we used \eqref{eq_rogers_tert} for the final approximation. \citet{rogersdorland2000} and \citet{rogers2005} found that the scale \(\Delta\) is a good approximation for the radial width of the ITG tertiary mode in the case of no equilibrium gradients. 

We find that the standard simplified picture of the ITG tertiary instability outlined above does not quite describe the observed tertiary modes. Namely, (i) we find a strong temperature-gradient-driven instability at ZF minima where \(\px^3 \zf{\phinorm} > 1\) and \eqref{eq_itgtert} predicts no instability; and (ii) the instability at the ZF maxima is significantly (an order-of-magnitude) slower than predicted by \eqref{eq_itgtert}. This is detailed in Appendix~\ref{appendix_convection_cells}.

\subsection{KH Tertiary Instability}
\label{appendix_khtertiary}

Recall that, as we showed in Section~\ref{sect_hasegawamima}, \eqref{curvy_phi} reduces to the Hasegawa-Mima equation in the case of \(\vt = 0\). This reduction is naturally achieved in the vicinity of a point \(x_0\) where \(\px^2 \zf{\phinorm}(x_0) = \px^2 \zf{\deltaT}(x_0) = 0\) and \(\px \zf{\deltaT}(x_0) = \vt\). Then the temperature equation \eqref{eq_curvy_psi_zf} decouples from \eqref{eq_curvy_phi_zf} due to the cancellation (or "flattening") of the equilibrium temperature gradient by the zonal temperature. In that case, \(\dw{\deltaT} = 0\) is a solution to \eqref{eq_curvy_psi_zf} and \eqref{eq_curvy_phi_zf} reduces to
\begin{equation}
\label{eq_khtertiary}
(\partial_t + \zf{u}_y \py ) \left( 1 - \delsq\right) \dw{\phinorm} - (1 -\px^2 \zf{u}_y) \partial_y \dw{\phinorm} =-a\chi \nabla^4 \dw{\phinorm},
\end{equation}
which is the linearised Hasegawa-Mima equation for tertiary modes. This equation has a Kelvin-Helmholtz-like (KH) tertiary instability \citep{kim2002, numata2007, stonge2017, zhu2018_tertiary}. This KH instability is also localised around the radial locations of zero zonal shear and has a threshold roughly given by the necessary (but, in general, not sufficient) condition that \(\px^2 \zf{u}_y - 1\) must change sign in the region of instability, known as the Rayleigh-Kuo criterion \citep{kuo49, zhu2018_tertiary}. 

\subsection{Tertiary Instabilities of the Zonal Staircase}
\label{appendix_convection_cells}

We now turn to a numerical investigation of the tertiary instability in the low-collisionality regime. We will consider the parameters and staircase profile of SimH (\(\vt = 0.36, \chi = 0.1\)). The ITG turbulence trapped in the convection zones at the ZF extrema can differ substantially between ZF maxima and minima. For example, it is evident from Figures~\ref{fig_staircase_profiles} and~\ref{fig_snapshot_t_dimits} that the ZF minima harbour turbulence with a larger poloidal wavenumber compared to the turbulence at the ZF maxima. 

Let us describe this difference quantitatively for SimL. Consider the poloidal Fourier transforms of our fields at a fixed radial location:
\begin{equation}
\label{eq_poloidal_fourier}
\phinorm(x, y) = \sum_{k_y} \hat{\phinorm}(x, k_y) e^{i k_y y}, \qquad \deltaT(x, y) = \sum_{k_y} \hat{\deltaT}(x, k_y) e^{i k_y y}.
\end{equation}
Figure~\ref{fig_convection_zones_spectra} shows that the perturbations located at the ZF minima have a \(k_y\) spectrum peaked at a finite wavenumber, whereas the perturbations around the ZF maxima saturate at the largest available poloidal scale in the box.

An asymmetry between the ZF extrema was noticed already by \citet{mcmillan2011}, who, in their GK simulations, found that the ZF minima are less effective than the ZF maxima at stabilising turbulence. This difference has also been extensively studied using the Hasegawa-Mima equation \citep{zhu2018}. We find that these differences are a consequence of the mechanisms the drive the tertiary modes. By inspecting equations \eqref{eq_curvy_phi_zf} and \eqref{eq_curvy_psi_zf}, one can instantly identify the culprit of the asymmetry: the sign of \(\px^3 \zf{\phinorm}\) is different in the maxima (negative) and minima (positive) of the ZF. According to our analysis of the tertiary instabilities in Appendices~\ref{appendix_itgtert} and \ref{appendix_khtertiary}, this has two distinct effects on the tertiary instability at the ZF extrema: one is related to the modification of the magnetic drift term and the second one to the existence of the KH tertiary at the ZF minima.

In order to investigate the tertiary instabilities that operate in the saturated state that we observe, we solve \eqref{curvy_phi} and \eqref{curvy_psi} numerically with an imposed static ZF profile extracted from the quasi-static zonal staircase. We do not impose a static zonal temperature profile because the observed flattening of the equilibrium temperature gradient suggests that the zonal temperature is an agent of saturation, rather than an instability mechanism.

\subsubsection{Tertiary Instability at ZF Minima}
\label{appendix_zfminima}

\begin{figure}
	\centering
	\begin{center}
		\begin{tabular}{ cc } 
			\includegraphics[scale=0.27]{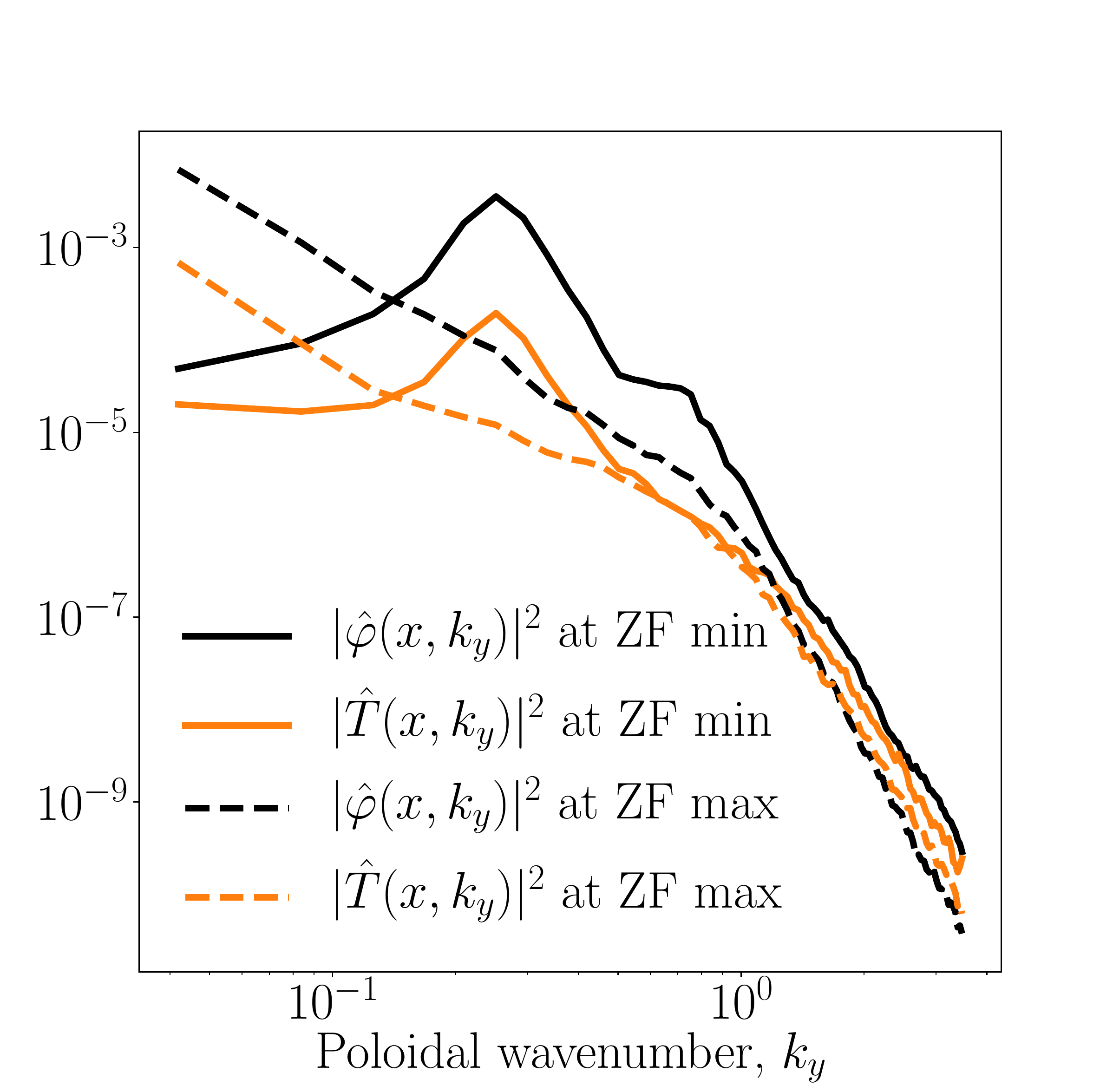} & \includegraphics[scale=0.27]{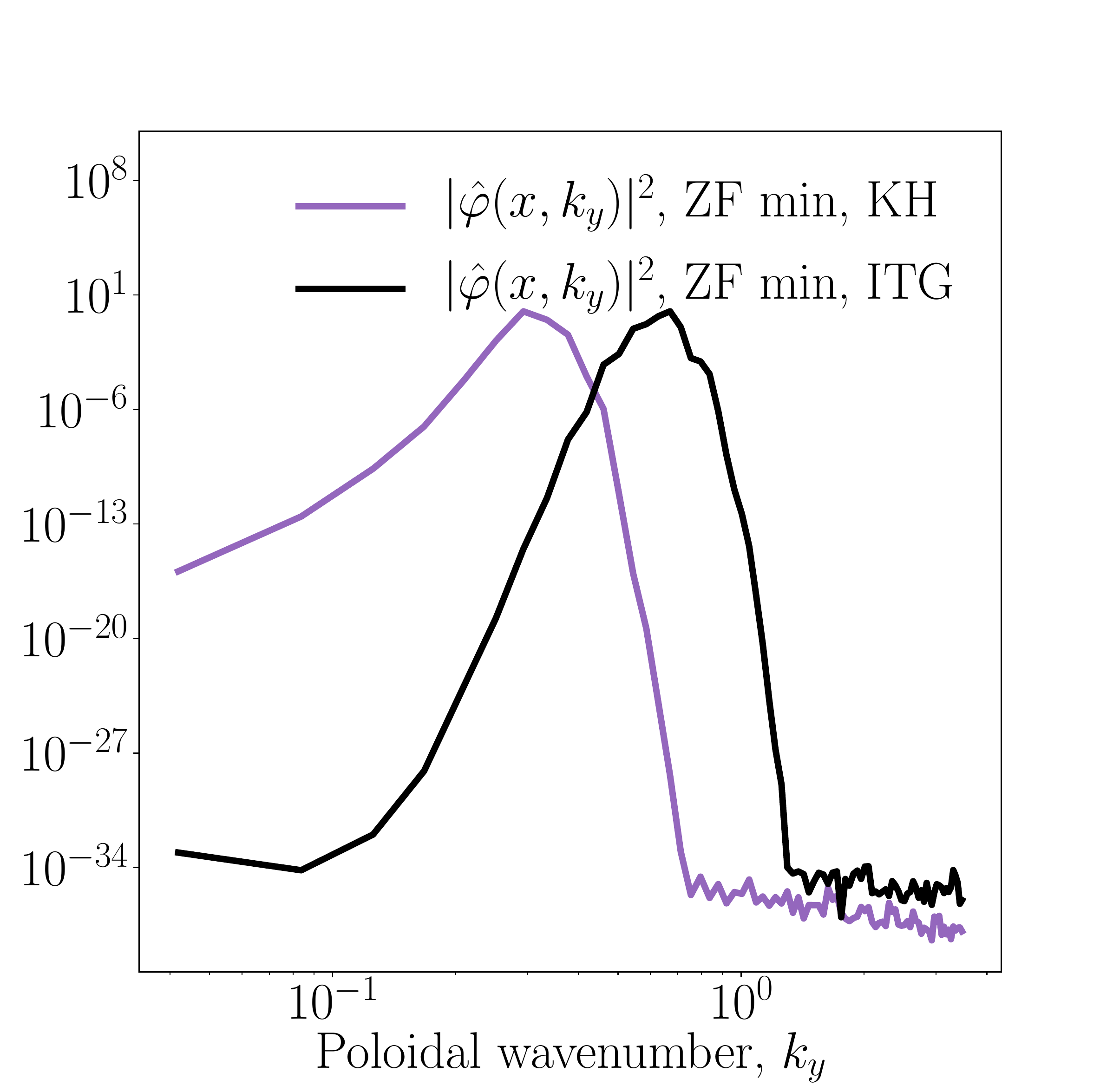}
		\end{tabular}
	\end{center}
	\caption{\textbf{Left:} Time-averaged poloidal spectra \(\langle|\hat{\phinorm}(x, k_y)|^2\rangle\) (black) and \(\langle|\hat{\deltaT}(x, k_y)|^2\rangle\) (orange), as defined by \eqref{eq_poloidal_fourier}, at a fixed radial location \(x\) for a ZF minimum (solid) and a ZF maximum (dashed). The time average is performed over the entire time period shown in Figure~\ref{fig_zf_r_t} (top panel). A peak in the spectrum is observed around \(k_y \approx 0.25\) for the turbulence at the ZF minima. This agrees with the wavelength evident in Figure~\ref{fig_snapshot_t_dimits}. In contrast, the ITG turbulence at the ZF maxima saturates at the largest available poloidal length in the box. \textbf{Right:} Poloidal spectra \(\langle|\hat{\phinorm}(x, k_y)|^2\rangle\) of the unstable modes at a ZF minimum for the KH instability that develops in the Hasegawa-Mima equation (purple) and the fast ITG instability without zonal temperature perturbations (black). The spectra have been normalised to the maximal value of \(1\). We see that both modes are approximately monochromatic, peaked around \(k_y \approx 0.29\) for the KH mode and \(k_y \approx 0.66\) for the ITG mode. The spectrum of saturated tubulence at the ZF minima peaks around \(k_y \approx 0.25\) (left panel). This supports the case for the poloidal wavenumber at the ZF minima to be determined by the KH instability there.}
	\label{fig_convection_zones_spectra}
	\label{fig_convection_zones_spectra_comparison}
\end{figure}

Figure~\ref{fig_zfextrema_phiandtgrad} shows the behaviour of the tertiary instability at the ZF maxima and minima. Let us first discuss the ZF minima. There we observe a very fast ITG-like initial instability that features both \(\dw{\phinorm}\) and \(\dw{\deltaT}\) growing at the rate \(\gamma_3^\text{fast} \approx 0.1\). This is of the same order as the largest linear ITG growth rate \(\gamma_\text{max} \approx 0.2\) for the parameters of this simulation. The strong radial transport caused by the ITG eddies drives a zonal temperature perturbation that opposes the equilibrium temperature gradient, thus flattening it. This causes a "quasilinear" saturation of the ITG instability around \(t = 50\) when the zonal temperature gradient reaches the level of the equilibrium gradient and the instability is largely quenched (see also Appendix~\ref{sect_zonaltempsat}). Afterwards, \(\dw{\phinorm}\) continues to grow exponentially, albeit at a much lower growth rate of \(\gamma_3^\text{quenched} \approx 0.003\). The nonzonal temperature \(\dw{\deltaT}\) does not grow during this quenched-growth phase and remains about an order of magnitude lower than the electrostatic potential. As we discussed in Appendix~\ref{appendix_khtertiary}, the flattening of the temperature gradient (\(\vt \approx \px \zf{\deltaT}\)) and the comparatively low levels of temperature perturbations (\(\dw{\deltaT} \ll \dw{\phinorm}\)) imply that \(\dw{\phinorm}\) satisfies the Hasegawa-Mima tertiary equation \eqref{eq_khtertiary}. We find that the Rayleigh-Kuo criterion \(\px^3 \zf{\phinorm} \gtrsim 1\) is satisfied at the ZF minima and so there is an unstable KH mode. Its growth rate (\(\gamma_3^\text{HM} \approx 0.008\)) and poloidal spectrum are found to be similar to those of the quenched tertiary following the temperature-profile flattening (see Figure~\ref{fig_convection_zones_spectra_comparison}). We have found that if we do not evolve the zonal temperature and, thus, do not allow a flattening of the temperature gradient, the initial fast growth is never quenched. 

In the vicinity of the ZF minima, we find \(\px^3 \zf{\phinorm} \gtrsim 1\) and \eqref{eq_itgtert} predicts no ITG mode there. In fact, we do find an ITG instability (the fast initial one in Figure~\ref{fig_zfextrema_phiandtgrad}) peaked exactly at \(x_0\) where \(\px^2 \zf{\phinorm}(x_0) = 0\), even though \(\px^3 \zf{\phinorm}(x_0) > 1\). We have verified that this is an ITG instability and not a KH one by setting \(\vt = 0\) or by freezing the nonzonal temperature perturbations and solving the corresponding Hasegawa-Mima equation numerically. Either of these reduces the tertiary growth rate by an order of magnitude, down to the level of the KH instability, indicating that the fast mode is indeed an ITG instability. Even artificially scaling the ZFs in order to get \(\px^3 \zf{\phinorm} > 2\) in the ZF minima results in a fast ITG instability there. This tertiary mode is not the one described in Appendix~\ref{appendix_itgtert} and familiar from the work of \citet{rogersdorland2000}. A further investigation of it is left for future work. Its effect in the quasi-stationary zonal-staircase-dominated state is to push the system in the vicinity of the ZF minima towards a state in which the zonal-temperature perturbations cancel the equilibrium temperature gradient (see Appendix~\ref{sect_zonaltempsat}).

\subsubsection{Tertiary Instability at ZF Maxima}
\label{appendix_zfmaxima}
\begin{figure}
	\centering
	\includegraphics[scale=0.27]{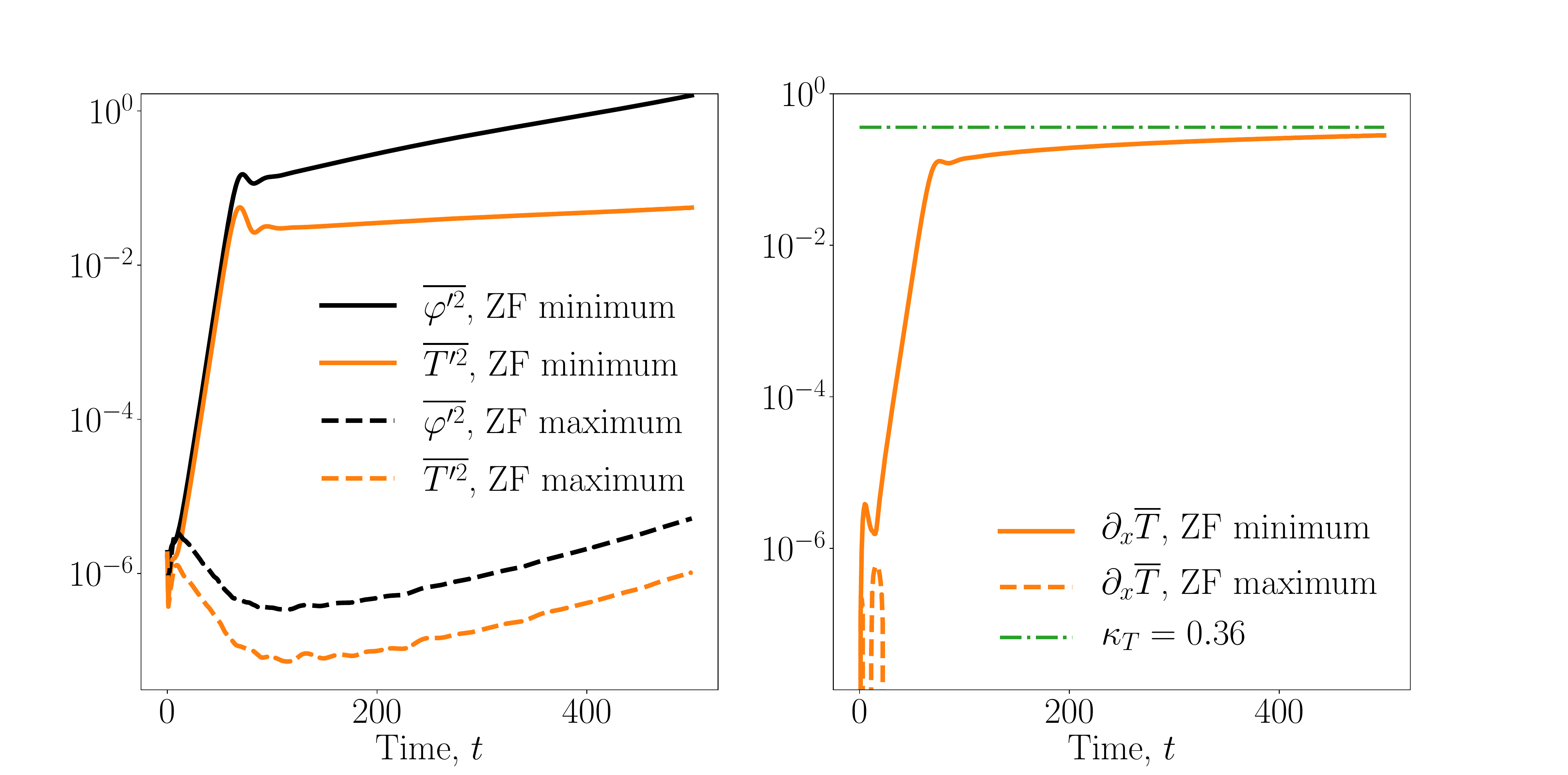}
	\caption{ Time evolution of the perturbations in the ZF extrema of the zonal staircase of SimH. \textbf{Left:} \(\zf{\dw{\phinorm}^2}\) (solid black) and \(\zf{\dw{\deltaT}^2}\) (solid orange) at the ZF minimum at \(x \approx 24.5\). Dashed lines show the same quantities at the ZF maximum at \(x \approx 47\). See Figure~\ref{fig_staircase_profiles} (top panel) for the ZF profile. \textbf{Right:} Zonal temperature gradient at the ZF minimum (the ZF maximum has a very low and negative zonal temperature gradient, not shown). After the initial fast instability, the zonal temperature gradient settles at \(\px \zf{\deltaT} \approx \vt = 0.36\). }
	\label{fig_zfextrema_phiandtgrad}
\end{figure}

\begin{figure}
	\centering
	\includegraphics[scale=0.47]{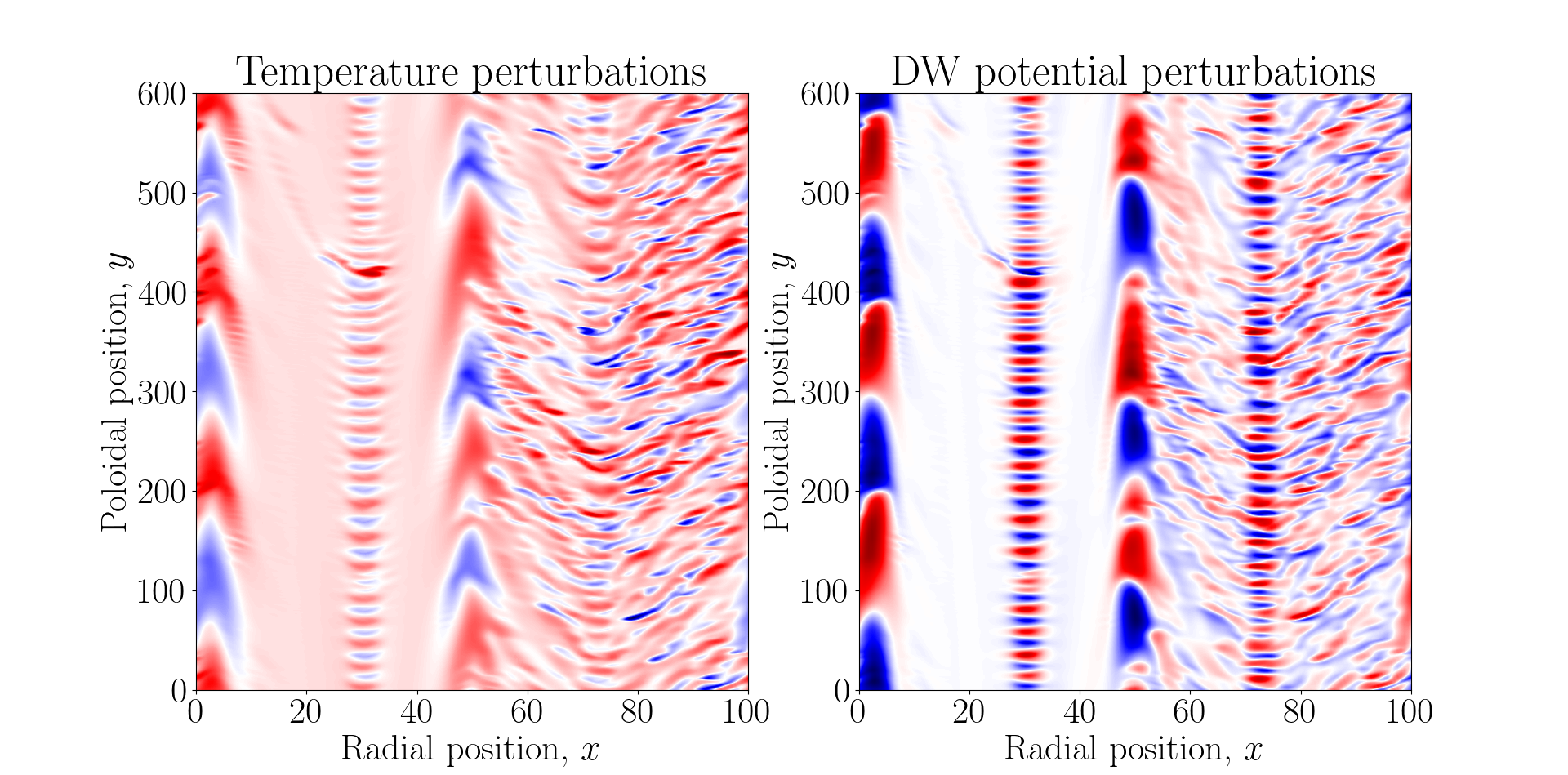}
	\caption{ Snapshots of \(\deltaT\) and \(\phinorm\) perturbations for SimH parameters, viz., \(\vt = 0.36\), \(\chi = 0.1\), but for \(L_y = 600\). Evident is a discrepancy in the poloidal wavenumbers of the nonzonal perturbations at the ZF maxima (\(x \approx 5, 50\)) and ZF minima (\(x \approx 30, 75\)). The dominant wavenumber at the ZF minima is \(k_y \approx 0.25\), just as in the shorter box (see Figure~\ref{fig_convection_zones_spectra}), and \(k_y \approx 0.03\) at the ZF maxima. The poloidal wavelength at the ZF maxima is shorter than the poloidal size of the box. }
	\label{fig_supertallbox}
\end{figure}

In contrast to the ZF minima, the ZF maxima are hosts to a wimpy ITG instability with a growth rate an order of magnitude smaller than that near the ZF minima: \mbox{\(\gamma_3^\text{wimpy} = 0.006\)} (see Figure~\ref{fig_zfextrema_phiandtgrad}, left panel). Setting \(\vt = 0\) eliminates this instability, confirming that it is, indeed, an ITG instability. The ZF satisfies \(\px^3 \zf{\phinorm} < 0\) at the ZF maxima, so there is no unstable KH mode there; this is confirmed numerically. The poloidal wavelength of the turbulence at the ZF maxima is significantly longer than that at the ZF minima. In fact, in SimL and SimH, it is determined by the poloidal box size \(L_y\). By increasing \(L_y\), we are able to obtain a saturated state with a well-defined wavelength at the ZF maxima that is smaller than the poloidal box size: see Figure~\ref{fig_supertallbox}.

The suppression of the ITG instability at the ZF maxima might be due to the localisation width \(\Delta\), na\"ively given by \eqref{eq_deltatert}. Using the observed values of \(\px^3\zf{\phinorm}\) and \(\px\zf{\deltaT}\) in \eqref{eq_deltatert}, we find \(\Delta \approx 1.8\) for the fastest mode of \eqref{eq_itgtert}. This suggests that these ITG modes should have radial wavenumbers of the order of \(k_x \sim 2 \pi / \Delta \approx 3.5\), which are deep in the stable region for the parameters considered (see Figure~\ref{fig_linear}, right panel). We leave the detailed investigation of these modes for future work. Their effect in the quasi-stationary zonal-staircase-dominated state is to seed turbulence in the shear zones through the emission of travelling structures (ferdinons) when the zonal shear in the shear zones has been sufficiently weakened by viscosity (see Sections~\ref{sect_zfdecay} and \ref{sect_ferds}). 

\subsection{Zonal Temperature Saturation and Equilibrium Gradient Flattening}
\label{sect_zonaltempsat}

Is it inevitable that the temperature profile is flattened in the regions of vigorous instability? Intuitively, one might expect this to be a consequence of the high level of radial transport by the nonzonal (\(\dw{\phinorm}\)) eddies. In that case, the flattening ought to depend on some measure of the strength of the radial \exb velocity. Let us translate this intuition into equations. 

\begin{figure}
	\centering
	\includegraphics[scale=0.27]{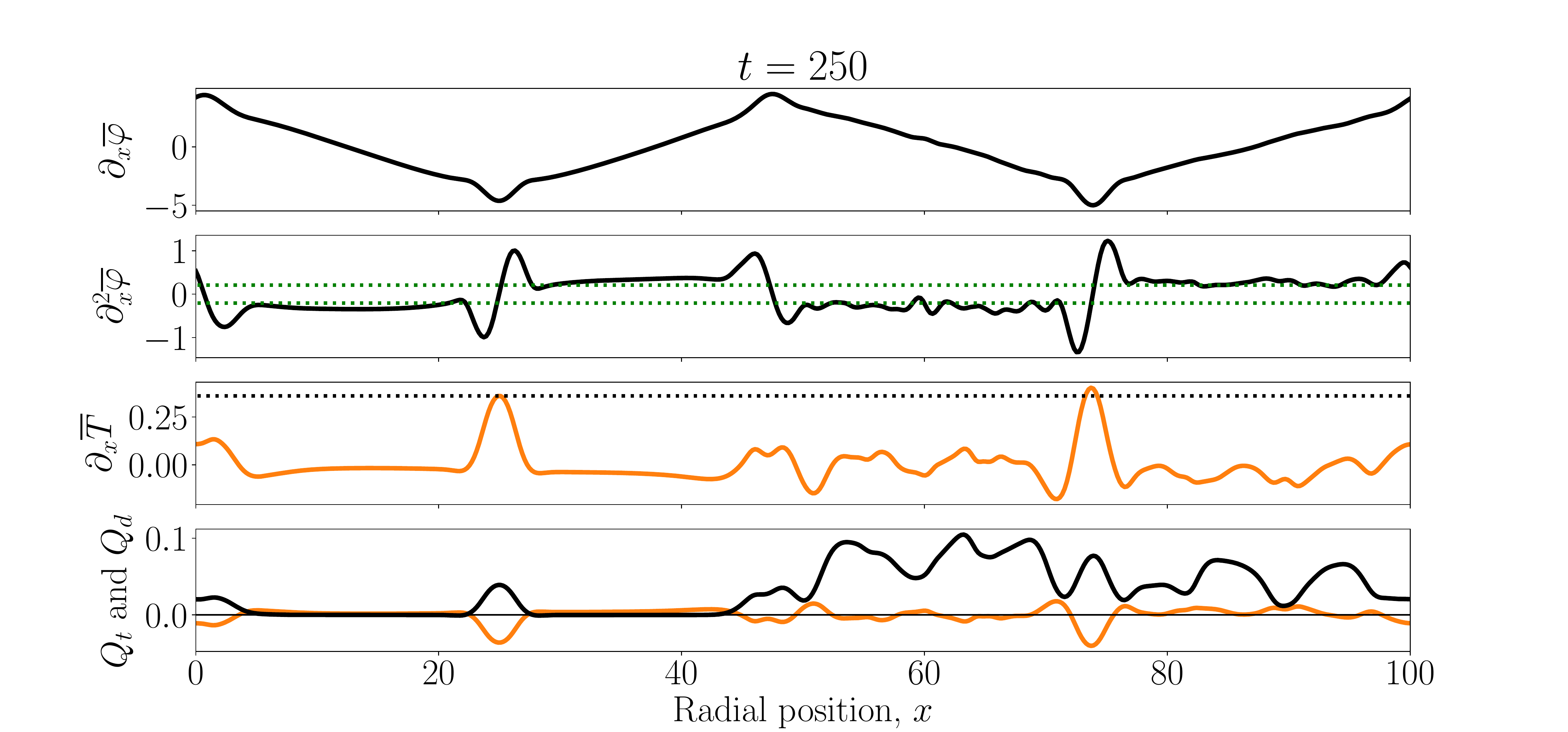}
	\caption{ Same as the bottom panel of Figure~\ref{fig_staircase_profiles}, but with plots of the turbulent \(Q_t =  -\zf{\deltaT \py \phinorm}\) (black) and diffusive \(Q_d = -\chi \px \zf{\deltaT}\) (orange) heat fluxes as well. We can see that the balance \eqref{eq_heatflux_integrationconst} between \(Q_t\) and \(Q_d\) holds in the saturated staircase (left half of the domain), but does not in the turbulent regions (right half of the domain), where the turbulence generates a significant turbulent heat flux. }
	\label{fig_zf_quasistatic_burst}
\end{figure}

The zonal part of \eqref{curvy_psi} is
\begin{align}
\label{eq_zonaltemp}
&\pt \zf{\deltaT} = \px  \zf{\left( \deltaT\py \phinorm + \chi \px \deltaT\right)}= -\px \left(Q_t + Q_d\right),
\end{align}
where \(Q_t =  -\zf{\deltaT \py \phinorm}\) and \(Q_d = -\chi \px \zf{\deltaT}\) are the local turbulent and diffusive radial heat fluxes, respectively. This is a conservation equation for the zonal temperature perturbations. The fact that the zonal temperature stays constant \(\pt \zf{\deltaT} = 0\) during the quiescent periods implies a balance of the local radial heat fluxes:
\begin{equation}
\label{eq_heatflux_integrationconst}
\px \left(Q_t + Q_d\right) = 0 \ \implies Q_t = -Q_d + Q \ \implies \px \zf{\deltaT} = \frac{Q_t - Q}{\chi},
\end{equation}
where the integration constant \(Q\) is the total box-averaged radial turbulent heat flux defined by \eqref{eq_heatflux_def}. Indeed, this balance is observed in the quiescent zonal staircase (see Figure~\ref{fig_zf_quasistatic_burst}).

\begin{figure}
	\centering
	\includegraphics[scale=0.27]{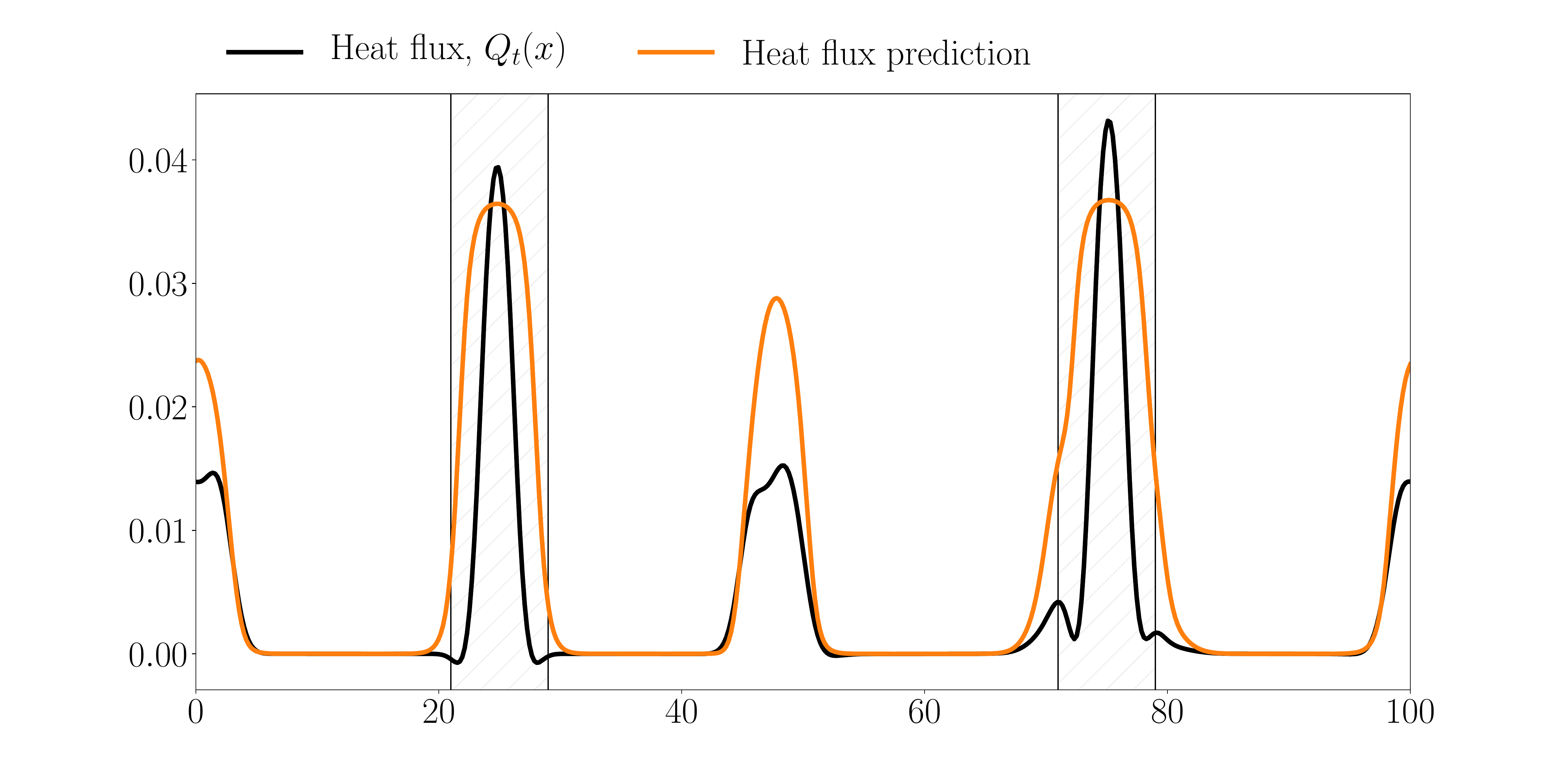}
	\includegraphics[scale=0.27]{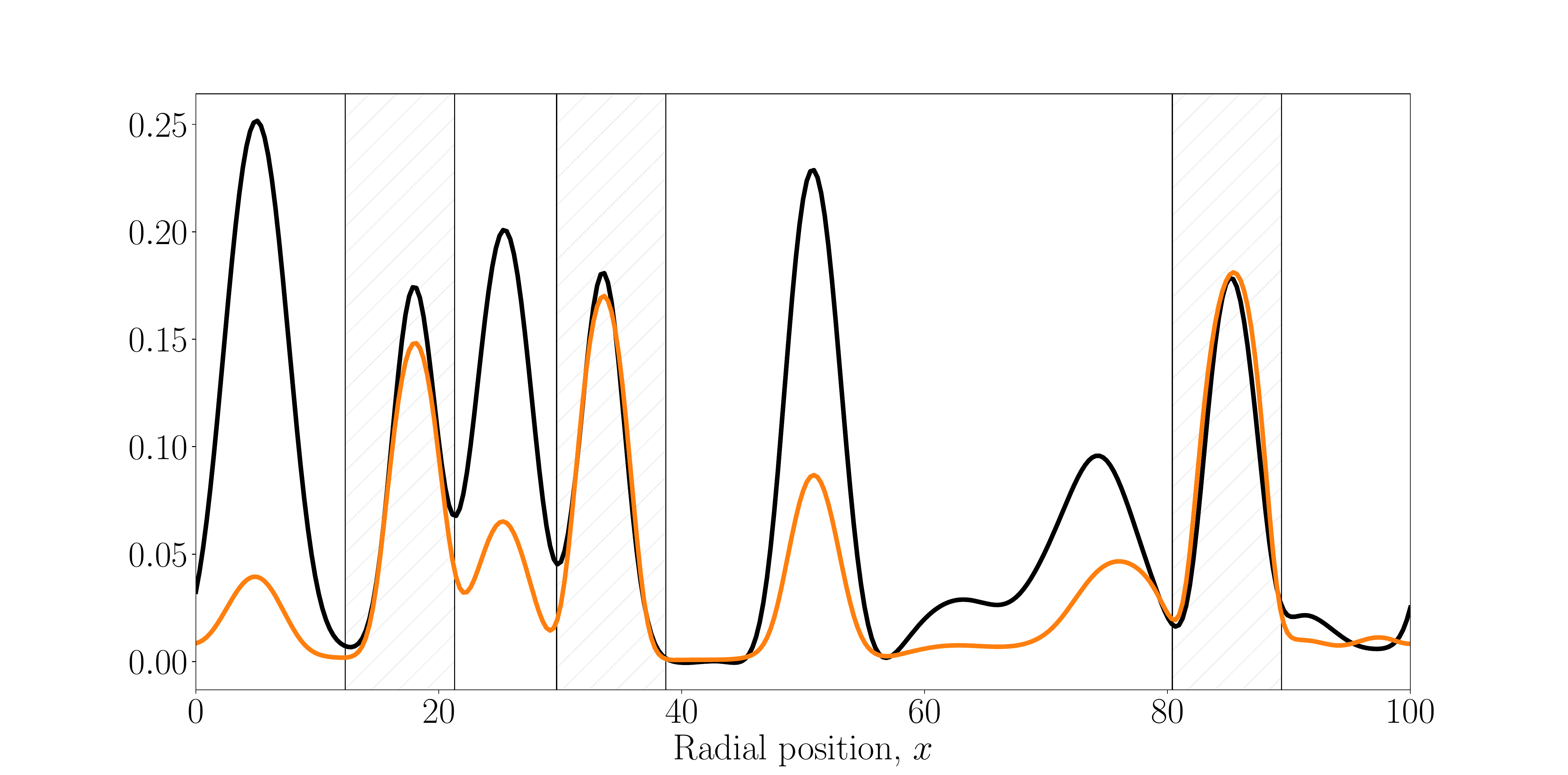}
	\caption{Comparison of the estimate \eqref{eq_hf_prediction} (orange) and observed turbulent heat flux (black) of the saturated zonal staircase. To choose \(k_\perp\), we estimated \(k_x\) and \(k_y\) using the observed radial width \(\delta\) of the convection zones and the poloidal spectrum of the turbulence there. The shaded areas highlight the ZF minima and have a width \(\delta\). \textbf{Top:} Data from SimH with \(\delta = 8\) (so \(k_x = 2\pi / 8 \approx 0.8\)) and \(k_y = 0.25\), corresponding to the spectral peak in Figure~\ref{fig_convection_zones_spectra}. This gives \(k_\perp \approx 0.8\). \textbf{Bottom:} Same as SimH, but with increased collisionality \(\chi = 1\). We estimated \(\delta = 9\) (so \(k_x \approx 0.7\)) and \(k_y = 0.25\), corresponding to the peak of the poloidal spectrum at the ZF minima in that simulation. This gives \(k_\perp \approx 0.7\). The agreement is better for the higher value of collisionality.}
	\label{fig_hf_prediction_chi01_z06}
\end{figure}

Consider the steady state (\(\pt = 0\)) solution of \eqref{eq_curvy_psi_zf}. Since we are interested in the behaviour of the temperature perturbations in the ZF extrema, where the zonal shear vanishes, we drop the effect of zonal shear (formally speaking, we transform to a frame moving with the local ZF velocity). We find
\begin{align}
\label{eq_dwtemp_tertiary}
&0 = \pt \dw{\deltaT} = \chi \nabla^2 \dw{\deltaT} - (\vt - \px \zf{\deltaT}) \py \dw{\phinorm}.
\end{align}
Substituting \eqref{eq_heatflux_integrationconst} into \eqref{eq_dwtemp_tertiary} gives us
\begin{equation}
\label{eq_dwtemp_subst}
(\chi \vt - Q_t + Q) \py \dw{\phinorm} = \chi^2 \nabla^2 \dw{\deltaT}.
\end{equation}
Multiplying both sides of \eqref{eq_dwtemp_subst} by the radial \exb velocity \(u_x = -\py \dw{\phinorm}\) and performing a poloidal average yields
\begin{equation}
(\chi \vt - Q_t + Q) \zf{u_x^2} = \chi^2 \zf{\left(\frac{\nabla^2 \dw{\deltaT}}{\dw{\deltaT}}\right)\dw{\deltaT}\py \dw{\phinorm}} \approx \chi^2 Q_t k_\perp^2,
\end{equation}
where we have approximated the thermal diffusion as that of a locally monochromatic temperature perturbation \(\dw{\deltaT}\) with a wavenumber \(k_\perp\). Then
\begin{equation}
\label{eq_hf_prediction}
Q_t = \frac{Q + \chi\vt}{1 + \chi^2k_\perp^2 / \zf{u_x^2}}.
\end{equation}
Substituting \eqref{eq_hf_prediction} into the rightmost expression in \eqref{eq_heatflux_integrationconst} yields the following prediction for the zonal temperature gradient
\begin{equation}
\label{eq_zonaltemp_prediction}
\px \zf{T} = \frac{\vt - Q\chi k_\perp^2 / \zf{u_x^2}}{1 + \chi^2k_\perp^2 / \zf{u_x^2}}.
\end{equation}
Therefore, for \(Q > 0\), steady-state saturation implies \(\px \zf{T} < \vt\). In the case of strong radial transport, \(\zf{u_x^2} \gg \chi^2k_\perp^2\), we obtain
\begin{equation}
Q_t \approx Q + \chi \vt, \quad \px \zf{\deltaT} \approx \vt.
\end{equation}
Thus, perturbations with a strong radial \exb velocity flatten out the temperature gradient by generating a strong zonal temperature perturbation, as intuitively expected. Indeed, \eqref{eq_hf_prediction} agrees with the numerical data for the heat flux at the ZF minima (see Figure~\ref{fig_hf_prediction_chi01_z06}), where we find the fast ITG tertiary instability that saturated via flattening of the equilibrium temperature gradient, as discussed in Appendix~\ref{appendix_zfminima}. 



\clearpage

\bibliographystyle{jpp}

\bibliography{bib}

\end{document}